\documentclass{aa}  

\usepackage{graphicx}
\usepackage{txfonts}

\usepackage[pdfpagelabels=false]{hyperref}	
\hypersetup{colorlinks=true,linkcolor=blue,citecolor=blue,filecolor=blue,urlcolor=blue,}
\usepackage{longtable}
\usepackage{lscape}
\usepackage{color}
\newcommand{\Msun}{{M}_{\odot}}

\begin{document}

   \title{Trading oxygen for iron I: \\ the [O/Fe] -- specific star formation rate relation of  galaxies}

\titlerunning{[O/Fe] -- specific star formation rate relation}

   \author{M. Chru{\'s}li{\'n}ska
          \inst{1},
R. Pakmor
          \inst{1},
J. Matthee \inst{2},     
T. Matsuno \inst{3}
          }

   \institute{Max Planck Institute for Astrophysics, Karl-Schwarzschild-Str. 1, D-85748 Garching, Germany\\
  \email{mchruslinska@mpa-garching.mpg.de}\\
         \and Department of Physics, ETH Z{\"u}rich, Wolfgang-Pauli-Strasse 27, Z{\"u}rich, 8093, Switzerland \\
\and Kapteyn Astronomical Institute, University of Groningen, Landleven 12, 9747 AD Groningen, The Netherlands              
             }

   \date{Received July 29, 2023; accepted XX XX, 2023}

 
  \abstract{
  Our current knowledge of star-forming metallicity of galaxies relies primarily on gas-phase oxygen abundance measurements. However, this may not allow one to accurately describe differences in stellar evolution and feedback, that are driven by variations in iron abundance. $\alpha$-elements (such as oxygen) and iron are produced  by sources that operate on different timescales and the link between them is not straightforward.\\
We explore the origin of the [O/Fe] - specific SFR (sSFR) relation, linking chemical abundances to galaxy formation timescales. This relation is followed by star-forming galaxies across redshifts according to cosmological simulations and basic theoretical expectations. Its apparent universality makes it suitable for trading the readily available oxygen for iron abundance. We show that the relation is determined by the relative iron production efficiency of core-collapse and type Ia supernovae and the delay time distribution of the latter - uncertain factors that could be constrained empirically with the [O/Fe] - sSFR relation.\\
We compile and homogenise a literature sample of star-forming galaxies with observational iron abundance determinations to place first constraints on the [O/Fe] - sSFR relation over a wide range of sSFR. The relation shows a clear evolution towards lower [O/Fe] with decreasing sSFR and a flattening above log$_{10}$(sSFR/yr)>-9. The result is broadly consistent with expectations, but better constraints are needed to inform the models. We independently derive the relation from old Milky Way stars and find a remarkable agreement between the two, as long as the recombination-line absolute oxygen abundance scale is used in conjunction with stellar metallicity measurements.
  }

   \keywords{
    Stars: abundances, formation - Supernovae: general - Galaxies: abundances, evolution, star formation
    }

   \maketitle
%

\section{Introduction}

Oxygen is the most abundant metal in the Universe and it is relatively easily observable via strong optical emission lines \citep[][]{Tremonti04,KewleyEllison08,MaiolinoMannucci19,Kewley19}.
Those lines were used to determine gas-phase abundances of oxygen relative to hydrogen (12 + log$_{10}$(O/H)) for large samples of star forming galaxies up to z$\lesssim$3. 
They are now accessible at even higher redshifts with JWST \citep[e.g.][]{Jones20,Arellano-Cordova22,Curti23,Katz23,Nakajima23}.
\\
However, it is the iron abundance that drives the differences in lives and fates of massive stars at different metallicities and regulates their impact on the surroundings \citep[][]{Garcia21,Eldridge22,Chruslinska22,Vink22}.
Their evolution, final core and explosion properties are strongly affected by radiation-driven winds, which scale with the iron abundance due to its dominant role in setting the opacity in stellar atmospheres \citep[][]{Abbott82,Pauldrach93,Vink01,Kudritzki02,Vink05,Sander20}.
By removing mass and angular momentum from stellar binaries/multiples, stellar winds further affect the orbit and evolution of such systems.
Together with the effects of binary interactions, iron abundance has a decisive role in shaping the high energy part of the galaxy spectra (in particular, UV continuum with absorption and wind features, ionising radiation emitted by stellar population, e.g. \citealt{Leitherer14,StanwayEldridge18,Gotberg19,Gotberg20,Vink23}), which affects the HII region emission-line \citep[e.g.][]{Steidel16,Strom18} and star formation rate diagnostics \citep{Lee02,MadauDickinson14}.
The determination of iron abundances in star-forming material is much more challenging than that of oxygen, and is currently available for only a limited number of galaxies. We review the methods used to determine oxygen and iron abundances in star-forming material, and discuss the related challenges in Section \ref{sec: metallicity - techniques}.
\\
Those two elements are produced abundantly by sources that operate on different timescales and the link between them is not straightforward \citep[][]{MatteucciGreggio86,WheelerSnedenTruran89}. 
Oxygen is promptly released to the interstellar medium via core-collapse supernovae (CCSN).
Those come from massive star progenitors reaching core-collapse stage within a few to $\approx$50 Myr \citep[][]{Woosley02,Heger03,Janka07,Schneider21}.
Iron is also generously produced by type Ia supernovae (SN Ia).
While their exact formation scenario is a matter of ongoing debate, those events are linked to theromnuclear explosions involving carbon-oxygen white dwarf(s) and come with a broad range of delays with respect to star formation of at least $\sim$ 40 Myr (i.e. the minimum time required to form a white dwarf, \citealt{Greggio05,MaozMannucci12,WangHan12,MaozMannucciNelemans14,LivioMazzali18}).
As a consequence, in young and highly star-forming environments, iron production is expected to lag behind that of the oxygen.
Especially in such environments, the composition of the star forming material may considerably deviate from the conventional solar abundance pattern.
This is clear from the abundances recorded in old, metal-poor stars in the Milky Way and its satellites: their oxygen to iron ratios can exceed the reference solar value by more than 5 times \citep[e.g.][]{Gratton00,ZhangZhao05,Tolstoy09,Amarsi19}. 
Evidence of super-solar oxygen/$\alpha$-element to iron abundance ratios was also found in local  \citep[e.g.][]{Izotov06,Hernandez17,Izotov18,Kojima21,Gvozdenko22,Senchyna22} and high redshift star-forming galaxies \citep[e.g.][]{Steidel16,Strom18,Sanders20,Topping20,Cullen21,Strom22} and in stellar populations of elliptical galaxies that formed in the high redshift Universe \citep[e.g.][]{Thomas10,Conroy14}.
This implies that in certain environments the wind mass loss of hot, massive stars may be severely overestimated if oxygen abundance is used as a proxy of their iron-group metallicity. This in turn has important consequences for their expected evolution, explosion and compact object properties (therefore, also feedback and chemical enrichment), observable properties of the stellar population (galaxy spectra, ionising radiation and certain emission line ratios) and related transients (various types of supernovae, long gamma ray bursts, gravitational wave sources).
Therefore, using oxygen abundance as a proxy for iron abundance (after scaling to solar pattern, as commonly done in many areas of astronomy) can lead to important and as yet largely unaccounted for systematic errors. 
\\
\newline
Motivated by the need to establish a link between the readily available oxygen abundance and the essential, but typically unknown star-forming iron abundance that would be applicable across a wide range of galaxy properties, we explore the relation between the star forming [O/Fe]\footnote{[O/Fe]=log$_{10}$(O/Fe) - log$_{10}$(O/Fe)$_{\odot}$ is the logarithm of the oxygen to iron abundance ratio relative to the reference solar abundance ratio of the two elements.} ratio and the specific star formation rate (sSFR).
One can expect that the two quantities are strongly related: sSFR=$\frac{\rm SFR}{\rm M_{*}}$ is the ratio between the production rate of stars (a subset of which quickly explodes as CCSN) and the stellar mass accumulated over time (available for the continuous production of delayed SN Ia). Therefore, to first degree sSFR sets the ratio between the rate of CCSN and SN Ia happening at a given time, and so [O/Fe].
Tight [O/Fe]--sSFR relation has indeed been found in the EAGLE cosmological simulations \citep{MattheeSchaye18} and within the semi-analytic gas-regulated galaxy evolution model \citep{Kashino22}.
We further discuss the origin and the main factors that are expected to shape the relation in Section \ref{sec: expectations}. 
 In section \ref{sec: simulations} we complement this qualitative discussion with a simple analytical description that allows to reproduce the average [O/Fe] -- sSFR relation resulting from the EAGLE \citep{Schaye15} and Illustris-TNG cosmological simulations \citep{Pillepich18} and explain the origin of the differences between them.
\\
The [O/Fe]--sSFR relation can be seen as a star-forming analogue of the well-known relic stellar [$\alpha$/Fe] -- [Fe/H] relation \citep[][]{Tolstoy09,Amarsi19}, where both sSFR and [Fe/H] serve as some proxy for the galaxy's age.
 We exploit this connection and roughly reconstruct the Milky Way's [O/Fe]--sSFR relation from its old disk stars in Section \ref{sec: obs data, MW}.
 \\
In Section \ref{sec: obs data} we collect the available observational data characterising other star forming galaxies and select a subset of those that can be brought to a common baseline by accounting for systematic offsets (Section \ref{sec: baseline}). We use them to empirically derive the [O/Fe] -- specific star formation rate relation (Section \ref{sec: results}) and confront the result with theoretical expectations (Section \ref{sec: results vs models}).
In Section \ref{sec: discussion} we reflect on the prospects of obtaining further constraints. We discuss the potential use of the [O/Fe] -- sSFR relation to discriminate between the different theoretical SN Ia delay time distributions, to infer the uncertain minimum delay at which they contribute significantly to enrichment, and to constrain the average CCSN iron yields.
\\
Where relevant, we explicitly use either [X/H]=log$_{10}$(X/H)-log$_{10}$(X/H)$_{\odot}$ or 12 + log$_{10}$(X/H) notation to refer to the iron (X=Fe) or oxygen (X=O) abundance.
We use Z$_{\rm O}$ (Z$_{\rm Fe}$) to refer to oxygen (iron) abundance in a general sense. 
We assume solar reference abundances of 12+log$_{10}$(O/H)$_{\odot}$=8.83, 12+log$_{10}$(Fe/H)$_{\odot}$=7.5 and log$_{10}$(O/Fe)$_{\odot}$=1.33 dex from \cite{GrevesseSauval98} (GS98 hereafter).

\section{Theoretical expectations}\label{sec: expectations}

\begin{figure}[h!]
\centering
\includegraphics[width=0.52\textwidth]{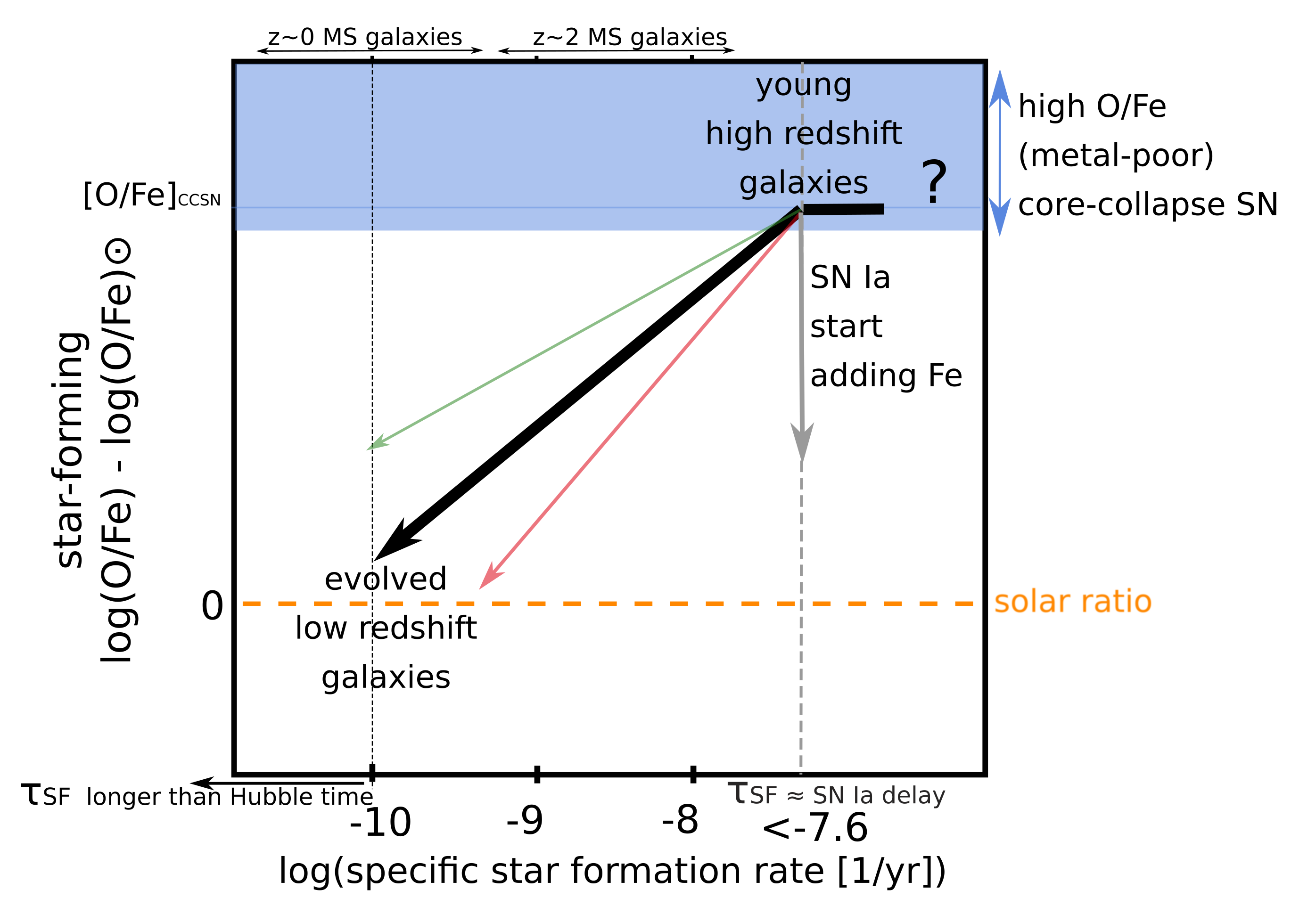}
\caption{ 
Schematic illustration of the expected characteristics of the star-forming [O/Fe] -- specific star formation rate relation of galaxies.
}
\label{fig: simple picture}
\end{figure}

\subsection{sSFR of typical galaxies at different redshifts and their locations along the relation}
The specific star formation rate, sSFR=$\frac{\rm SFR}{\rm M_{*}}$=$\frac{1}{\tau_{\rm SF}}$, defines a characteristic timescale $\tau_{\rm SF}$ on which a galaxy grows to its current stellar mass if that growth happens at its current SFR.
High sSFR values correspond to galaxies that are either young or forming stars more rapidly compared to their average star-forming activity in the past. 
For galaxies with regular star formation histories $\tau_{\rm SF}$ can serve as some proxy for the age of their stellar population (the age increases towards the left side of the Figure \ref{fig: simple picture}).
\\
The typical sSFR of star forming galaxies can be determined from the redshift-dependent star formation -- mass relation (SFMR, also called `main sequence' of galaxies, e.g. \citealt{Brinchmann04,Salim07,Speagle14,Popesso23}).
The range of sSFR of main sequence (MS) galaxies at $z\sim0$ and $z\sim2$ is roughly indicated at the top of Figure \ref{fig: simple picture}.
For a given stellar mass, the SFRs of galaxies are higher at higher redshifts: a $M_{*}\sim10^{10} \Msun$  main-sequence galaxy at redshift $z\gtrsim$3 has log$_{10}$(sSFR)$\sim$-8.5 and this value drops to log$_{10}$(sSFR)$\approx$-10 by $z$=0 \citep{Boogaard18,Popesso23}.
Therefore, the (upper) right corner of the Figure \ref{fig: simple picture} is expected to be occupied by early Universe galaxies.
Conversely, the typical low redshift galaxies are expected to occupy the lower left corner of the relation. 
If the slope of the star formation -- mass relation is close to unity, then all MS galaxies at a given redshift have similar sSFRs  and occupy similar locations on the [O/Fe]--sSFR plane.
In reality there is a $\sigma_{\rm SFR}\sim$0.3 dex scatter in SFR at fixed galaxy stellar mass around the SFRM and regular star-forming galaxies at any epoch always occupy a range of sSFR \citep[e.g.][]{MattheeSchaye19}.
The redshift evolution of the  normalisation of the SFMR is found to be relatively steep up to $z\lesssim$2-3, but much more gradual at higher redshifts \citep{Weinmann11}.
This means that the typical sSFR of MS galaxies does not increase much beyond this point if the current SFMR estimates are extrapolated to higher redshifts. 
Consequently, extremely high log$_{10}$(sSFR)$\gtrsim$-7.6 galaxies are either very low mass ($M_{*} < 10^{6} \Msun$ at $z\sim$3 -- the SFMR is essentially unconstrained at such low stellar masses even at low redshifts) and/or undergo a strong starburst phase.

\subsection{High sSFR: SN Ia free regime}

If the $\tau_{\rm SF}$ timescale is short compared to the typical delay of SN Ia ($\tau_{\rm SF} < \tau_{\rm Ia;min}$, top-right corner of the Figure \ref{fig: simple picture}), the enrichment is dominated by core-collapse supernovae.
CCSNe pollute the interstellar medium with material with a relatively high super-solar [O/Fe] ratio \citep[e.g.][]{Tominaga07,HegerWoosley10,Nomoto13,Sukhbold16,Grimmett18,LimongiChieffi18,Curtis_2019,Ebinger_2020}.
The most massive stars are the first to evolve and undergo core-collapse. They are expected to eject material at higher [O/Fe] ratio than the lower mass CCSN progenitors, which may already lead to some evolution in the [O/Fe]-sSFR plane. 
Note that this is an extremely brief phase: all (single) stars massive enough to give rise to CCSN ($\gtrsim$7-8 $\Msun$ at birth) are expected to explode or collapse within $<$40-50 Myr following their formation. 
Whether some `typical' [O/Fe] pattern is expected in this regime is unclear: $\tau_{\rm SF}<$40 Myr implies log$_{10}$(sSFR)$\gtrsim$-7.6. As discussed above, such sSFR are not found for typical main sequence galaxies.
The early enrichment is dominated by the most massive (rare), very  metal-poor stellar progenitors.
Such progenitors have been proposed to lead to a variety of explosions with very different properties from the regular CCSN (e.g. pair-instability supernovae \citealt{ElEid86,Langer07,Woosley17}) and may eject matter with a different [O/Fe] \citep[][]{HegerWoosley02,Nomoto05,HegerWoosley10,Grimmett18,Takahashi18}, possibly leading to a large scatter in the rightmost part of the [O/Fe] - sSFR relation.
\\
At slightly longer $\tau_{\rm SF}$ galaxy's [O/Fe] may approach some population-averaged CCSN [O/Fe]$_{\rm CCSN}$ ratio. 
This expectation is guided by the existence of a plateau found for the stellar [$\alpha$/Fe] -- [Fe/H] relation at low [Fe/H] for the Milky Way and local dwarf galaxies (where [Fe/H] serves as a proxy for the age of the galaxy playing an analogous role to sSFR in Figure \ref{fig: simple picture}, e.g. \citealt{Tolstoy09,Miglio18,Amarsi19}).
Theoretical CCSN yields (and hence the value of [O/Fe]$_{\rm CCSN}$) are very uncertain, especially for the metal-poor CCSNe progenitors.
Observational constraints on possible [O/Fe]$_{\rm CCSN}$ are therefore highly desirable.

\subsection{The onset of SN Ia and a possible turnover}
\begin{figure}[h!]
\centering
\includegraphics[scale=0.54]{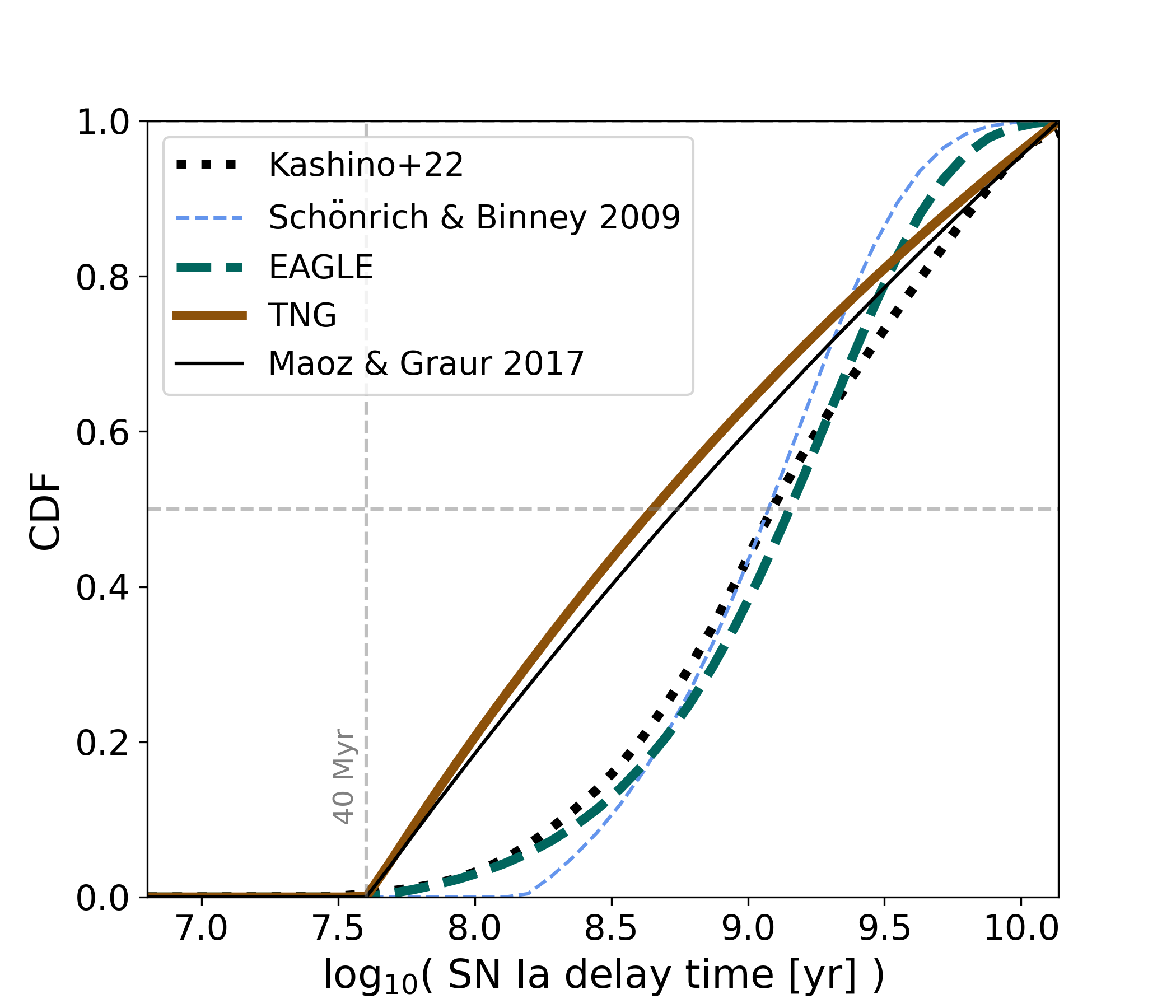}
\caption{ 
Cumulative distribution functions of the example SN Ia delay time distributions (DTD). The distributions were normalised to unity when integrated over the Hubble time.
Solid lines correspond to single power-law-shaped DTD: brown - Illustris-TNG cosmological simulations, black - fitted to observed volumetric SN Ia rate by \citet{MaozGraur17}.
Dashed lines correspond to exponential DTD: thick turquoise line - EAGLE cosmological simulations, thin blue line -  Galaxy chemical evolution model from \citet{SchonrichBinney09} with DTD parameters tuned to reproduce observed oxygen abundances.
Black dotted line corresponds to DTD used by \citet{Kashino22}, which follows the theoretical analytic formulation by \citet{Greggio05,Greggio10} allowing for a mixed contribution of different proposed SN Ia progenitors. Intersection with the horizontal dashed line indicates the median of each DTD. Vertical dashed line at 40 Myr marks the evolutionary timescale of an 8 $\Msun$ star (i.e. roughly the minimum time needed to form a white dwarf).
}
\label{fig: SN Ia DTD examples}
\end{figure}

SN Ia start to contribute to the chemical enrichment of the interstellar gas at times longer than $\tau_{\rm Ia;min}$. The minimum theoretically feasible delay is set by the evolutionary timescale of the most massive white dwarf progenitor (leading to $\tau_{\rm Ia;min}\gtrsim$40-50 Myr, or log$_{10}$(1/$\tau_{\rm Ia;min}$)$\lesssim$-7.6).
However, it could be longer than that depending on the assumed SN Ia formation scenario \citep[e.g.][]{Greggio10,MaozMannucciNelemans14}.
The mapping between the birth stellar mass and its final fate (i.e. white dwarf, neutron star/CCSN or a black hole), which sets the timescales in the above considerations, is not straightforward.
It is expected to depend on metallicity of the progenitor star and can be altered by binary interactions.
In particular, in presence of binary interactions a fraction of CCSN can originate from lower mass stars and happen with delays longer than 50 Myr \citep{Zapartas17}.
SN Ia are thought to eject material with high iron and negligible oxygen abundances compared to average CCSN yields \citep{Nomoto97,Iwamoto99,Lach20}. They act to reduce the galaxy's [O/Fe], possibly leading to a break/change in the slope of the [O/Fe]-sSFR relation at sSFR=1/$\tau_{\rm SF}$, corresponding to the timescale at which the SN Ia contribution becomes significant.
Figure \ref{fig: SN Ia DTD examples} compares some of the SN Ia delay time distributions (DTD) used in the literature, which predict considerably different SN Ia contributions at early times following the star formation.
The possible turnover point would be expected at $\tau_{\rm SF}$ relatively close to $\tau_{\rm Ia;min}$ in the case of the conventionally assumed power-law SN Ia DTD, but it could be considerably longer than $\tau_{\rm Ia;min}$ if the DTD has a different form.

\subsection{Intermediate and low sSFR}

At intermediate and low sSFRs, the relation between the [O/Fe] and sSFR is to first order governed by the relative rates of CCSN and SN Ia and their corresponding oxygen and iron yields - i.e. factors determined by stellar evolution and supernovae explosion properties. 
This is supported by the semi-analytical considerations from
\citealt[][]{Kashino22}, who show that within the gas-regulator galaxy evolution framework \citep{Lilly13} the [O/Fe] -- sSFR relation is independent of the model parameters related to large-scale processes in galaxy evolution (mass-loading factor, star formation efficiency) and is effectively determined by the abundance ratio produced by CCSN and SN Ia at any given time.
Note that such `instantaneous' [O/Fe] ratio might decrease faster than the abundance ratio in the star-forming material, where metals may accumulate and enter the star-forming phase with some delay rather than being reused immediately.
In general, the slope of the [O/Fe]-sSFR relation can also be affected by inflows of metal-poor (or high [O/Fe]) material and/or feedback processes removing some fraction of the enriched matter from the star forming material.
Metal retention/feedback can in principle depend on the galaxy mass \citep[e.g.][]{} and induce a secondary dependence of the [O/Fe]-sSFR relation on this property.
\\
In the simplest scenario where the population-averaged supernova metal yields and formation efficiencies are constant, other stellar sources contribute negligible O and Fe, potential inflowing material has zero metallicity, and star-forming [O/Fe] approaches the instantaneous production ratio, the [O/Fe]--sSFR relation can be readily derived from Equation 12 from \cite{Kashino22}.
This can be rewritten as follows:
\begin{multline}\label{eq: OFe}
\rm [O/Fe] = [O/Fe]_{CCSN} \\
\rm - log_{10}\left( 1 + \frac{m^{Ia}_{Fe}}{m^{CCSN}_{Fe}} \frac{N_{Ia0}}{k_{CCSN}} \frac{\int_{0}^{t} SFR(t`) f_{Ia}(t-t`) \,dt`}{SFR} \right)
\end{multline}
where $\rm [O/Fe]_{CCSN}$ is the average CCSN oxygen to iron abundance ratio relative to solar, $\rm m^{CCSN}_{Fe}$ and $\rm m^{Ia}_{Fe}$ is the average iron mass ejected per CCSN or SN Ia event, respectively, $\rm k_{CCSN}$ is the CCSN formation efficiency (i.e. number of CCSN formed per unit stellar mass formed),
$\rm N_{Ia0}$ is the SN Ia formation efficiency (i.e. the Hubble time-integrated number of SN Ia formed per unit stellar mass) and $\rm f_{Ia}$ describes the SN Ia delay time distribution (DTD) normalised to unity when integrated over the Hubble time.
The last term in the parenthesis in Equation \ref{eq: OFe}, i.e. the ratio of the time integral over the star formation history of the galaxy ($\propto M_{*}$) modulated by the SNIa DTD to its current SFR masks the dependence on sSFR.
\\
For convenience, we define
\begin{math}
C_{\rm Ia/CC} \coloneqq \rm \frac{m^{Ia}_{Fe}}{m^{CCSN}_{Fe}} \frac{N_{Ia0}}{k_{CCSN}}
\end{math}. For a fixed SN Ia DTD, the slope of the relation is only sensitive to the relative iron yield of SN Ia and CCSN.
Increasing the SN Ia formation efficiency or $\rm m^{Ia}_{Fe}$ would have the same effect as lowering $\rm k_{CCSN}$ or $\rm m^{CCSN}_{Fe}$ by the same amount and would act to steepen the relation.
Increasing the oxygen yield per CCSN (i.e. increasing $\rm [O/Fe]_{CCSN}$ at fixed $\rm m^{CCSN}_{Fe}$) would only change the overall normalisation of the relation without affecting its shape.
In reality all $\rm k_{CCSN}$, $\rm m^{CCSN}_{Fe}$ and especially $\rm [O/Fe]_{CCSN}$ may vary with birth metallicity of stellar progenitors.
With the example discussed in Section \ref{sec: simulations} we show that in such a case the [O/Fe]--sSFR relation can still be well described with fixed values of those parameters. However, their interpretation is much less straightforward.
No clear dependence of SN Ia formation or metal yields on metallicity is predicted by current models (but see \citealt{Cooper09,Toonen12}). However, in principle such a dependence could be induced by environmental variations of the stellar IMF \citep{WeidnerKroupa05,Chruslinska21}.
\\
Finally, the shape of the [O/Fe] -- sSFR relation is sensitive to the SN Ia delay time distribution.
From Figure \ref{fig: SN Ia DTD examples} it is clear that the DTD used in the literature predict very different time evolution of the SN Ia rate.
The power-law DTD (TNG and \citealt{MaozGraur17} examples shown in Figure  \ref{fig: SN Ia DTD examples}) allows for a broader range of delay times at which SN Ia contribute compared to more concentrated exponential DTD (as in the EAGLE and \cite{SchonrichBinney09} examples shown in Figure  \ref{fig: SN Ia DTD examples}). In the latter cases SN Ia start contributing at longer times following the star formation, but their contribution is rising more steeply. With such a DTD the expected turnover/change of slope in the [O/Fe]-sSFR relation would happen at lower sSFR but [O/Fe] would then decrease more steeply than in the power-law DTD scenario.
This is evident in the Figure \ref{fig: simulations}, where we compare the [O/Fe]-sSFR relations resulting from the Illustris-TNG and EAGLE simulations. We further discuss those examples in Section \ref{sec: simulations}.

\subsection{Examples from the cosmological simulations} \label{sec: simulations}

\begin{figure}
\centering
\includegraphics[scale=0.5]{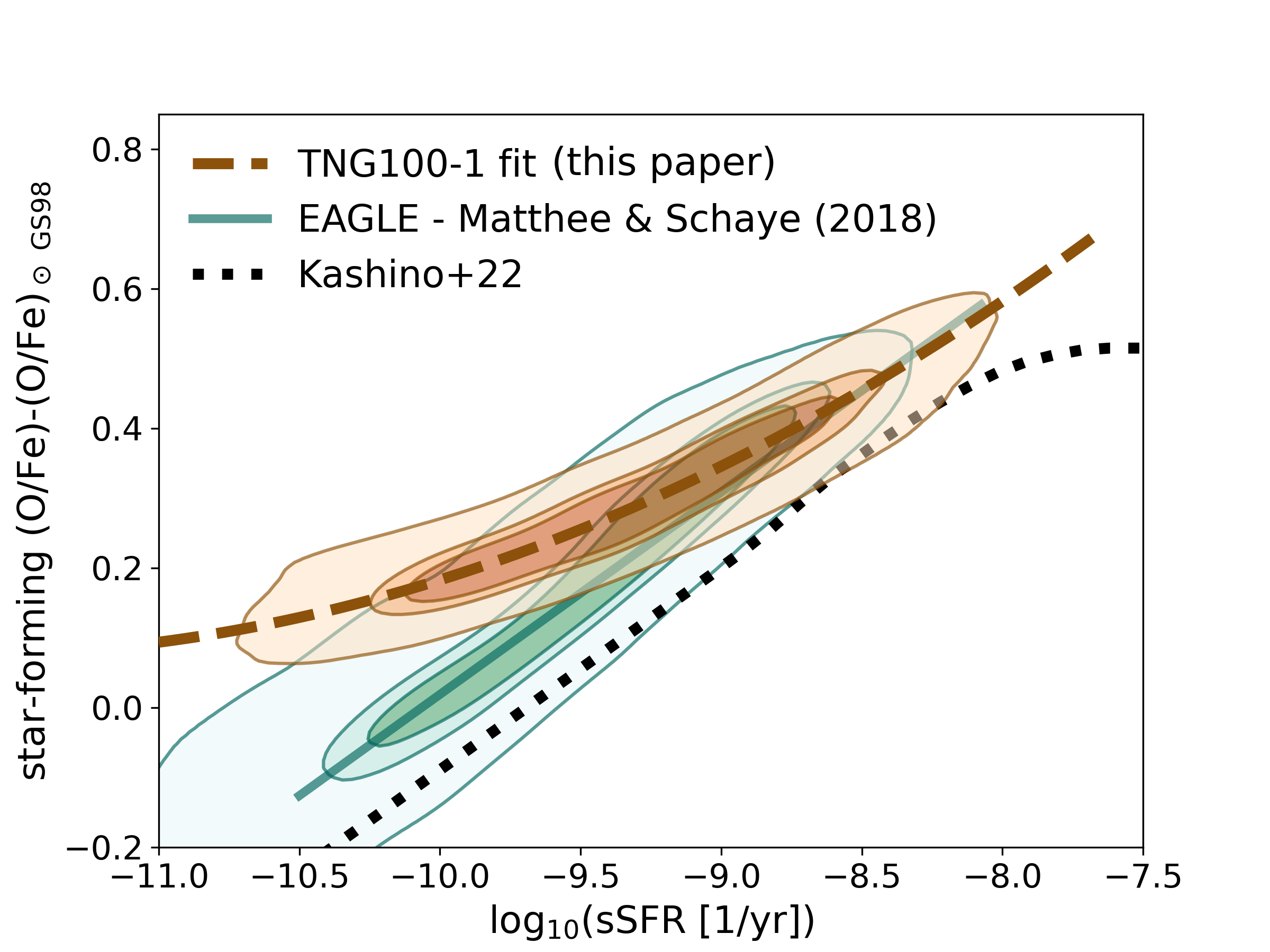}
\caption{ 
Star-forming [O/Fe] versus specific star formation rate relation for the simulated galaxies from the EAGLE cosmological simulations (turquoise contours), Illustris-TNG 100 cosmological simulations  (brown contours). The contours enclose 50, 68 and 95 \% of all central galaxies with log$_{10}$(M$_{*}$/$\Msun$)=9 - 10.5 at redshifts between 0 and 8.
Solid turquoise line shows the fit to EAGLE relation from \cite{MattheeSchaye18}, while the dashed brown line shows our fit to Illustris-TNG 100 relation.
Black dotted line shows the relation obtained by \cite{Kashino22} for `main-sequence' galaxies modeled within the gas-regulator framework.
}
\label{fig: simulations}
\end{figure}

The existence of a tight [O/Fe] -- sSFR relation has been previously shown by \cite{MattheeSchaye18} with the use of the EAGLE cosmological simulations \citep{Crain15,Schaye15,McAlpine16} and by
\cite{Kashino22} within the gas-regulator galaxy evolution semi-analytical model \citep{Lilly13}.
We show the corresponding relations, along with the results extracted from the Illustris-TNG cosmological simulations \citep{Pillepich18,Nelson19_TNG} in Figure \ref{fig: simulations}.
We plot several density contours indicating the locations of the EAGLE (using the Ref-L0100N1504 run) and Illustris TNG (using the TNG100-1 run) galaxies selected from the full simulation snapshots between redshifts 0 and 8. We select galaxies following the same criteria as \cite{MattheeSchaye18} (central star-forming galaxies in the mass range log$_{10}$(M$_{*}$/$\Msun$)=9-10.5).
\cite{MattheeSchaye18} conclude that the simulated galaxies follow a fundamental plane linking SFR, M$_{*}$ and [O/Fe] that is well described by the following fit at least up to redshift of 2:
\begin{math}
   \rm [O/Fe] =  -0.282 \ log_{10}(M_{*}/10^{10} \Msun)+ 0.29 \ log_{10}(SFR/\Msun \  yr) +0.023
\end{math}
where [O/Fe] assumes GS98 reference solar scale and the constant value was adjusted accordingly \footnote{\cite{MattheeSchaye18} fit a relation that separates the mass and SFR dependencies but the difference with respect to using only sSFR is not significant.
This is reflected in the similarity of the fitted $M_{*}$ and SFR coefficients. The average EAGLE relation can effectively be described by [O/Fe] = 0.29 log$_{10}$(sSFR) + const.
}.
The corresponding relation for log$_{10}$M$_{*}$=9.75  is shown as a thick solid turquoise line in Figure \ref{fig: simulations}.
The fit to TNG-100 galaxies shown in Figure \ref{fig: simulations} is described by:
\begin{math}
 \rm [O/Fe] = 0.0368 \ log_{10}(sSFR)^{2} + 0.8616 \ log_{10}(sSFR) + 5.1213
\end{math}
\\
The overall shape of the relation followed by the EAGLE galaxies at log$_{10}$(sSFR)$\lesssim$-9 is very similar to the one found by \cite{Kashino22}, but significantly differs from the relation followed by the Illustris TNG-100 galaxies.
\\
\begin{figure*}[h!]
\centering
\includegraphics[scale=0.4]{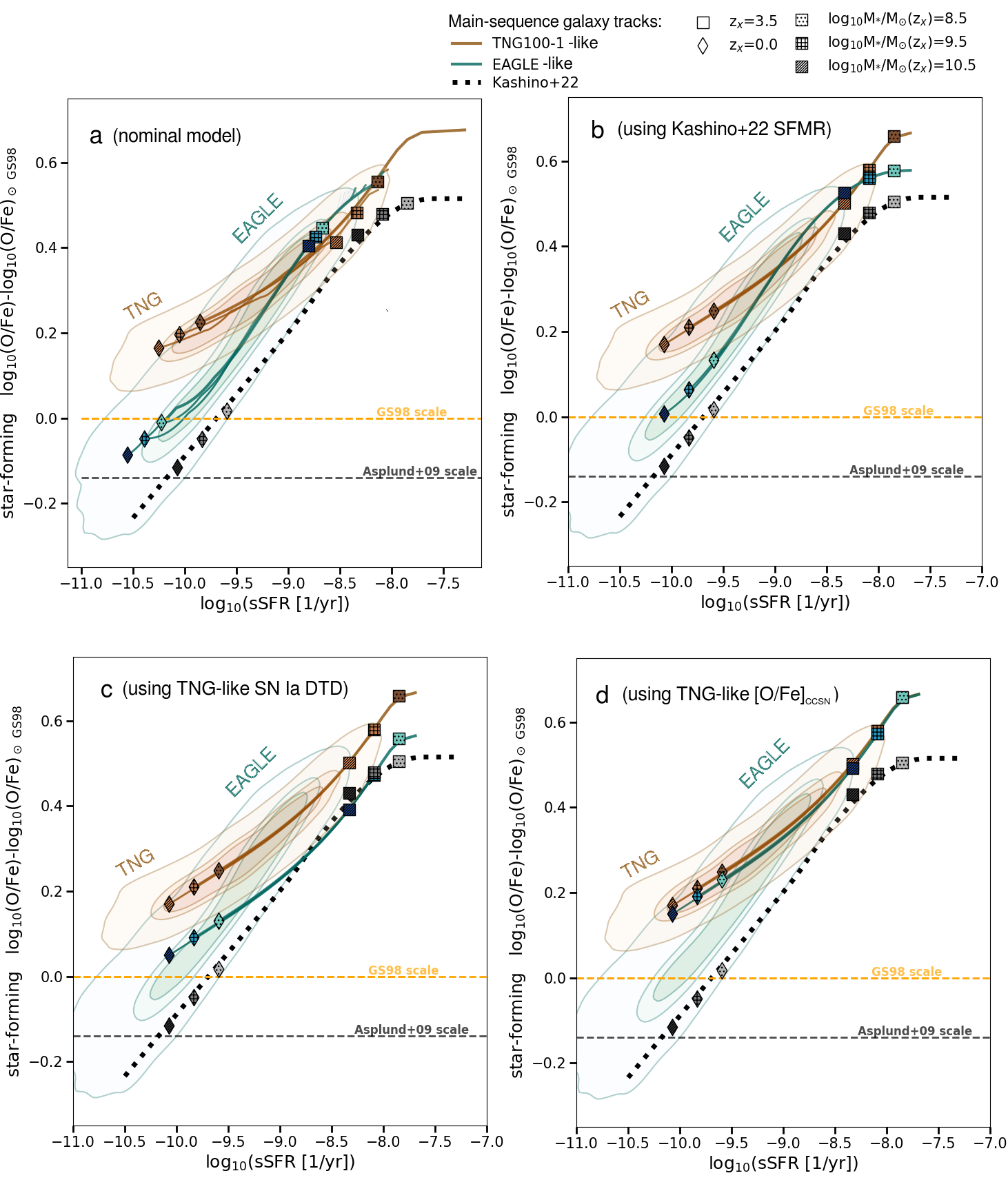}
\caption{ 
Each panel shows the star-forming [O/Fe] -- log$_{10}$(sSFR) relation for the simulated galaxies from the EAGLE cosmological simulations (turquoise contours), Illustris-TNG 100 cosmological simulations  (brown contours) and obtained by \citet{Kashino22} (black dotted line). The contours enclose 50, 68 and 95 \% of all central galaxies with log$_{10}$(M$_{*}$/$\Msun$)=9 - 10.5 at redshifts between 0 and 8.
Thick solid lines show evolutionary tracks of MS galaxies along the relation calculated with equation \ref{eq: OFe} for different parameter choices. Squares/diamonds show the locations of galaxies of different masses (indicated by the symbol shading) at redshift 3.5/0.
The tracks in panel a) were calculated using the SFMR, DTD, m$^{\rm Ia}_{\rm Fe}$, k$_{\rm CCSN}$ and N$_{\rm Ia0}$ from the EAGLE (turquoise) or TNG (brown) simulations (see appendix \ref{sec, app: model}).
 The remaining parameters are chosen to
 match the relation resulting from the EAGLE or TNG simulations, respectively.
In panel b) all tracks were caluclated with the same SFMR as in \citet{Kashino22}. In panel c) we additionally change the DTD for turquoise tracks to the one used in the TNG simulations. In panel d) we further change [O/Fe]$_{\rm CCSN}$ used to calculate the turquoise tracks to be the same as used to calculate the brown tracks.
}
\label{fig: mock simulations}
\end{figure*}
As anticipated in the previous section, those differences stem mostly from different assumptions about the SN Ia DTD (shown in Figure \ref{fig: SN Ia DTD examples}).
To illustrate this we consider mock 'main sequence' galaxies, i.e. galaxies that follow the evolving SFMR through cosmic time. 
Similarly to \cite{Kashino22}, we use this to determine average galaxy star formation histories and calculate their evolutionary tracks in the [O/Fe]-sSFR plane following Equation \ref{eq: OFe}. 
In panel a) in Figure \ref{fig: mock simulations} we show the result of this procedure when we use k$_{\rm CCSN}$ and N$_{\rm Ia0}$, m$^{\rm Ia}_{\rm Fe}$, SN Ia DTD and SFMR as in the EAGLE and TNG100-1 simulations (see appendix \ref{sec, app: model} for more details).
The CCSN metal yields (and therefore m$^{\rm CCSN}_{\rm Fe}$ and [O/Fe]$_{\rm CCSN}$) used in the simulations vary with metallicity of the stellar population.
We therefore cannot extract a single value for those parameters from the simulation settings. Instead, we choose their values in a way that allows to match the locations of galaxies from the corresponding simulations (indicated by density contours in Figure \ref{fig: mock simulations}).
It can be seen that the corresponding evolutionary tracks of our mock main-sequence galaxies reproduce the [O/Fe]--sSFR relations resulting from the simulairontions extremely well, despite the very simplistic assumptions that we made.
This conclusion is not affected by the choice of the SFMR, which only shifts the locations of galaxies of different masses along the relation. This can be seen in panel b, where in all cases we use the same SFMR as \cite{Kashino22} (see Section 5.2.1 and Figure 15 therein).
In panel c) we show that when we additionally change the SN Ia DTD to the one used in the TNG simulations, the mock galaxies previously tracing EAGLE-like relation start to follow the [O/Fe]--sSFR relation which resembles the one from the TNG simulations.
If we further change [O/Fe]$_{\rm CCSN}$ to the same value as in TNG, the two sets of tracks overlap almost entirely (see panel d).
The small difference in slope which remains between the brown and turquoise tracks in panel d) results from the small difference in the ratio of the supernovae efficiencies $\frac{N_{\rm Ia0}}{k_{\rm CCSN}}$ used in the EAGLE and TNG simulations.

\subsubsection{The relative delay and efficiency of iron production in SN Ia and CCSN determines the relation}
We draw two conclusions from the comparison performed in Section \ref{sec: simulations}:
\begin{enumerate}
\item Despite the overall strong metallicity-dependence of CCSN yields (present in the yield tables used in both the EAGLE and TNG simulations), the average relation can be well reproduced using fixed values of m$^{\rm CCSN}_{\rm Fe}$ and [O/Fe]$_{\rm CCSN}$ 
\item Feedback model (which differs considerably between the two simulations) and other processes that are not captured by the simple description given by Equation \ref{eq: OFe} but are accounted for in the simulations (e.g. galaxy mergers, gas recycling) do not have a major effect on the average [O/Fe]--sSFR relation.
\end{enumerate}
This gives us a certain degree of confidence that the average [O/Fe]--sSFR relation may serve as a diagnostic of: 
\begin{enumerate}[i)] 
\item The relative iron production efficiency in SN Ia and CCSN
\item SN Ia delay time distribution and 
\item Some sort of a cosmic average O/Fe abundance ratio produced per CCSN.
\end{enumerate}

\section{Observational data}

Several methods are used to determine metallicity of the star forming material. 
As we summarize in Section \ref{sec: metallicity - techniques}, 
different observational probes are suitable to infer iron-based ($Z_{\rm Fe}$) and oxygen-based ($Z_{\rm O}$) metallicity.
Furthermore, different approaches are used at different redshifts and/or for different objects - especially in the case of iron.
Both factors complicate the observational picture of the [O/Fe] evolution as a function of cosmic time or galaxy properties and particular attention needs to be paid to systematic uncertainties.
We highlight the identified sources of such systematic offsets that have been quantified in the literature and that are specific to the methods introduced in Section \ref{sec: metallicity - techniques}.
In section \ref{sec: baseline} we summarize our attempt to correct for those and to bring the results compiled from the literature (see Section \ref{sec: obs data} and Appendix \ref{sec, app: obs data}) to a common baseline before combining them in Section \ref{sec: results}.

\subsection{Star-forming metallicity determination techniques}\label{sec: metallicity - techniques}

\subsubsection{Stellar-based methods}\label{sec,method: stars}
The most intuitive approach to learn about the metallicity at which the stars are forming is to infer atmospheric abundances from spectra of individual massive (therefore recently formed) stars. 
Alternatively, one can rely on atmospheric abundance estimates derived from lower mass stars that either have very accurate age determinations or are members of young/open clusters. 
These methods are mostly restricted to the Milky Way and its closest satellites, with the notable exception of AB-type blue supergiant (BSG) and red supergiant (RSG) based measurements.  Especially the former are very bright, which makes it possible to obtain their high signal-to-noise ratio spectra in galaxies within a few Mpc \citep[e.g.][]{Przybilla06,Bresolin07,Kudritzki08,Urbaneja08,Hosek14,Bresolin16,Kudritzki16,Davies17,Bresolin22,Liu22}.  
The observed spectra contain absorption lines from many elements and can be compared to a grid of line-blanketed models to determine metallicity \citep[e.g.][]{Kudritzki08,Hosek14}.
Typically many iron-group (Fe, Cr) but also some $\alpha$-element lines (e.g., Mg, Ca, Si, Ti) are included in the analysis.
Therefore, the derived bulk metallicity may not be straightforwardly linked to iron or oxygen abundances. 
Nevertheless, it has been used in the literature as a proxy for both and compared with gas phase oxygen abundances \citep[e.g][]{Bresolin16,Bresolin22} or stellar-based iron measurements \citep[e.g.][]{Hosek14}.
Solar abundance pattern is assumed in the spectral models, which by construction does not allow to detect deviations from the solar ratios.
Potential departures from the solar ratios are also difficult to infer from comparison with gas-phase oxygen abundances due to the known systematic uncertainty in the absolute value of such measurements: the BSG/RSG determinations typically fall within the range spanned by oxygen derivations obtained with different methods \citep[e.g.][]{Bresolin16,Bresolin22}.
Encouragingly, \cite{Garcia14} find that the iron abundance of IC 1613 needs to be higher than $\sim$1/10 solar estimated from its HII region oxygen abundances in order to match stellar spectra and wind properties, more in line with RSG based metallicities in this system. This suggests that RSG/BSG metallicities may reasonably probe iron abundances.
\\
Finally, we note that while the surface iron abundance of massive stars is expected to be a good representation of the abundance of that element in the star forming material, surface oxygen abundance may be affected by evolutionary processes (in particular, strong rotation and mixing of the material processed in nuclear reactions may substantially lower birth oxygen abundance, e.g. \citealt{Brott11I,Maeder14}).

\subsubsection{Methods based on the UV emission of stellar populations}\label{sec,method: UV}
For more distant objects, rest-frame UV galaxy spectra can be used to obtain iron-based metallicity estimate \citep[e.g.][]{Heckman98,Rix04,Crowther06,Sommariva12,Cullen19}.
This part of the spectrum is dominated by massive OB-type stars, and therefore reflects the star-forming metallicity.
The UV continuum contains absorption features that can be linked to elements present in stellar photospheres (especially highly ionized iron) and stellar winds. The strength of the stellar winds and the associated line profiles are expected to be a strong function
of metallicity \citep[predominantly iron abundance due to its dominant contribution to opacity in radiation-driven winds, e.g.][]{Kudritzki87,Vink01,VinkDeKoter05,Crowther06}. 
In practice, the metallicity is obtained by comparing the observed spectra with predictions of stellar population synthesis (SPS) and the result is inevitably model sensitive.
The two SPS models that are the most commonly used to derive the properties
of high redshift galaxies (Starburst99 \citealt{Leitherer14} and BPASS \citealt{Stanway16,Eldridge17,StanwayEldridge18}) are known to yield UV-continuum based metallicities that are systematically lower for BPASS models than for Starburst99 \citep[S99, with the average offset of $\sim$0.1 dex, e.g.][]{Chisholm19,Cullen19} \footnote{However, the differences can be much larger. For instance, \cite{Cullen21} find $\sim$0.6 dex lower UV continuum based metallicity obtained using BPASS models than when using Starburst99 models for their low mass stack. }.
BPASS models include the effects of binary evolution (in particular
stripping of stellar outer layers in mass transfer) that tend to produce harder
spectra for the same stellar population metallicity than single star based models (even when including rotation) \footnote{Consequently, metallicities derived with methods sensitive to the ionising part of UV (e.g. using HII region emission line rations as discussed later in this section) rather than the continuum can be expected to be lower
when obtained with Starburst99 than with BPASS models (i.e. the opposite to what is found for the UV continuum).}. Such harder ionisation fields as obtained when accounting for binary evolution effects help to reproduce the observed line ratios in high redshift galaxies, but even harder spectra than predicted by any existing SPS may be required to match the emission of the most metal poor objects \citep[e.g.][]{Steidel16,Xiao18,Nanayakkara19,Strom22,Eldridge22,Senchyna22,Katz23}.
Furthermore, current models struggle with simultaneously reproducing wind and photospheric features in the UV spectra of young, highly star forming galaxies \citep[see Section 6.2 in][for the extensive discussion]{Senchyna22}.
\citealt{Senchyna22} caution that UV stellar metallicity estimates relying primarily on wind line features (more easily detected at high redshifts than the photospheric features) may significantly bias the result towards higher values 
\citep[see also][]{Wofford21}.
\\
Efficient absorption of UV-wavelengths in the Earth's atmosphere means that obtaining UV spectra of local galaxies requires the use of space telescopes.
Furthermore, typical local galaxies are UV-faint (i.e. have low SFR and declining star formation histories). 
Both factors result in this method being rarely applied in the local Universe \citep[but see][for the recent efforts]{Senchyna22}.
Conversely, rest-frame UV spectra of high redshift galaxies are much brighter and conveniently shifted to optical wavelengths for z$\gtrsim$2 objects. Applying this technique still requires very good data quality (high-S/N continuum emission detection), and the UV-based iron metallicity is currently only available for a limited number of galaxies \citep{Rix04,Sommariva12,Steidel16,Topping20,Cullen19,Cullen21,Calabro21,Matthee22}.
\\ \newline
\underline{Known source of systematic uncertainty: } \\
Choice of SPS model: we assume the average difference between  Z$_{\rm Fe}$ derived with S99 and BPASS models $\Delta_{\rm SPS}$=0.1 dex
(Z$_{\rm Fe}^{\rm S99}$=$\Delta_{\rm SPS}$+Z$_{\rm Fe}^{\rm BPASS}$), unless the exact offset is known.

\subsubsection{Gas-phase HII region-based methods: oxygen}
\label{sec,method: gas O}
Oxygen abundance is commonly inferred using optical emission lines from HII regions \citep[see e.g.][for the recent reviews]{MaiolinoMannucci19,Kewley19}. HII regions are ionised by neighboring
massive stars, therefore suitable to learn about the star-forming metallicity. 
The most direct approach requires detection of faint recombination lines and is mostly limited to nearby HII regions \citep[e.g.][]{Peimbert67,Esteban14}.
The most commonly used alternative method relies on the fact that the electron temperature T$_{e}$ of the gas is a sensitive probe of metallicity. 
For the same ionisation source, an HII region with higher metallicity has lower T$_{e}$ due to more efficient cooling by metal line emission than its low metallicity counterpart.
The line cooling is dominated by oxygen due to its high abundance and low excitation energies relative to other metals.
T$_{e}$ can thus be determined based on ratios of temperature-sensitive collisionally-excited oxygen emission lines (e.g. [O III] 4363, 4959 and 5007).
However, the suitable auroral lines are also challenging to detect. 
As a consequence, methods relying on strong-line proxies of auroral lines are commonly used.
Estimates obtained with different strong-line calibrations lead to discrepant results \citep[e.g.][]{KewleyEllison08,Kewley19, MaiolinoMannucci19}:
the so-called ‘direct’ method (i.e. where the suitable auroral lines are detected and T$_{e}$ can be inferred directly) and methods based on empirical calibrations typically lead to lower oxygen abundances than theoretical calibrations.
Notably, there is also a known discrepancy between the measurements based on auroral and  the O II recombination lines. The latter typically leads to $\sim$0.24 dex  higher oxygen abundance inferred for the same region \citep[this offset is often called the Abundance Discrepancy Factor ADF, e.g.][]{Peimbert67,Esteban14, Kewley19,Chen23}.
It is currently unclear which of the methods leads to the correct absolute oxygen abundance value and what is the origin of the differences between them \citep{Chen23}.
\cite{Steidel16} and \cite{Sanders20} find that in order to reproduce the observed HII region line ratios with photoionisation model grids using the `direct' method oxygen abundances as input, they need to increase the empirically derived value.
They suggest that adding ADF=0.24 dex can solve this issue (at least when BPASS models are used to supply the ionisation field).
This suggests that oxygen abundances inferred with recombination lines may be better suited for combining/comparing with stellar-based metallicity measurements (but see \citealt{Bresolin22}). 
However, as discussed earlier, because stellar and gas-phase abundance measurements typically trace different elements, they may not be (and are not expected to be) consistent with each other in certain environments.
While the exact oxygen abundance value is not important for establishing the existence of trends in metallicity evolution with redshift or with galaxy properties (e.g. the mass-metallicity relation; \citealt{Tremonti04}) as long as consistent calibration is used for the entire sample, it is relevant for the discussion of the relative enrichment in different elements.
\\ \newline
\underline{Known sources of systematic uncertainty: } \\
i) Method used to translate the observed optical emission line ratios to oxygen abundance. While nearly all $Z_{O}$ estimates used in this study are based on the empirical `direct' calibrations, they are still subject to the Abundance Discrepancy Factor (ADF) - i.e. systematic uncertainty between the absolute $Z_{\rm O}$ value derived on recombination lines and collisionally excited lines ($Z_{\rm O}^{\rm  RL}$=$Z_{\rm  O}^{\rm  CEL}$+ADF).
We assume ADF=0.24 dex, unless the exact offset is known.
\\
ii) Oxygen depletion onto dust grains: it is commonly assumed that 
dust depletion may lead to an underestimation of the true abundance of this element in the interstellar medium by up to $\sim$0.1 dex  when derived from the gas phase. Therefore, we consider $\Delta_{\rm d}$=+0.1 dex dust correction uncertainty in $Z_{\rm O}$.

\subsubsection{Gas-phase HII region-based methods: iron}
\label{sec,method: gas Fe}
Iron lines are rarely detectable for HII regions, but
in principle its gas-phase abundance can
be derived from collisionally excited Fe III–Fe V lines, as has been done for the local low mass and very metal poor galaxies\citep[ e.g.][]{Izotov99,Izotov06,Izotov18,Kojima20}.
However, there are significant systematic uncertainties associated with the applied iron ionization correction factors \citep[ICF, correcting the estimate for its abundance in the `unseen' ionisation states][]{StasinskaIzotov03,Rodriguez05,Izotov06,Kewley19}.
In particular, $\sim$0.2 dex difference between the Fe$^{2+}$ ICF resulting from the models of \cite{StasinskaIzotov03} and \cite{Rodriguez05} was found.
Furthermore, iron is subject to severe depletion onto dust grains \citep{Izotov06,Rodriguez05,Roman-Duval2021}.
Therefore, except for the most metal poor galaxies that typically have very little dust,
gas-phase iron abundance estimates require substantial and uncertain depletion correction to reflect the true abundance of that element in the ISM.
Finally, we note that the ADF uncertainty which affects the oxygen abundance (and T$_{e}$) derived with this method can also affect the iron abundance, because both quantities are used to estimate the iron abundance \citep[e.g.][]{Izotov06}.
It is unclear how Z$_{\rm Fe}$ (and so [O/Fe]) is affected by this uncertainty and to our knowledge it has not been quantified in the literature.
\\ \newline
\underline{Known sources of systematic uncertainty:}\\
Ionisation correction factors: 
all HII region-based Z$_{\rm Fe}$ that are quoted in our study use ICF from \cite{StasinskaIzotov03} and may overestimate Z$_{\rm Fe}$ by $\Delta_{\rm ICF, Fe}$=0.2 dex.

\subsubsection{Indirect iron abundance determination}
\label{sec,method: indirect}
When neither iron lines nor UV continuum is observed,
some constraints on the iron abundance can be inferred indirectly from HII region photoionisation models by considering metallicity of the input ionising source (supplied through SPS model) independently of the oxygen abundance when fitting for the observed line ratios \citep[e.g.][]{Strom18,Sanders20,Runco21,Strom22}.
This approach takes advantage of the fact that while gas cooling is dominated by oxygen, gas heating is to large extent determined by the abundance of iron due to its decisive role in setting the shape the ionising radiation field coming from massive stars.
Since this method probes stellar rather than the gas-phase iron content, it does not require dust depletion corrections.
The result is only sensitive to the ionising part of the model spectra, while the other stellar-based methods used to infer iron abundances rely on the UV continuum lines and wind features.
While its downside is that it is indirect, it is a complimentary approach that allows for important consistency tests of the results derived with a given SPS model.

\subsection{Milky Way - based [O/Fe]--sSFR relation}\label{sec: obs data, MW}

\begin{figure*}
\centering
\includegraphics[scale=0.55]{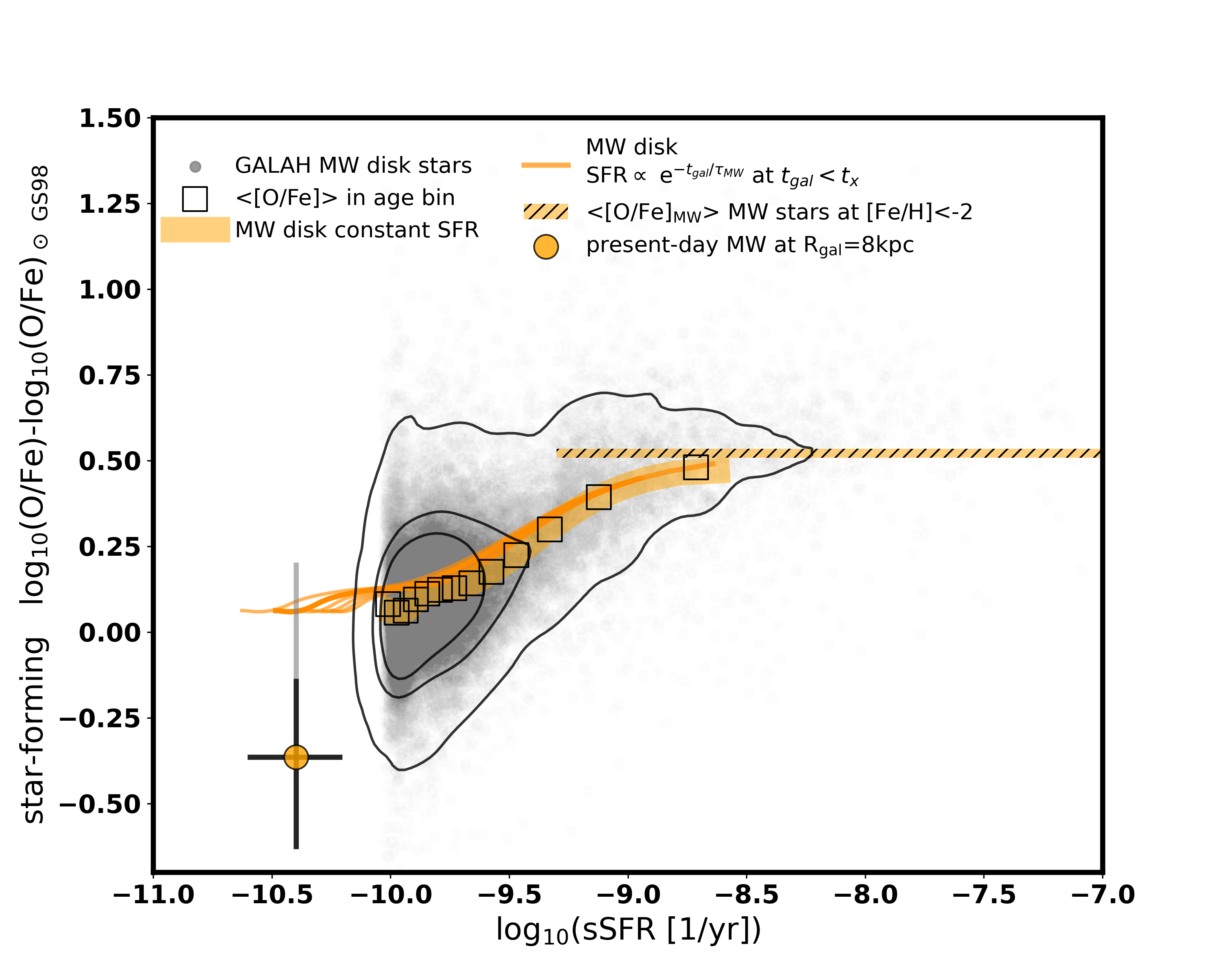}
\caption{ 
Milky Way (MW)-based star-forming [O/Fe] versus specific star formation rate relation.
The big data point corresponds to the present-day MW, where the light gray extension of the errorbars indicate the systematic uncertainty (see text and Table \ref{tab:data}).
Gray points: [O/Fe] of the MW disk main sequence turn-off stars from GALAH DR3 with sSFR estimated from stellar ages assuming constant MW disk star formation history.
Contours enclose 50\%, 68\% and 95 \% of stars in the diagram.
Black empty squares indicate the average [O/Fe] of those stars grouped in 14 age bins (see Table \ref{tab: MW disk O/Fe in age bins}).
Orange curves show part of the MW evolutionary track in the diagram reconstructed using binned ages and average [O/Fe] for different assumptions about the disk star formation history (see text for the details).
The horizontal hatched bar at [O/Fe]$\approx$0.52 dex shows the average abundance ratio of the MW thick disk/halo dwarf stars with [Fe/H]<-2 from \cite{Amarsi19} (x-axis value is arbitrary).
}
\label{fig: MW constraints}
\end{figure*}
Milky Way (MW) studies alone can provide constraints on different parts of the [O/Fe]-sSFR diagram.
We discuss the MW data that can be used in this context below and show the resulting relation in Figure \ref{fig: MW constraints}.
We put those results in the context of other star forming galaxies in Figure \ref{fig: data}.
\\
i) Low sSFR: present-day Milky Way
\\
Star forming spiral galaxies at low redshifts typically show negative radial metallicity gradients \citep[e.g.][]{Sanchez14,Carton18,Hernandez19}. The MW is no exception from this rule.
To estimate its present-day [O/Fe], we combine the oxygen abundance gradient determination from Galactic HII regions obtained by 
\cite{Arellano-Cordova20} and the iron abundance gradient determination from MW open clusters reported by \cite{Spina22}.
Extragalactic metallicity estimates are typically representative of the metallicity at $\sim$1-1.5$R_{e}$ effective radius \citep{KewleyEllison08}. For the local spiral galaxies the value at $\sim$1.5 $R_{e}$ is often quoted as representative of the integrated metallicity \citep{Bresolin16,Bresolin22}.
Therefore, for comparison with other objects we use the [O/Fe] calculated at the Galactic radius $R_{gal}=6.8$ kpc ($\approx$1.5 $R_{e}$ for MW \citealt{Arellano-Cordova20}). We also indicate the [O/Fe] ratio at $R_{gal}$=8 kpc, i.e. around the solar location (where it is best constrained).
\\
ii) Low/intermediate sSFR: MW disk evolutionary track\\ 
Photospheric abundances and age determinations available for a large sample of Galactic disc stars allow for a crude reconstruction of (part of) the Milky Way's evolutionary track in the [O/Fe] - log$_{10}$(sSFR) plane.
To this end, we use [O/Fe] and stellar ages of main-sequence turn-off stars from the third data release of the Galactic Archaeology with HERMES (GALAH) survey \citep{Buder21} and estimate sSFR from stellar ages by assuming a MW disk star formation history as detailed below.
Oxygen abundances for this sample were derived from the OI 777nm triplet using the non-LTE grids of departure coefficients \citep{Amarsi20}. 
Stellar ages are provided in one of the value-added catalogues and estimated from T$_{\rm eff}$, log$_{10}$(g), [Fe/H], [$\alpha$/Fe], and photometric and astrometric information using the Bayesian Stellar Parameter Estimation code BSTEP \citep{Sharma18}.
We follow the quality cut and selection criteria of turn-off stars in the Milky Way disk from \cite{Hayden22} \footnote{
Namely, we remove any object whose stellar parameter, metallicity or [O/Fe] are flagged and select stars with $S/N>45$, $\chi^2_{sp}<4$, $T_{\rm eff}<6200\,\mathrm{K}$, $\sigma(T_{\rm eff})<150$, $-1<[{\rm Fe/H}]<0.5$, $3.5<\log_{10} g<4.1$, $age>1.75\,\mathrm{Gyr}$, and $\sigma(age)/age<0.2$. }
\\
We group the data in 14 age bins and calculate the average [O/Fe] in each bin (see Table \ref{tab: MW disk O/Fe in age bins}).
Stellar ages peak at around 6 Gyr and span roughly between $t_{*}\approx$2 Gyr and $t_{*}\approx$12.5 Gyr. We adopt the latter value as the limit on the formation time $T_{form}$ of disk stars. 
The average [O/Fe] increases from 0.08 dex to 0.48 dex during this time.
Assuming a constant disk SFR, the sSFR in each bin can be obtained by simply inverting the difference between $T_{form}$ and the median stellar age (i.e. sSFR = 1/t$_{gal}$ and t$_{gal}=T_{form} - t_{*}$, black squares in Figure \ref{fig: MW constraints}).
Realistic Milky Way disc star formation history estimates feature a burst/phase of rapid star formation followed by a decline to approximately constant SFR at a current level \citep{Aumer09,Fantin19,Bonaca20}. This means that the true sSFR can be expected to be lower than the one calculated with the constant disk SFR (i.e. the Milky Way track in the [O/Fe]- log$_{10}$(sSFR) plane in Figure \ref{fig: MW constraints} is likely leftwards of the one calculated with the constant SFR).
To illustrate this, we also show the [O/Fe]- log$_{10}$(sSFR) tracks for which sSFR is obtained assuming an exponentially declining SFH (SFR(t$_{gal}$) $\propto e^{-t_{gal}/\tau_{MW}}$ for t$_{gal}$<$t_{x}$ and SFR=const at t$_{gal}$>$t_{x}$). We treat $\tau_{MW}$ and $t_{x}$ as free parameters and use a range of values for which the calculated sSFR does not extend below the present-day Milky Way constraints.
We caveat that the sSFR assigned to the oldest stars ($>$10 Gyr, with the highest [O/Fe]) is particularly uncertain, both because the average uncertainty in their age estimate is $\gtrsim$1 Gyr (compared to $\sim$0.4 Gyr for stars with ages $<5$Gyr), and because their sSFR is sensitive to the assumed $T_{form}$ (if $T_{form}$ is lower than assumed and the stars are younger, their assigned sSFR would be higher). Overall, the above considerations indicate that the MW-based [O/Fe]--sSFR relation may level off somewhere at log$_{10}$sSFR$\gtrsim$-9.
While our estimate of the MW `evolutionary track' should be taken with a grain of salt, it shows that a more careful analysis (beyond the scope of this study) can potentially provide valuable constraints. 
\\
iii) High sSFR: metal-poor stars
\\
Old, metal-poor stars are expected to hold a stable record the SN Ia-free [O/Fe]$_{\rm CCSN}$ ratio.
Therefore, they can shed light on the enrichment level expected in the high sSFR part of the relation.
To estimate this, we select MW thick disk/halo dwarf stars with oxygen and iron abundance determinations from \cite{Amarsi19} and calculate the average [O/Fe] of stars with [Fe/H]<-2. We use 3D-LTE iron abundance and 3D non-LTE oxygen abundance estimate ([Fe/H]3L and [O/H]3N reported in table 7 in \citealt{Amarsi19}), as recommended by the authors. 
The resulting log$_{10}$(O/Fe)$_{\rm CCSN}\approx$1.85 dex (which corresponds to [O/Fe]$_{\rm CCSN} \approx$0.52 dex on our reference GS98 solar scale) is shown as a hatched horizontal bar in Figure \ref{fig: MW constraints}.
The metallicity cut of [Fe/H]<-2 was chosen to ensure that the stars belong to the flat part of the [O/Fe] (or [$\alpha$/Fe]) - [Fe/H] relation, see Figure \ref{fig: Amarsi19}). As discussed in Section \ref{sec: expectations}, such a flattening is expected in the regime where the SN Ia contribution is subdominant.
It is unclear to what range of sSFR it corresponds (the range of sSFR spanned by the horizontal bar in Figure \ref{fig: MW constraints} is chosen arbitrarily).

\subsection{Other star forming galaxies}\label{sec: obs data}

We collect literature estimates of the star-forming iron and oxygen abundances for galaxies spanning a wide range of sSFR.
Their metal abundances were obtained with a range of methods described in Section \ref{sec: metallicity - techniques}. 
The compiled data and the references to original papers are given in Table \ref{tab:data}. 
In Figure \ref{fig: data} we plot those results as reported (i.e. only correcting for solar scale differences). We indicate the known sources of systematic uncertainty that are relevant for each of those estimates in column 5 and/or in the comments in Table \ref{tab:data} and show them as gray extensions to error bars in Figure \ref{fig: data}.
We summarize how we correct for those offsets and how we select the sample to constrain the relation in Section \ref{sec: baseline}.
We briefly discuss the estimates shown in Figure \ref{fig: data} below and refer the interested reader to Appendix \ref{sec, app: obs data} for more details. 

\subsubsection{Low sSFR: local star forming galaxies}

Objects with log$_{10}$(sSFR)$<$-9.5 in Figure \ref{fig: data} correspond to local star forming spiral and dwarf galaxies and probe a relatively typical low redshift galaxy population.
Blue crosses in Figure \ref{fig: data} mark galaxies with BSG-based metallicity estimates from \cite{Bresolin16,Bresolin22}, assuming that those metallicity measurements can serve as a proxy of the iron abundance.
As discussed in Section \ref{sec,method: stars}, this is not strictly correct, as those estimates provide the bulk metal mass fraction $Z$ obtained from matching solar-scaled spectral model to multiple observed lines (mostly, but not only iron-group). It is unclear what is the error associated with this assumption.
As we discuss in the appendinx \ref{sec,app: with BSG}, whether we use this sample or not does not affect our main conclusions.
However, currently it is the only method that could allow to extend the star-forming [O/Fe] estimates to relatively massive (and metal rich) typical low-redshift spiral galaxies.
See Appendix \ref{sec, app: local galaxies} for the discussion of the remaining low sSFR galaxies.

\subsubsection{Intermediate sSFR: high redshift star forming galaxies}

Galaxies that occupy intermediate sSFR range (roughly -9.5$\lesssim$log$_{10}$(sSFR)$\lesssim$-8) are expected to be mostly high redshift MS objects.
This is where the [O/Fe] is expected to show strong evolution.
Several authors obtain iron abundance estimates using rest-UV spectra (individual or stacked) and `direct' gas-phase oxygen abundances using rest-optical spectra  of such galaxies at redshifts $\gtrsim$2 \citep[][see section \ref{sec,method: UV} and Appendix
\ref{sec, app: high z} for further details]{Steidel16,Topping20,Cullen21}.
 All of those studies probe intermediate galaxy stellar mass range log$_{10}$(M$_{*}$/$\Msun$)$\sim$9 - 10 and also provide the SFRs, which allows us to put their estimates on the [O/Fe]--sSFR plane.
Note that SFRs estimated for individual galaxies in \cite{Cullen21} (purple stars in Figure \ref{fig: data}) fall above the average $z\sim3.4$ MS (and have high log$_{10}$(sSFR)>-8).
\\
We also indicate the $z\sim2.2$ estimate from \citet{Kashino22} where the iron-based metallicity is obtained from stacked rest-UV galaxy spectra. However, the oxygen abundance is not measured directly but inferred from the `direct' method mass-metallicity relation at a similar redshift obtained by \cite{Sanders20}.
Finally, we show the results from \cite{Strom18,Sanders20,Strom22}, who estimate [O/Fe] with photoionisation models using only rest-frame oxygen optical emission lines as constraints (i.e. without any UV constraints, see section \ref{sec,method: indirect}). 
 The results obtained by \cite{Sanders20} show a substantial scatter and come with large errors/upper limits on the iron abundance (lower limits on [O/Fe]). Strikingly, some of them point to [O/Fe]$>$1 and even despite the large uncertainties, clearly stand out from all the other estimates. 
 In particular, [O/Fe] estimate obtained by \cite{Sanders20} for the  KBSS-LM1 stack earlier analysed by  \cite{Steidel16} is $>$1 dex higher than the value obtained by the latter authors.
 The reason behind this is unclear. 
In contrast, the estimates obtained by \cite{Strom18,Strom22} are well within the range of values covered by the other observational results summarized in this section. 
\cite{Strom18} also analyse the KBSS-LM1 stack with their method. The value that they obtain [O/Fe] $\approx$0.55 dex is somewhat higher (i.e. Fe abundance is lower), but consistent with the one reported by \cite{Steidel16}.
Note that \cite{Strom18,Strom22} include a prior requiring [O/Fe]$<$0.73 dex ([O/Fe]$<$0.59 dex  on our solar scale) and therefore by construction avoid such high values as quoted by \cite{Sanders20}.
 Given that the iron abundance is not directly constrained in those studies and it is unclear how to properly compare them with other estimates, we do not include those results in further analysis.
 Figure \ref{fig: data_excl_indirect} in the Appendix \ref{sec, app: obs data} shows the [O/Fe]-sSFR plane when excluding the data with indirect Fe or O abundance estimates. 
This leaves only a few data points in the intermediate sSFR range, but the scatter is much reduced.
 
\subsubsection{High sSFR: very metal poor local dwarf galaxies}

\begin{figure*}
\centering
\includegraphics[scale=0.55]{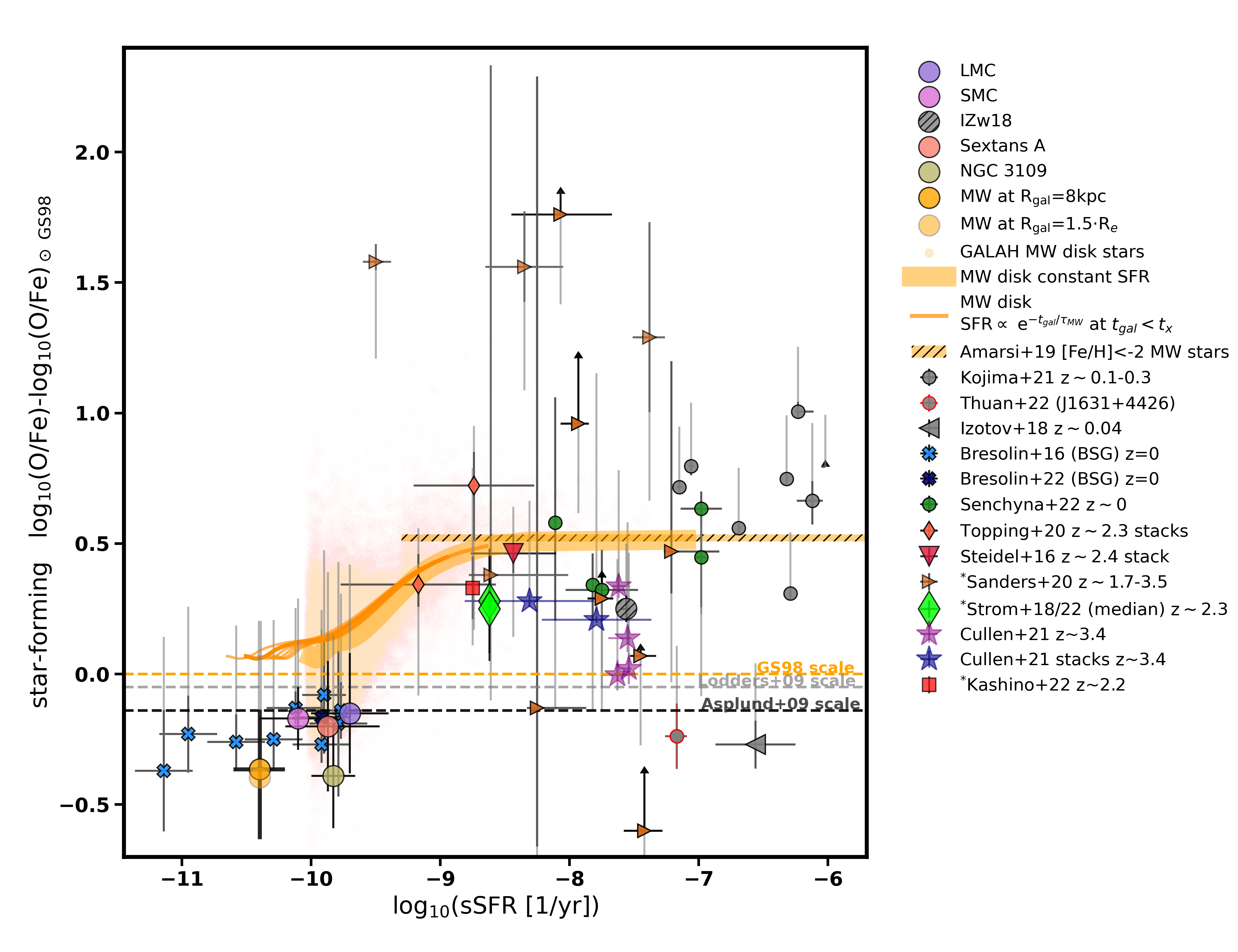}
\caption{ 
Observational estimates of the star-forming [O/Fe] versus specific star formation rate for the Milky Way (MW, estimated at $R_{gal}$=8 kpc and at 1.5$\cdot R_{e}$ effective radius), nearby dwarf galaxies (LMC, SMC, IZw 18, Sextans A, NGC 3109), 
local galaxies with blue supergiant based metallicity estimates from \cite{Bresolin16,Bresolin22},  extremely metal poor dwarf galaxies from \cite{Senchyna22}, \cite{Kojima21}, \cite{Thuan22} and \cite{Izotov18}, and high redshift star forming galaxies/stacks from \cite{Steidel16}, \cite{Cullen21} and \cite{Topping20}.
For the remaining high redshift estimates (marked with $^{*}$ in the legend) either the iron or oxygen abundance was inferred indirectly.
Light gray extensions of the errorbars indicate known sources of systematic uncertainty in the abundance determination (see text and Table \ref{tab:data}).
Small orange points show the MW disk stars and orange curves indicate part of the MW disk evolutionary track in the diagram (see Section \ref{sec: obs data, MW} and Figure \ref{fig: MW constraints}). Horizontal hatched bar at [O/Fe]$\approx$0.52 dex indicates the average abundance ratio of the MW thick disk/halo dwarf stars with [Fe/H]<-2 from \cite{Amarsi19} (x-axis value is arbitrary).
\\ 
Only the offsets due to different reference solar abundances choices were corrected in this Figure.
Horizontal lines indicate zero points for different reference solar O and Fe abundance choices. There is a 0.14 dex offset between the \cite{GrevesseSauval98} (GS98, orange dashed line) scale used here and the commonly used \cite{Asplund09} solar scale (black dashed line).
}
\label{fig: data}
\end{figure*}

Galaxies gathered in this section (IZw 18 and galaxy samples from \citealt{Kojima21} and \citealt{Senchyna22} - gray symbols and dark green circles in Figure \ref{fig: data}) are characterised by very low metal content and high specific star formation rates (log$_{10}$(sSFR)$\gtrsim$-8). 
Such properties are typical of the early rather than the local Universe, where they can be viewed as outliers.
Those extreme local low-mass galaxies received a lot of attention precisely due to their potential to serve as testbed for future high redshift studies.
Direct method HII region oxygen abundances are available for all of those galaxes.
Iron abundance for the sample from \cite{Senchyna22} is based on UV-continuum constraints. For the remaining objects iron abundance is derived from HII region lines following the method described in \cite{Izotov06} (see section \ref{sec,method: gas Fe} and Appendix \ref{sec, app: VMPDG} for further details).
\\
 Given their properties (in particular their estimated young ages), especially the galaxies selected by \cite{Kojima21} are not expected to be enriched by iron from SN Ia yet. Therefore, such objects can help to constrain the typical core-collapse [O/Fe]$_{\rm CCSN}$ ratio (or the potential plateau level of the [O/Fe]-sSFR relation).
 While for most of the galaxies in this sample the estimated [O/Fe] falls within the expected range ($\gtrsim$0.5-0.8 dex), two of them
(J1631+4426 with abundances revised by \cite{Thuan22} and J0811+4730 from \cite{Izotov18}) have [O/Fe] comparable to LMC and SMC.
Early enrichment by very massive, rapidly rotating stars and rare Pair-Instability Supernovae were proposed as a possible explanation of their unexpected abundance ratios \citep{Isobe22, Goswami22}

\subsection{Selecting the final sample and bringing the data to the common baseline} \label{sec: baseline}

The results shown in the previous section were obtained with a variety of methods and different modelling assumptions. Comparing them at face value may easily lead to erroneous conclusions.
Here we discuss how we select the data that can be compared in a more consistent way in order to constrain the [O/Fe]--sSFR relation.
\\
Firstly, we choose to use HII region based oxygen abundance estimates for all objects in our analysis, even if massive-star based estimates are available (mostly the case for MW and Magellanic Clouds). We do this for two main reasons: i) as discussed in section \ref{sec,method: stars}, stellar atmospheric oxygen abundance may be significantly affected by processes related to stellar evolution ii) `direct' HII region based oxygen metallicity estimates are now available for galaxies across redshifts and with different properties. The latter is advantageous, as it can reduce the impact of additional (possibly unidentified) systematics that can be introduced by combining results obtained with very different techinques.\\
We consider the following sources of systematic offsets between the results obtained in different studies and correct for them as outlined below:
\\
1. \underline{Reference solar abundances.}\\
The correction is straightforward as long as the assumed solar reference abundances are reported in the original studies. There can be $>$0.14 dex difference in [O/Fe] value depending solely on the choice of solar reference abundances (see horizontal lines in Figure \ref{fig: data}). Therefore, while the specific choice is not relevant for the conclusions, it is important to convert all measurements to a consistent solar scale.
All abundances used in our study are converted to \cite{GrevesseSauval98} solar scale.
\\
2. \underline{Uncertainty in the absolute value of the $Z_{\rm O}$ derived from the} \underline{HII region optical emission lines} (see section \ref{sec,method: gas O}).\\
We correct for the systematic shift of ADF=0.24 dex to `direct' method esitmates reported on auroral/recombination line scales. If the exact offset is known, we use the ADF reported by the authors. We further consider $\Delta_{\rm d}$=+0.1 dex uncertainty due to dust depletion. Such dust correction has been explicitly added only to estimates reported by \citealt{Senchyna22}.
Not all data can be corrected for those offsets in a consistent way. As discussed in  \ref{sec,method: gas Fe}, it is unclear how to correct measurements where both $Z_{\rm Fe}$ and $Z_{\rm O}$ are based on HII region emission lines for ADF. For this reason, we do not use IZw18, J0811+4730 and the sample from \cite{Kojima21} to characterise the [O/Fe]--sSFR relation. 
In any case, with the exception of IZw18, all these objects fall outside the sSFR range occupied by regular MS galaxies and cannot serve to constrain the relation in the part where it is expected to show the strongest and orderly evolution.
Note that the oxygen abundance derived by \cite{Strom18,Strom22} is the only gas-phase $Z_{\rm O}$ reported here that is not based on the `direct' method and may be subject to additional systematic differences with respect to other measurements. However, as discussed in the previous section, we exclude all estimates with indirect $Z_{\rm Fe}$ (this includes \cite{Strom18,Strom22}) or $Z_{\rm O}$ determinations from further analysis.
\\
3. \underline{Uncertainty in the absolute value of the $Z_{\rm Fe}$ derived from rest-} \underline{frame UV galaxy spectra} (see section \ref{sec,method: UV}). \\
$Z_{\rm Fe}$ is rarely derived with multiple SPS models and we assume $\Delta_{\rm SPS}$=0.1 dex difference between the estimates relying on BPASS and S99 SPS models, unless the exact offset is known.
Examples discussed in \cite{Cullen21} and \cite{Senchyna22} show that this offset can differ a lot from case to case and assuming fixed $\Delta_{\rm SPS}$ is certainly a simplification.
We exclude the low M$_{*}$ stack from \citealt{Cullen21} from further analysis, as the difference between BPASS and S99 SPS models is $\Delta_{\rm SPS}>$0.6 dex in this case and so [O/Fe] is very poorly constrained.
The sample from \cite{Senchyna22} shown in Figure \ref{fig: data} relies on yet different set of SPS models (Charlot \& Bruzual in prep., C\&B hereafter).
The comparison of [O/Fe] derived using S99 and C\&B SPS shown in Figure \ref{fig: Senchyna22} (see also Table 7 therein) shows that they are consistent within errors, except for the object HS1442+4250 (where S99 SPS leads to 0.36 dex lower $Z_{\rm Fe}$).
For easier comparison with other results used in this study, we report [O/Fe] based on S99 SPS in Table \ref{tab:data}.
We estimate $\Delta_{\rm SPS}^{*}$ for this sample by taking the difference $\Delta_{\rm S99-C\&B}$ between S99-based and C\&B-based $Z_{\rm Fe}$ values given in Table 7  in \cite{Senchyna22}.
The results are given in Table \ref{tab:data}.
Note that for all objects except J082555, C\&B-based $Z_{\rm Fe}$ is higher than  S99-based $Z_{\rm Fe}$.
Therefore, the offset between S99 and C\&B is in the opposite direction than between S99 and BPASS-based $Z_{\rm Fe}$, which we indicate be reporting $\Delta_{\rm SPS}^{*}$=$^{+0.1}_{-\Delta_{\rm S99-C\&B}}$ in the last column of Table \ref{tab:data}.
\\
We do not apply $\Delta_{\rm SPS}$ to the estimate from \cite{Steidel16}, as the reported value is averaged over the results obtained with BPASS and S99 models.
\\
We can bring the data remaining in the sample that we choose for further analysis to several common baselines. In particular, we consider:
\begin{enumerate}[I)]
\item High [O/Fe] baseline obtained by including $\Delta_{\rm d}$ in all $Z_{\rm O}$ estimates, bringing all $Z_{\rm O}$ estimates to recombination line scale and all $Z_{\rm Fe}$ estimates derived with S99 SPS to BPASS scale.
\item Intermediate [O/Fe] baseline obtained by 
bringing all $Z_{\rm O}$ estimates to recombination line scale, all $Z_{\rm Fe}$ estimates derived with BPASS SPS to S99 SPS scale and subtracting $\Delta_{\rm d}$ from dust-corrected $Z_{\rm O}$ estimates.
This combination minimizes the number of data points which require applying $\Delta_{\rm SPS}$ (only data from \citealt{Topping20}) and $\Delta_{\rm d}$ corrections (only the sample from \citealt{Senchyna22}).
We consider additional variation (intermediate + $\Delta_{\rm d}$) by including $\Delta_{\rm d}$ in all $Z_{\rm O}$ estimates.
\item Low [O/Fe] baseline obtained by subtracting $\Delta_{\rm d}$ from dust-corrected $Z_{\rm O}$ estimates, bringing all $Z_{\rm O}$ estimates to collisionally excited line scale and all $Z_{\rm Fe}$ estimates derived with BPASS SPS to S99 scale.
\end{enumerate}
As discussed in Section \ref{sec,method: stars}, it is unclear whether BSG-based metallicity estimates reported in the literature can be used as a measure of $Z_{\rm Fe}$. Therefore, we exclude the sample from \cite{Bresolin16,Bresolin22} from our main analysis. This leaves us with 18 objects in the final sample. We summarize the results obtained when including the results from \cite{Bresolin16,Bresolin22} in the Appendix \ref{sec,app: with BSG}.
We note that for the NGC3109 (big light green circle in Figure \ref{fig: data}) $Z_{\rm Fe}$ estimate is also based on BSG \citep{Hosek14}. However, in this case BSG metallicity is also interpreted as such by the authors and we decide to include it in further analysis. 
\\
We caution that while we focus on issues related to abundance determinations, there are uncertainties associated with the SFR and M$_{*}$ measurements as well.
These include, among others, systematics due to the choice of the SFR tracer, IMF\footnote{For common IMF choices sSFR is largely unaffected by this assumption because it affects M$_{*}$ and SFR estimates in similar way.}, or SPS model (used to estimate stellar mass in certain methods). 
Calibrations of the commonly used SFR proxies are metallicity ($Z_{\rm Fe}$) dependent, which means that some of the uncertainties affect both [O/Fe] and sSFR. 
SFR and M$_{*}$ for a given galaxy are often estimated in a separate analysis than its metal abundances. As a result, even though their derivation may rely on the same type of input information\footnote{For instance, SPS model and IMF assumptions are used in both common galaxy stellar mass determination method and UV-continuum/indirect iron abundance derivations}, sSFR and abundances are not necessarily obtained with the same set of assumptions. We do not attempt to correct for such factors in this study.

\section{Results: the observed [O/Fe] - specific SFR relation}\label{sec: results}

\begin{figure*}
\centering
\includegraphics[scale=0.38]{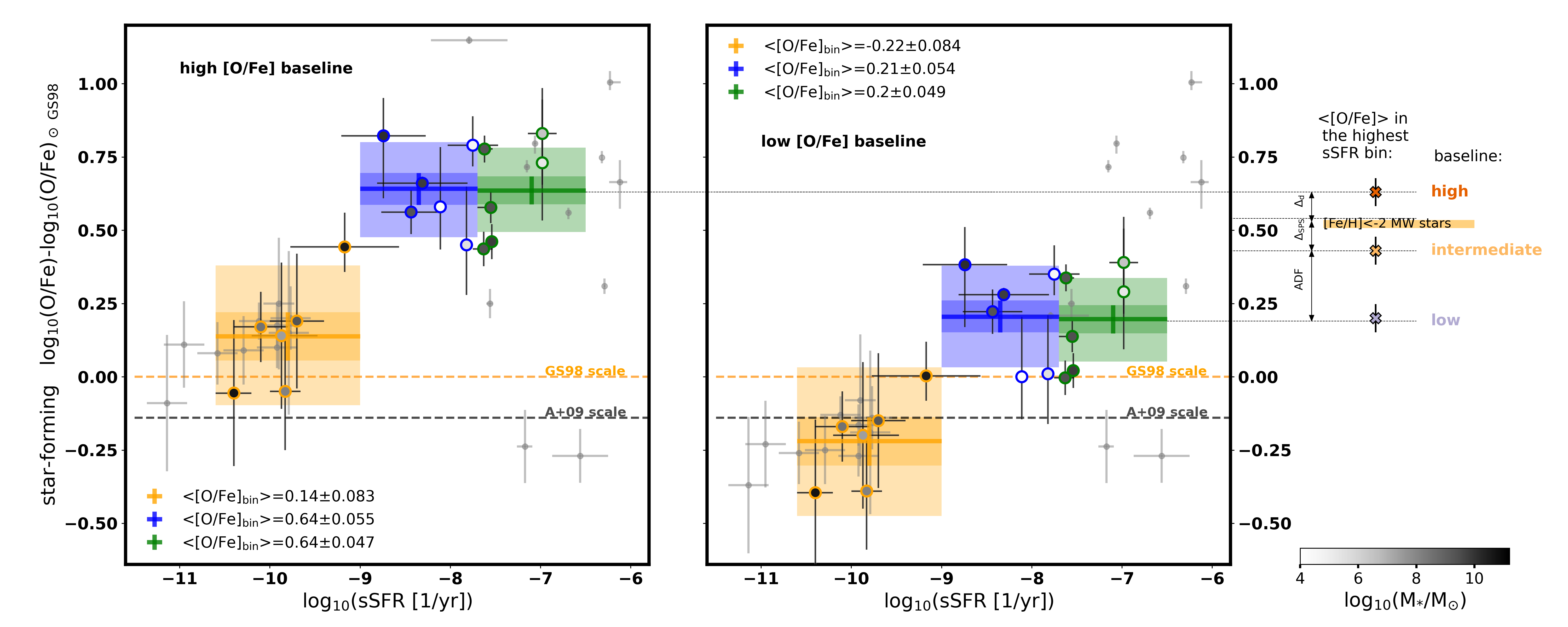}
\caption{ 
Star-forming [O/Fe] versus specific star formation rate relation.
Left/right panel: data points (big circles) were shifted to a common high/low [O/Fe] baseline. The inner color of big circles indicates the stellar mass, the outer color indicates the bin to which the data point belongs.
Values listed in the figure indicate the median [O/Fe] in each of the three equally-populated log$_{10}$(sSFR) bins. 
Dark (light) area of each bin spans between 16-84 (0.13-99.87) percentiles of 10$^{5}$ draws of the average [O/Fe].
Small gray data points were not used in the analysis as it is not clear how to consistently correct them for systematic offsets. Data with indirect O or Fe estimates are not shown.
Thick horizontal lines indicate zero points for reference solar O and Fe abundance from \cite{GrevesseSauval98} (GS98) and \cite{Asplund09} (A+09).
The absolute [O/Fe] values are uncertain but there is a clear evolution towards lower [O/Fe] with decreasing log$_{10}$(sSFR) and no apparent secondary dependence on log$_{10}$(M$_{*}$) within the current sample.
On the right hand side of the figure we summarize the average <[O/Fe]$_{\rm bin}$> values found in the highest sSFR bin for different choices of the common baseline described in Section \ref{sec: baseline}. The differences are equal to the sum of the average systematic offsets between the baselines. The yellow horizontal bar on the right shows the average <[O/Fe]$_{MW}>$ of metal-poor dwarf stars in the Milky Way from \cite{Amarsi19}. Baselines where $Z_{\rm O}$ measurements are placed on the collisionally excited line scale give <[O/Fe]$_{\rm bin}$> which are inconsistent with this value.
}
\label{fig: summary_bins_no_BSG}
\end{figure*}

The results obtained when bringing the data selected as described in Section \ref{sec: baseline} to the common `high [O/Fe]' and `low [O/Fe]' baselines are compared in Figure \ref{fig: summary_bins_no_BSG}.
To better illustrate the evolution in [O/Fe]--sSFR plane, we group the data in three log$_{10}$(sSFR) bins and calculate the average <[O/Fe]$_{\rm bin}$> in each bin.
We choose the bins so that they contain the same number of data points (6).
We assign split normal distribution to each data point (with mean and dispersion is set by its reported value and uncertainties) and draw 10$^{5}$ values for each data point in a given bin to assess the uncertainty of <[O/Fe]$_{\rm bin}$>.
The resulting median <[O/Fe]$_{\rm bin}$> and 0.13 th and 99.87 th percentiles  (light colored areas in Figure \ref{fig: summary_bins_no_BSG}) for each bin and baseline choice are reported in Table \ref{tab: bins}.
\\
There is a clear trend for [O/Fe] to decrease towards lower sSFR, independent of the choice of common baseline. Our results can be broadly summarised as follows:
\begin{itemize}
    \item There is no evidence of [O/Fe] evolution between the two highest sSFR bins (corresponding to -6.5>log$_{10}$(sSFR)>-9) regardless of the baseline.
    This may hint at the existence of the expected flattening in the [O/Fe]--sSFR relation at high log$_{10}$(sSFR). In view of the small size of the current sample and the considerable uncertainties, we do not place a better constraint on the possible turnover point.
    \item There is a clear increase in [O/Fe] between the lowest sSFR bin and the two remaining bins. In all cases, the offset is larger than the area spanned by the 99.8 and 0.13 percentiles  of <[O/Fe]$_{\rm bin}>$ (light colored areas in Figure \ref{fig: summary_bins_no_BSG}).
    \item There is no clear secondary dependence on the galaxy stellar mass within the current sample.    
\end{itemize}
The absolute <[O/Fe]$_{\rm bin}>$ value in each bin is uncertain:
\begin{enumerate}[i)]    
    \item  <[O/Fe]$_{\rm bin}>$ differs by 0.44 dex for the highest sSFR bins when the `low' and `high' [O/Fe] baselines are compared.
    This is a direct consequence of the considered systematic uncertainties (see Section \ref{sec: baseline} and the inset on the right hand side of Figure \ref{fig: summary_bins_no_BSG}). 
    
    \item The differences between <[O/Fe]$_{\rm bin}>$ obtained in the lowest sSFR bin are $\approx$0.1 dex smaller than in the highest sSFR bins when extreme baselines are compared. This is not surprising, as there are no low sSFR data points in our sample for which $Z_{\rm Fe}$ was estimated with UV-spectra relying on SPS models for interpretation. Therefore, the lowest sSFR bin is not affected by $\Delta_{\rm SPS}$ systematics, for which we assumed the average value of 0.1 dex.

    \item <[O/Fe]$_{\rm bin}>$ found in the high sSFR bin on the `low [O/Fe]' baseline is 0.3 dex below the average level of enrichment of the Milky Way metal-poor stars <[O/Fe]$_{MW}$>$\approx$0.52 dex. 
    Any common baseline choice where $Z_{\rm O}$ measurements are placed on the collisionally excited line scale lead to high sSFR <[O/Fe]$_{\rm bin}>$ lower than <[O/Fe]$_{MW}$>. 
    As we discuss further in \ref{sec: results,high sSFR}, <[O/Fe]$_{\rm bin}>$ at high sSFR is not expected to fall below <[O/Fe]$_{MW}$>. Therefore, $Z_{\rm O}$ assuming collisionally excited line abundance scale may underestimate the oxygen-based metallicity with respect to stellar measurements.
    If the above interpretation is correct, the systematic uncertainty on the absolute <[O/Fe]$_{\rm bin}$> values presented in our paper is reduced by ADF=0.24 dex.
\item When only `intermediate [O/Fe]' or `high [O/Fe]' baselines are considered (as motivated above), the MW-based relation and the MW disk evolutionary track in the [O/Fe]--sSFR plane (see Section \ref{sec: obs data, MW}) are consistent with the observational [O/Fe] -- sSFR relation inferred here from the properties of star forming galaxies across redshifts (see Figure \ref{fig: results_vs_models} for the direct comparison).
\end{enumerate}
Point ii) means that the slope of the [O/Fe]--sSFR relation is influenced by the systematics related to the choice of SPS models. In particular, evolution between the bins is steeper when we shift all UV-spectra based  $Z_{\rm Fe}$ to the BPASS scale (as in high [O/Fe] baseline) than if we use the S99 SPS scale (as in low and intermediate [O/Fe] baselines), because the former SPS models tend to lead to lower $Z_{\rm Fe}$ values (higher [O/Fe]) than the latter (see Table \ref{tab: bins}).
Nonetheless, the slopes of the [O/Fe]--sSFR relation for different baseline choices are consistent within  uncertainties (see Figure \ref{fig: slopes}). 

\begin{table}[]
\centering
\begin{tabular}{cccc}
\hline \\
 baseline & <[O/Fe]>  & <[O/Fe]> &  <[O/Fe]>  \\
 &  bin 1  & bin 2 & bin 3  \\ \hline
high                          & 0.14$\pm$0.24  & 0.64$\pm$0.16 & 0.64$\pm$0.14 \\
intermediate+$\Delta_{\rm d}$ & 0.12$\pm$0.25  & 0.55$\pm$0.16 & 0.54$\pm$0.14 \\
intermediate                  & 0.021$\pm$0.25 & 0.45$\pm$0.16 & 0.44$\pm$0.13 \\
low                           & -0.22$\pm$0.24 & 0.2$\pm$0.17  & 0.2$\pm$0.14 \\ \hline
\end{tabular}%
\caption{
Average <[O/Fe]> values in three log$_{10}$(sSFR) bins for data selected as described in Section \ref{sec: baseline}.
Column 1 indicates the choice of the common baseline. 
Bin 1 spans between log$_{10}$(sSFR [yr$^{-1}$])=-11 and -9, bin 2 between log$_{10}$(sSFR [yr$^{-1}$])=-9 and -7.7 and bin 3 between log$_{10}$(sSFR [yr$^{-1}$])=-7.7 and -6.5.
The bins were selected to contain equal number of data points (6).
The errors span between 0.13 and 99.87 percentiles of the averages found when sampling the data within statistical errors (assuming split normal distribution for each data point) and the main value indicates the median.
}
\label{tab: bins}
\end{table}

\subsection{Expected level of [O/Fe] enrichment at high sSFR and the oxygen abundance scales}\label{sec: results,high sSFR}
In the inset on the right hand side of Figure \ref{fig: summary_bins_no_BSG} we compare <[O/Fe]$_{\rm bin}>$ found in the highest sSFR bin on different baselines with <[O/Fe]$_{\rm MW}$>=0.52 dex found for dwarf MW stars with [Fe/H]<-2 and abundances derived in non-LTE analysis performed by \cite{Amarsi19}.
  As discussed in earlier sections, those stars follow the flat part of the Galactic [O/Fe]--[Fe/H] relation (see also Figure \ref{fig: Amarsi19}). This flattening is commonly interpreted as a consequence of the early Galaxy chemical evolution being dominated by CCSN \citep[e.g.][]{MatteucciGreggio86,WheelerSnedenTruran89,KobayashiKarakasLugaro20}. 
  In this view, 
  <[O/Fe]$_{\rm MW}$>  probes the same feature as <[O/Fe]$_{\rm bin}$> in the high sSFR bin(s) (see Section \ref{sec: expectations}).
In principle, <[O/Fe]$_{\rm bin}$> could be higher than <[O/Fe]$_{\rm MW}$> because the sample of \cite{Amarsi19} does not probe very low metallicities ([Fe/H]$\lesssim$-3), where [O/Fe] could be influenced by explosions of the more massive and metal poor CCSN progenitors. Such CCSN can eject material with higher oxygen abundance \citep[e.g.][]{Nomoto13}.
Given that the CCSN oxygen yields are predicted to considerably vary with the mass of CCSN progenitor, the plateau [O/Fe] value can also be affected by variations in the stellar IMF. IMF may become top-heavy in metal-poor and high SFR conditions  \citep[][]{Bromm03,WeidnerKroupa05,Marks12,Jerabkova18}. This would further increase the [O/Fe]$_{CCSN}$ level, unless the excess massive low metallicity stars do not explode in CCSN, but rather collapse without any significant metal ejecta \citep[][]{Fryer99,Fryer06,Sukhbold16,Schneider21}. 
However, if such environmental-dependence of the stellar explosion properties and IMF exists, there is no obvious reason for it to be significantly different in the early evolution of the MW (recorded in the properties of old, metal-poor stars) and in young galaxies included in our sample.
\\
All common baseline choices where Z$_{O}$ is placed on the collisionally excited line abundance scale lead to [O/Fe] in the high sSFR bin that is below <[O/Fe]$_{\rm MW}$>.
The above mismatch argues against the use of this abundance scale in combination with stellar metallicity measurements (unless the average ADF=0.24 dex correction is overestimated or $Z_{\rm Fe}$ values are overestimated for the high sSFR part of our galaxy sample).
  <[O/Fe]$_{\rm bin}>$ derived for the `intermediate [O/Fe] ' baseline is also below <[O/Fe]$_{\rm MW}$>, but consistent with this estimate within 3-$\sigma$-equivalent percentiles.
  Including the systematic offset $\Delta_{\rm d}$=0.1 dex associated with the oxygen dust depletion in all measurements on this baseline brings <[O/Fe]$_{\rm bin}$>=0.54 dex to near perfect agreement with <[O/Fe]$_{\rm MW}$>. We use this `intermediate + $\Delta_{\rm d}$' baseline as a reference for further comparison with models.
 
\section{Comparison with theoretical expectations} \label{sec: results vs models}

\begin{figure*}
\centering
\includegraphics[width=1.0\textwidth]{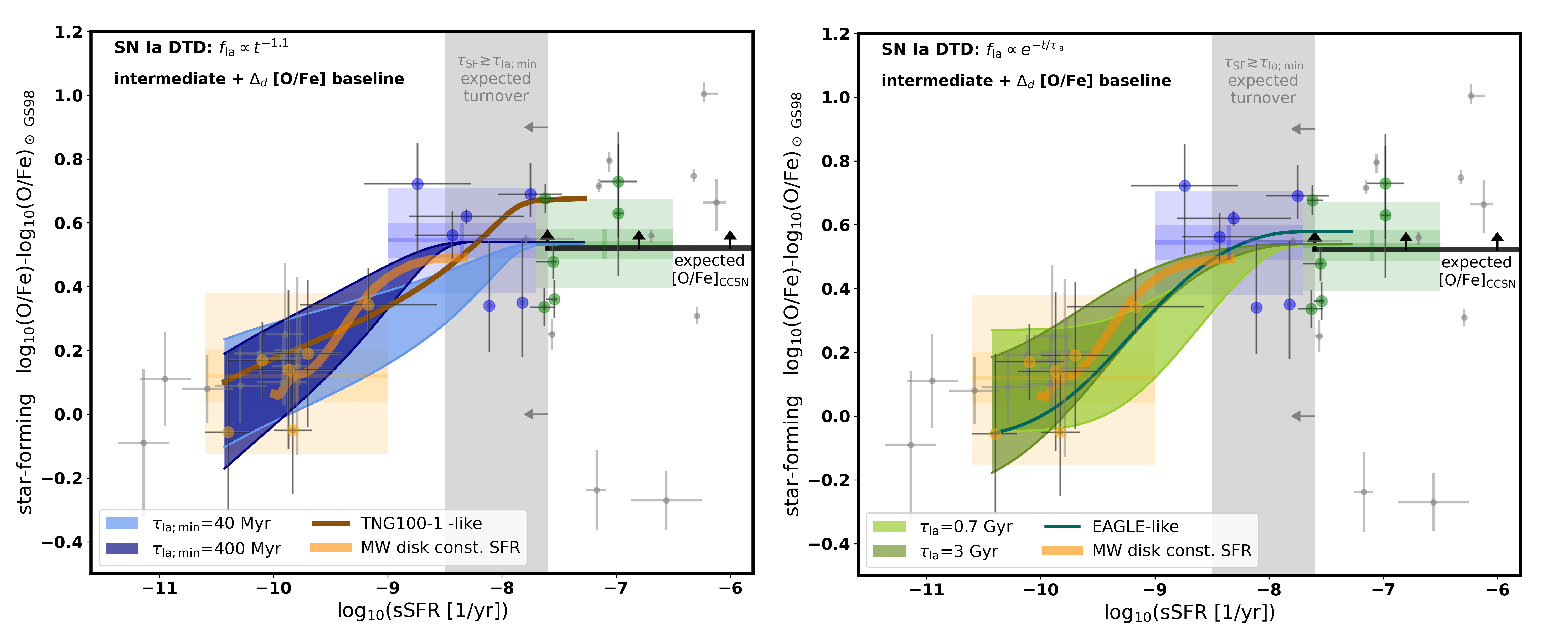}
\caption{ 
Observational [O/Fe]--sSFR relation compared with a broad range of theoretical expectations. Data points (big circles) were shifted to a common intermediate [O/Fe] baseline with additional offset due to possible oxygen dust depletion ($\Delta_{\rm d}$) included. 
This baseline provides the closest match to [O/Fe] of metal-poor dwarf stars in the Milky Way from \cite{Amarsi19} on the high sSFR end (which we use as a lower limit for the expected [O/Fe]$_{\rm CCSN}$).
The average [O/Fe] found in three sSFR bins is shown in the background (see caption of Figure \ref{fig: summary_bins_no_BSG} and Table \ref{tab: bins}).
Small gray data points were not used in the analysis as it is not clear how to consistently correct them for systematic offsets.
Gray vertical band indicates the range of sSFR in which we expect the turnover based on the literature SN Ia DTD (see Figure \ref{fig: SN Ia DTD examples}).
Dark colored ranges show the [O/Fe]--sSFR relations calculated with equation \ref{eq: OFe} for different SN Ia DTD and parameter choices. 
Left panel: power-law DTD with slope $\alpha_{\rm Ia}$=-1.1 and different minimum SN Ia delay times $\tau_{\rm Ia; min}$.
Right panel: exponential DTD for two different values of the $\tau_{\rm Ia}$ parameter).
The colored ranges span between the relations calculated with C$_{\rm Ia/CC}$=0.74 (upper edges) and C$_{\rm Ia/CC}$=2.5 (bottom edges) -- see text for the details.
We also plot the relations followed by TNG100-1-like galaxies (brown line, left panel) and EAGLE-like galaxies (turquoise line, right panel) and the average MW evolutionary track reconstructed with disk stars assuming constant star formation history (thick orange line, see Section \ref{sec: obs data, MW}).
}
\label{fig: results_vs_models}
\end{figure*}
In Figure \ref{fig: results_vs_models} we compare the observational [O/Fe] -- sSFR relation with a broad range of model relations.
We indicate the minimum level of [O/Fe] enrichment expected at high sSFR based on MW metal-poor stars
as discussed in Section \ref{sec: results,high sSFR} and bring the data to the `intermediate + $\Delta_{\rm d}$' common baseline which allows to match this value.
The gray vertical band roughly indicates the range of sSFR below which we expect the change of slope of the relation given the minimum time required to form a white dwarf (the right edge of the band) and the range of SN Ia DTD compared in Figure \ref{fig: SN Ia DTD examples}.
Current observations suggest that the possible turnover is located at log$_{10}$(sSFR [yr$^{-1}$])>-9 ($\tau_{SF}<$1 Gyr), which is broadly consistent with expectations, but does not constrain the models.
We note that we independently find the same log$_{10}$(sSFR)>-9 limit when considering old MW disk stars (Section \ref{sec: obs data, MW}) and star-forming properties of other galaxies (Section \ref{sec: results}).
We show the MW evolutionary track crudely inferred from MW disk stars assuming constant star formation history (thick orange line in both panels in Figure \ref{fig: results_vs_models}).
As discussed in Section \ref{sec: obs data, MW}, other plausible star formation histories tend to shift the MW track leftwards in the [O/Fe]--sSFR diagram, but overall do not strongly affect the result.
It can be seen that the MW-based relation is consistent with the current constraints inferred from the properties of other star forming galaxies.
This is astonishing, and suggests that by taking a similar astro-archeological approach and following a more careful analysis, one can obtain tight constraints on the overall [O/Fe]--sSFR relation characterising star forming galaxies across cosmic time.
\subsection{SN Ia delay times}
The two panels of Figure \ref{fig: results_vs_models} show [O/Fe]--sSFR relations obtained with equation \ref{eq: OFe} for different SN Ia DTD and C$_{\rm Ia/CC}$ choices.
In both cases $f_{\rm Ia}$ is normalized to unity when integrated over the Hubble time, i.e. 1=N$_{\rm Ia0}$ $\int_{\tau_{\rm Ia,min}}^{\tau_{\rm H}}$ f$_{\rm Ia}$(t`) dt` and [O/Fe]$_{\rm CCSN}$ is fixed to the average [O/Fe] value found in the high sSFR data bin.
\\
In the left panel, we use a power-law SN Ia DTD: $f_{\rm Ia}\propto t^{- \alpha_{\rm Ia}}$ at $t\geqslant \tau_{\rm Ia,min}$ and zero at  $t<\tau_{\rm Ia,min}$.
We assume $\alpha_{\rm Ia}$=1.1 (average between the slope derived by \citealt{MaozGraur17} and used in Illustris-TNG $f_{\rm Ia}$). Steeper slopes were also reported \citep[e.g.][]{Heringer19} but as long as $\alpha_{\rm Ia}$ is close to unity, the exact choice of its value has a minor effect on the relation compared to other factors (see additional examples in the appendix \ref{sec, app: results vs models}).
$t^{-1}$-like DTD is generically found in double-degenerate SN Ia progenitor scenarios (i.e. involving two WDs whose merger triggers the explosion, e.g. \citealt{IbenTutukov84,Webbink84}). In such scenarios the time until SN Ia explosion depends steeply on the separation of the progenitor binary, driven to merger via gravitational waves emission.
Such a DTD is consistent with a variety of observational estimates \citep{MaozMannucciNelemans14, Strolger20}.
These observations mostly probe delay times of 1-10 Gyr, and the time at which SN Ia begin to significantly contribute is not constrained. The possibility that a power-law DTD continues to shorter delay times cannot be ruled out, and short $\tau_{\rm Ia,min}\sim$40 Myr are favoured if a $t^{-1}$-like SN Ia DTD is fitted to the cosmic SN Ia rate \citep{MaozGraur17}.
However, double-degenerate SN Ia scenarios typically require at least a few 100 Myr to $\sim$1 Gyr after the formation of the most massive WDs to start producing SN Ia according to a power-law DTD and the earlier behaviour is uncertain \citep[e.g.][]{MaozMannucciNelemans14}.
In Figure \ref{fig: results_vs_models} we compare the relations for two minimum SN Ia delay times $\tau_{\rm Ia,min}$=40 and 400 Myr (light and dark blue areas, respectively).
\\
\newline
In the right panel, we use an exponential SN Ia DTD:
$f_{\rm Ia}\propto e^{-t/ \tau_{\rm Ia}}$ at $t>\tau_{\rm Ia,min}$ and zero  $t<\tau_{\rm Ia,min}$. \cite{Strolger20} consider both individual galaxies and cosmic SN Ia rates and star formation histories and show that such SN Ia DTD parametrisation is also consistent with observations.
A more concentrated DTD resulting from the exponential form is expected in some single degenerate SN Ia formation scenarios (i.e. involving mass accretion onto the WD from a close non-degenerate companion star, e.g. \citealt{Whelan73,Nomoto82}). 
It can be seen that this leads to a steeper [O/Fe] decline in the intermediate sSFR range and flattening on the low sSFR side earlier than power-law-like DTD due to the scarcity of SN Ia with very long delay times.
Again, we show the relations for two example characteristic timescales $\tau_{\rm Ia}$=0.7 and 3 Gyr (light and dark green areas, respectively) and assume $\tau_{\rm Ia,min}$=40 Myr.
To satisfy the cosmic SN Ia rate constraints, assuming an exponential SN Ia DTD, long $\tau_{\rm Ia} \sim$2 Gyr are required \citep[e.g.][]{Schaye15, Strolger20}. This leads to a turnover in the [O/Fe]--sSFR relation at distinctly lower sSFR than the power-law DTD with $\tau_{\rm Ia,min}$=40 Myr used to fit the same cosmic SN Ia rate constraints. 
While we cannot rule out any of the discussed models with current data, further constraints on the turnover sSFR can help to distinguish between such scenarios.
\\
Different SN Ia formation channels may operate in nature, leading to a more complex overall $f_{\rm Ia}$ than considered in this section \citep[e.g.][]{Greggio10,Nelemans13,MaozMannucciNelemans14,LivioMazzali18,Rajamuthukumar23}.
While this makes the interpretation in light of a particular SN Ia formation scenario challenging, with better constraints [O/Fe] -- sSFR relation can help to infer valuable information about the general properties of the SN Ia population. 
In particular, as long as the iron mass ejected per SN Ia is high compared to that produced per CCSN: i) the turnover point on the high sSFR side of the relation can be linked to the minimum timescale at which SN Ia start to significantly contribute to iron enrichment
and ii) the range of sSFR over which the relation shows steep evolution before it saturates on the low sSFR side carries information about the extent and importance of the long delay time tail of the SN Ia DTD.
If the functional form of f$_{Ia}$ is known (or inferred from other observations), then the evolution of the low sSFR part of the relation alone can shed light i).

\subsection{Iron yields}
Each of the colored areas shown in Figure \ref{fig: results_vs_models} spans between the relations calculated with 
\begin{math}
C_{\rm Ia/CC} = \rm \frac{m^{Ia}_{Fe}}{m^{CCSN}_{Fe}} \frac{N_{Ia0}}{k_{CCSN}} = 0.74
\end{math} (upper edges) and C$_{\rm Ia/CC}$ = 2.5 (bottom edges). 
SN Ia are expected to eject most of the mass in iron-group elements, with the typical iron mass $\rm m^{Ia}_{Fe}\approx$0.7 $\Msun$ \citep{Nomoto97,Mazzali07,MaozMannucciNelemans14,Kobayashi20} and the relative formation efficiency of CCSN to SN Ia is close to 10 \citep[e.g.][]{MadauDickinson14,MaozGraur17,Strolger20}. Both quantities appear known to within a factor of a few.
The iron mass that is ejected per CCSN event is by far the most uncertain ingredient of $C_{\rm Ia/CC}$.
Observational estimates span a broad range \citep{Muller17,Anderson19,Rodriguez21,Rodriguez22,Martinez22}
and indicate systematically higher iron masses produced by stripped-envelope supernovae ($\rm m^{CCSN}_{Fe} \gtrsim$0.07 $\Msun$, e.g. \citealt{Anderson19,Afsariardchi21,Rodriguez22}) than by normal, hydrogen-rich CCSN events (with the average $\rm m^{CCSN}_{Fe}\sim$0.03 - 0.045 $\Msun$, e.g. \citealt{Rodriguez21,Martinez22}).
In this view, the average $\rm m^{CCSN}_{Fe}$ depends on the relative mixture of different types of CCSN happening in the Universe. Predicted iron yields also vary significantly between the CCSN explosion models \citep[e.g.][]{WoosleyHeger07,PejchaThompson15,Sukhbold16,Curtis19,Ebinger20,Ertl20,Schneider21, Imasheva23, Sawada23}.
Assuming $\rm m^{Ia}_{Fe}$=0.74 $\Msun$ \citep[i.e. nucleosynthesis yields of SNe Ia in the commonly used W7 model][]{Nomoto97,Iwamoto99} and $\rm \frac{N_{Ia0}}{k_{CCSN}}$=1/10, the considered $C_{\rm Ia/CC}$ values correspond to a broad range of $\rm m^{CCSN}_{Fe}$=0.03--0.1 $\Msun$.
It can be seen that a full variety of presented model relations is broadly consistent with the current constraints.\\
The [O/Fe] value at which the relation saturates on the low sSFR side strongly depends on $C_{\rm Ia/CC}$ and can inform the relative iron production efficiency in SN Ia and CCSN.
Constraining this requires extending the sample of  galaxies with available iron abundances to log$_{10}$(sSFR)$\lesssim$-10.5, i.e. accounting for main sequence MW-like galaxies at low redshifts.

\section{Discussion and future prospects}
\label{sec: discussion}
Improving the constraints on the [O/Fe]--sSFR relation requires predominantly expanding the sample of galaxies with iron abundance determination and having a good handle on the related systematic uncertainties.
Contrary to iron abundances, sSFR and Z$_{\rm O}$ are already known for large samples of star forming galaxies.
Those samples will only grow with the instruments like MOONS, expected to conduct a SDSS-size survey of galaxies at z$\sim$1.5 and provide their Z$_{\rm O}$ measurements \citep[][]{Maiolino20}.
Furthermore, with JWST `direct' (collisional-line) gas-phase oxygen abundances can now be determined at z$\gtrsim$3 \citep[e.g.][]{Curti23,Nakajima23}.
While oxygen abundance determinations suffer from significant systematic uncertainties, these are relatively well characterised in the literature \citep{KewleyEllison08,Telford16, MaiolinoMannucci19,Kewley19} compared to issues associated with iron abundance determinations and can be dealt with.
Furthermore, we argue that the collisionally excited line oxygen abundance scale (conventionally used in `direct' gas-phase oxygen abundance measurements) is inconsistent with stellar metallicity measurements. This allows to reduce the biggest source of systematic uncertainty in the absolute oxygen abundance values relevant for this study.
\\
Currently, the sSFR range in which the relation shows the strongest evolution (and can be particularly constraining for the models) is probed by a single data point at log$_{10}$(sSFR)$\approx$-9.
This is partially due to the fact that galaxies with sSFR around this value are the most likely to be found at redshifts 0.5$\lesssim$z$\lesssim$2 \citep[e.g.][]{Popesso23}, where none of the currently available methods is suitable to measure Z$_{\rm Fe}$ (see Section \ref{sec: metallicity - techniques}).
However, at least at log$_{10}$(sSFR)$\gtrsim$-9, the sample can be expected to grow with ongoing efforts to obtain rest-frame UV spectra of main sequence galaxies z$\sim$2.5-4 (where the rest-frame UV is conveniently shifted to optical).
This regime is particularly important to pinpoint the high sSFR turnover of the [O/Fe]--sSFR relation, interesting as a potential probe of the poorly-constrained short delay time end of the SN Ia DTD.
However, we emphasize that the turnover point and the overall [O/Fe]--sSFR relation are not sensitive to SN Ia DTD \emph{per se}, but to SN Ia iron production DTD (which is an important distinction if SN Ia iron yields were to correlate with some intrinsic SN Ia progenitor properties and favour certain delay times).
Z$_{\rm Fe}$ obtained for galaxies in the intermediate sSFR range relies on SPS models for the interpretation of the observed rest-frame UV spectra. As discussed in Section \ref{sec: metallicity - techniques}, different SPS models can lead to substantial differences in Z$_{Fe}$ and their origin needs to be better understood. To this end, studies of local metal-poor galaxies are desirable, where more details in the UV spectra are available and can be compared with the models \citep[e.g.][]{Senchyna22}, and where ideally several methods can be used to infer metallicity to test their consistency.
\\
Constraints are also lacking on the low log$_{10}$(sSFR)$\lesssim$-10.5 side of the relation, occupied by low redshift main sequence galaxies with masses comparable to the Milky Way. This regime is also important for differentiating between models, as discussed in the previous section.
 In principle, the low sSFR sample could be expanded with BSG/RSG based methods, provided that they can be adapted to target the iron-group element abundance instead of providing metallicity estimates relying on both iron-group and $\alpha$-element lines. While it would be beneficial to obtain Z$_{Fe}$ of such galaxies using the same method as applied for high redshift objects, this would require the use of space telescopes to access the UV emisson and even then obtaining tight constraints might not be feasible given their low SFR (i.e. they can be expected to be UV-faint).
 \\
 Finally, the average [O/Fe] at which the relation flattens at log$_{10}$(sSFR)$\gtrsim$-7.6 is expected to reflect some cosmic average [O/Fe] enrichment from massive, low metallicity stars with no/negligible contribution from SN Ia. We argue in Section \ref{sec: results,high sSFR} that <[O/Fe]$_{MW}$> provides a lower limit on this value, which can be challenged with future observations. The scatter/prevalence of galaxies that are outliers in terms of [O/Fe] in this sSFR regime can in turn yield valuable constraints on the efficiency of formation of rare, massive-star related explosions predicted to eject material with [O/Fe] ratios significantly different than produced by regular CCSNe. In particular, pair instability supernovae \citep[e.g.][]{HegerWoosley02,Takahashi18} originating from massive ($>$ a few 100 $\Msun$) metal-poor progenitors exploding within just a few Myr after the star formation are predicted to eject material with [O/Fe]$<$0. 
 Similarly low [O/Fe] ratio is predicted in some of the hypernovae models \citep[e.g.][]{Grimmet21}.
 A signature massive pair instability supernova abundance pattern has recently been found in a Milky Way halo star \citep{Xing23}. Such explosions could potentially explain the surprisingly low [O/Fe]$<$0 reported by \cite{Kojima21} for two of their extremely metal poor local galaxies with log$_{10}$(sSFR)$\gtrsim$-7.6  \citep[e.g.][]{Isobe22,Goswami22}.

\section{Conclusions}
To date, our knowledge of star-forming metallicity relies mostly on gas-phase oxygen abundances.
This is unfortunate, since the main element responsible for the differences in the evolution and fate of stars formed with different metallicities is iron.
It is the varying abundance of iron that determines the strength of radiation-driven winds, feedback (mechanical and chemical), and ionising radiation input from the stellar population. This makes it the most critical element for properly describing and understanding the evolution and properties of star-forming galaxies, stars, stellar afterlives and related transients.
Oxygen and iron abundances are known to evolve on different timescales and observations of old stars reveal that their relative abundances can differ by a factor of $\gtrsim$5 when compared to solar ratio. This means that the former is not a good proxy for the latter.
Furthermore, determining the star-forming iron abundances for large, representative samples of galaxies is likely to remain a challenge for the foreseeable future.
\\
To remedy the situation, we investigate the [O/Fe] -- sSFR relation, expected to be tightly followed by star forming galaxies on theoretical grounds as discussed in Section \ref{sec: expectations}. Due to its apparent universality, the relation can provide a reasonable and simple way of translating the readily available oxygen abundances to iron abundances. We further explore this possibility and derive the iron-based metallicity dependent cosmic star formation history in paper II (Chruslinska et al, in prep.).
In this study, we present the first observational determination of [O/Fe] -- sSFR relation over a wide range of sSFR:
\begin{itemize}
\item We compile a sample of star-forming galaxies with available iron abundances from the literature (Section \ref{sec: obs data} and Table \ref{tab:data}) and bring the data to a common baseline by correcting for the known systematic offsets related to Z$_{\rm O}$ and Z$_{\rm Fe}$ determinations.
\item The resulting relation shows a clear sign of evolution towards lower [O/Fe] with decreasing sSFR and a hint of flattening at log$_{10}$(sSFR)$>$-9, consistent with theoretical expectations (Figures \ref{fig: summary_bins_no_BSG} and \ref{fig: results_vs_models}).
\item We independently reconstruct the [O/Fe]--sSFR relation from old MW disk stars  (Section \ref{sec: obs data, MW}).
The MW relation is remarkably consistent with the one inferred from the properties of star forming galaxies, as long as the `direct' oxygen abundances are placed on recombination-line abundance scale.
\item The above conclusion argues against the use of collisionally excited line abundance scales in combination with stellar metallicity measurements (Section \ref{sec: results,high sSFR}).
\item The agreement between the MW-based relation and the relation obtained by populating the [O/Fe] -- sSFR plane with present-day properties of star forming galaxies reinforces the fact that the relation can be used both i) to follow the [O/Fe] evolution of regular star forming galaxies as they age, and ii) to deduce the typical star-forming [O/Fe] ratio of galaxies of a given sSFR.
\end{itemize}
We compare the [O/Fe] -- sSFR relations resulting from the EAGLE and Illustris-TNG cosmological simulations and show that the differences between them can be fully attributed to the differences in the assumed SN Ia delay time distribution and the relative SN Ia and CCSN formation efficiency and metal yields (Section \ref{sec: simulations}). 
Based on that, we conclude the following:
\begin{itemize}
\item The relation is driven by the differences in the timescales (probed by sSFR) on which stellar sources enrich the interstellar medium in oxygen (prompt CCSNe) and iron (both CCSN and delayed SN Ia).
\item 
Its main characteristics are determined by stellar evolution and interactions (setting the formation efficiency of different types of supernovae progenitors and timescales on which they explode/collapse), core-collapse physics and supernovae explosion properties (determining which stellar progenitors explode and the associated metal yields) and not strongly influenced by large scale processes (e.g. galaxy mergers, feedback).
\item With better constraints, the [O/Fe] -- sSFR relation can shed light on the uncertain SN Ia delay time distribution, CCSN metal yields and formation efficiency of rare explosions of the most massive, metal-poor stellar progenitors (Section \ref{sec: results vs models}).
\end{itemize}
In particular, the relation could help to constrain the minimum delay at which SN Ia start to contribute significantly to iron enrichment, which is difficult to constrain with other observations and strongly dependent on the SN Ia formation scenario.
Improving the constraints on the [O/Fe]-sSFR relation requires expanding the sample of galaxies with measured iron abundances and having better control of the systematic uncertainties. Especially the origin of the differences between the iron abundances derived (directly and indirectly) from different parts of the rest-frame UV galaxy spectra  (continuum, wind features, ionising part) and with different spectral population synthesis models needs to be better understood.

\begin{appendix}

\section{Cosmological simulations example} \label{sec, app: model}

\begin{table}[]
\resizebox{0.9\columnwidth}{!}{
\begin{tabular}{lcc}
Parameter                            & EAGLE                   & TNG100               \\ \hline
k$_{\rm CCNS} [\Msun^{-1}]$                       & 0.017 [1]                  & 0.0118 [2]              \\
N$_{\rm Ia0} [\Msun^{-1}]$                        & 2 $\cdot 10^{-3}$ [1]      & 1.3 $\cdot 10^{-3}$ [2] \\
m$^{\rm Ia}_{\rm Fe} \ [\Msun${]}    & 0.74  [1]                 & 0.74   [2]             \\
m$^{\rm CCSN}_{\rm Fe}  \ [\Msun${]} & 0.04                    & 0.04                 \\
{[}O/Fe{]}$_{\rm CCSN}$              & 0.68                    & 0.58                 \\
m$^{\rm CCSN}_{\rm O}  \ [\Msun${]}  & 0.96                    & 1.2                 \\
f$_{\rm Ia}$                         & $\propto$ e$^{-t/2 Gyr}$ [1] & $\propto$ t$^{-1.12}$ [3] \\ \hline
\end{tabular}
}
\caption{
Parameters of Equation \ref{eq: OFe} used to model the tracks of EAGLE-like and TNG-like MS galaxies in the [O/Fe] -- sSFR plane in Figure \ref{fig: mock simulations}. 
m$^{\rm CCSN}_{\rm Fe}$ and {[}O/Fe{]}$_{\rm CCSN}$ were chosen to match the relation followed by the simulated galaxies (see text), which defines m$^{\rm CCSN}_{\rm O}$. The remaining parameters come from the settings of the cosmological simulations: [1] \cite{Schaye15}, [2] \cite{Pillepich18}, [3] \cite{Naiman18}.
}
\label{tab: simulations parameters}
\end{table}

\begin{table*}[]
\begin{center}
\begin{tabular}{ c c c c c c c c c c}
EAGLE-like & SFMR      &       &             &       &       &       &       &       &      \\ \hline
a$_{\rm SFR}$ & 0.836 & 0.852 & 0.882 & 0.902 & 0.915 & 0.919 & 0.923 & 0.942 & 0.95 \\
b$_{\rm SFR}$ & -8.83 & -8.83 & -8.74 & -8.57 & -8.48 & -8.33 & -8.11 & -8.09 & -8   \\
redshift             & 0     & 0.1   & 0.5   & 1     & 1.49  & 2.01  & 3.02  & 3.98  & 5.04 \\ \hline \\
  TNG100-like  & SFMR      &       &            &       &       &       &       &       &      \\ \hline
a$_{\rm SFR}$ & 0.8   & 0.8   & 0.81  & 0.77  & 0.77  & 0.8   & 0.8   & 0.8   &      \\
b$_{\rm SFR}$ & -8.15 & -7.72 & -7.64 & -6.97 & -6.83 & -6.8  & -6.1  & -6    &      \\
redshift             & 0     & 0.75  & 1     & 1.75  & 2     & 3     & 4     & 5     &     \\ \hline
\end{tabular}%
\end{center}
\caption{ Parameters of the SFMR: log$_{10}$(SFR) = a$_{\rm SFR}$ log$_{10}$(M$_{*}$)+b$_{\rm SFR}$ used to model SFH of mock EAGLE-like and TNG-like MS galaxies.
}
\label{tab: SFMR simulations}
\end{table*}
In Table \ref{tab: simulations parameters} we summarize the choice of parameters of Equation \ref{eq: OFe} used to model the tracks of EAGLE-like and TNG-like main sequence galaxies in the [O/Fe] -- sSFR plane shown in Figure \ref{fig: mock simulations}. 
Table \ref{tab: SFMR simulations} summarizes the parameters of the the SFMR used to describe the star formation histories of EAGLE-like and TNG-like main sequence galaxies.
In case of TNG we use the fits from \citet{Donnari19} up to redshift=2 (Table 3 therein). To obtain the rough SFMR of the EAGLE galaxies and extend the TNG SFMR to redshift=5 we use the SFR and M$_{*}$ of simulated galaxies selected as described in Section \ref{sec: simulations} and perform least-square linear fit in log$_{10}$(SFR)-log$_{10}$(M$_{*}$) plane. We interpolate between the redshift bins. If needed, at higher redshifts we use the SFMR from redshift=5.

\section{Observational results used in this study: further discussion} \label{sec, app: obs data}

\begin{figure*}
\centering
\includegraphics[scale=0.55]{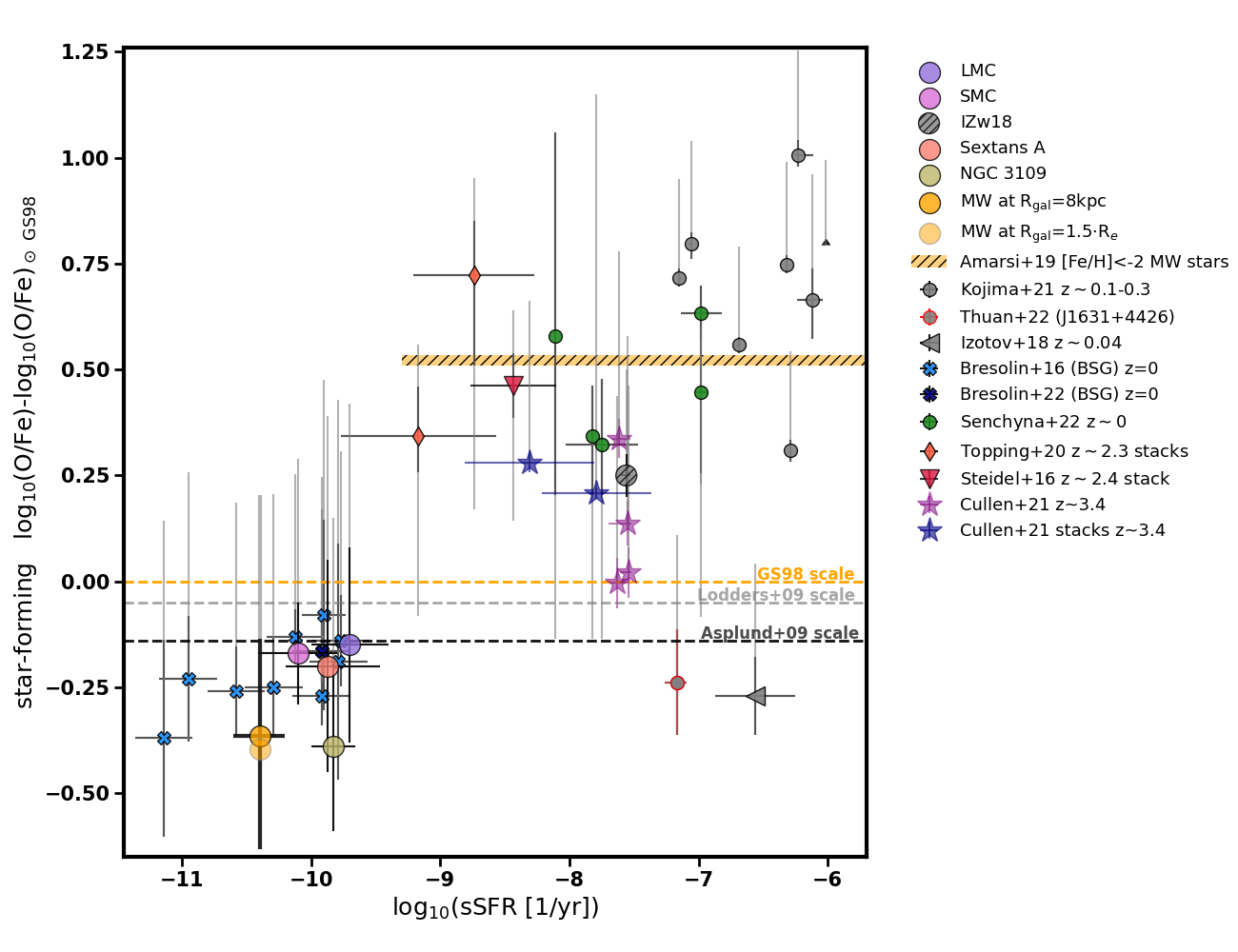}
\caption{ 
Same as Figure \ref{fig: data}, but excluding data points with `indirect' Fe or O determinations.
}
\label{fig: data_excl_indirect}
\end{figure*}
\subsubsection{Old Milky Way stars}

\begin{figure}
\centering
\includegraphics[scale=0.55]{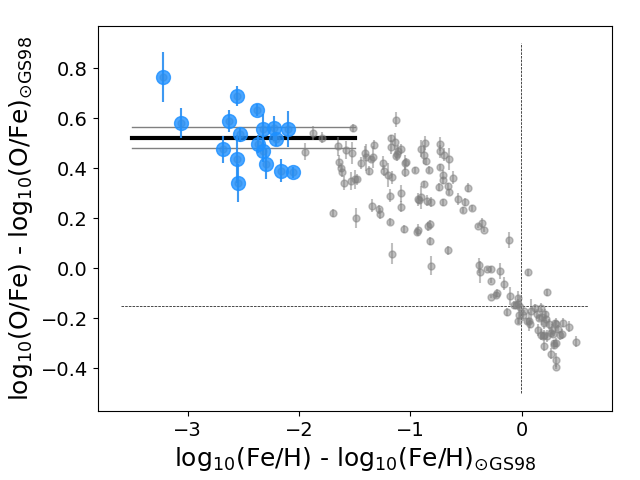}
\caption{ 
[O/Fe] versus [Fe/H] of MW disk and halo dwarf stars from \cite{Amarsi19}. Blue points indicate stars which were used to estimate the [O/Fe]$_{\rm CCSN}$ shown in Figure \ref{fig: data}. Their average <[O/Fe]$_{MW}$>=0.52 dex is indicated by the thick black line. Thin gray lines span between 13 and 99.8 percentiles of the average ([O/Fe]=0.48-0.56 dex).
\cite{Amarsi19} measure relative abundances with respect to the Sun and derive log$_{10}$(O/Fe)$_{\odot}$=1.179 dex. 
For consistency with the rest of the analysis, we convert those relative measurements to absolute abundances and place them on GS98 solar scale.  The thin dashed lines show the solar values from \cite{Amarsi19}.
}
\label{fig: Amarsi19}
\end{figure}

Figure \ref{fig: Amarsi19} shows the [O/Fe] -- [Fe/H] relation for the sample of MW disk/halo stars from \cite{Amarsi19}.
We highlight the stars used to determine the average <[O/Fe]$_{MW}$> value at the low metallicity [Fe/H]$<$-2 part of the relation. Table \ref{tab: MW disk O/Fe in age bins} summarizes the <[O/Fe]> of MW disk  stars grouped in 14 age bins as discussed in Section \ref{sec: obs data, MW}.

\subsubsection{Present-day Milky Way}

As discussed in \ref{sec: obs data, MW}, to obtain present-day [O/Fe] Milky Way we combine the HII region-based oxygen abundance gradient obtained by \cite{Arellano-Cordova20} and the iron abundance gradient determination from Galactic open clusters from \cite{Spina22}. We assume a present day Milky Way SFR = 2$\pm 0.5 \Msun$/yr and stellar mass of $M_{*}=5 \pm 1 \times 10^{10} \Msun$ \cite{Bland-Hawthorn16,Elia_2022}.
\\
We caution that the adopted {[Fe/H]} may somewhat underestimate the star-forming iron abundance for two main reasons: i) it is based on metallicities of open clusters, whose ages span a broad range from a few Myr to $\sim$1 Gyr (and therefore, not only recently formed objects) and ii) current [Fe/H] determinations may be underestimated for the youngest objects.
Puzzlingly, literature [Fe/H] abundance estimates of Galactic star forming regions and open clusters with ages $<$100 Myr (and especially $<$10 Myr) are lower than those of the older ones \citep[e.g.][and references therein]{Spina22}. \cite{Baratella20} show that this anomalous behaviour might be attributed
 to issues with abundance analysis approaches.
 In particular, chromospheric and magnetic activity effects on line formation that are not accounted for in atmospheric models of dwarf stars (used in abundance estimates of young clusters) but expected to be particularly strong in young stars may lead to underestimated [Fe/H].
 \cite{Baratella20} propose a new approach to bypass those issues and find [Fe/H] = -0.01 to 0.06 dex for a sample of stars from 5 open clusters younger than 150 Myr and located at R$_{gal}$=7.72 - 8.66 kpc.
 Using the [Fe/H] metallicity gradient from \cite{Spina22}, we obtain [Fe/H]=0.018 dex at R$_{gal}$=8 kpc. Therefore, there is no clear indication that our assumed value is underestimated.

\begin{table}[]
\centering
\resizebox{0.8\columnwidth}{!}{%
\begin{tabular}{ccc}
\multicolumn{2}{c}{age [Gyr] bin edges} & <[O/Fe]> in bin\\ \hline
1.75 & 2.67 & 0.081 \\
2.67 & 3.48 & 0.057 \\
3.48 & 4.29 & 0.062 \\
4.29 & 5.1  & 0.095 \\
5.1  & 5.91 & 0.11  \\
5.91 & 6.72 & 0.12  \\
6.72 & 7.53 & 0.13  \\
7.53 & 8.34 & 0.14  \\
8.34 & 9.15 & 0.18  \\
9.15 & 9.96 & 0.22  \\
9.96 & 10.8 & 0.3   \\
10.8 & 11.6 & 0.39  \\
11.6 & 12.4 & 0.48 \\ \hline
\end{tabular}%
}
\caption{
Average [O/Fe] of Milky Way disk main sequence turn-off stars from the third data release of the GALAH survey \citep{Buder21} grouped in 14 age bins. Stellar ages come from the catalogue by \citet{Sharma18}. Sample selection criteria follow \citet{Hayden22}. Abundances are converted to \citet{GrevesseSauval98} solar scale. 
}
\label{tab: MW disk O/Fe in age bins}
\end{table}

\subsubsection{
Local star forming galaxies}\label{sec, app: local galaxies}
Figure \ref{fig: data} includes the present-day [O/Fe] and sSFR estimates for nearby star forming dwarf galaxies: Small and Large Magellanic Clouds (SMC and LMC, respectively), Sextans A, NGC3109.\\
For the Magellanic Clouds we use the average direct method HII regions based $Z_{O/H}$ reported by \cite{Dominguez-Guzman22}.
In contrast to big spirals, galaxies such as SMC and LMC are typically rather chemically homogeneous, i.e. there is little difference between the abundances estimated in their different regions \citep{Dominguez-Guzman22}.
We use iron atmospheric abundance estimates from individual stars in NGC 330 (with the age of $\sim$40 Myr) for the SMC and from NGC 1850 (younger than $\lesssim$100 Myr) for the LMC derived by \cite{Song21}. Those are the youngest clusters that were included in their analysis.
The adopted LMC iron abundance (12+log$_{10}$(Fe/H)=7.19) is further supported by the recent analysis of red supergiants in NGC 1850 cluster by \cite{Sollima22} and consistent with that derived from $\sim$100-200 Myr old Cepheids \citep{Lemasle17}. Note that it is 0.13 dex lower than the commonly used OB-star based estimate from \cite{Rolleston02}, which brings the present-day LMC $[O/Fe]$ ratio closer to the value expected given its global properties \citep[e.g.][]{RusselDopita92,Pagel98}.
We use the global LMC and SMC SFR and M$_{*}$ values from \cite{Skibba12}.
\\
For Sextans A we adopt the average of the direct method HII region based oxygen abundance estimates from \cite{Magrini05} and the average iron abundance determined from 3 A-type supergiant star UVES spectra by \cite{Kaufer04}.  Their uncertainty estimates include systematic errors due to uncertainties in the stellar atmospheric parameters.
We use the range spanning between the FUV and V-band SFR estimates from \cite{Hunter10}.
and the stellar mass reported in \citet{Woo08} to estimate its sSFR\footnote{
But note that the Sextans A stellar mass appears particularly uncertain, with values used in the literature  ranging from
log(M$_{*}$/$\Msun$)=6.24 following \cite{Lee06},
log(M$_{*}$/$\Msun$)=7.38 in \cite{delosReyes19},
up to log(M$_{*}$/$\Msun$)=8.14 estimated by \cite{Weisz11}. Using the latter would result in Sextans A sSFR comparable to that of the MW.}.
For NGC3109 we use the blue supergiant based metallicity derived mainly from
Fe-group elements lines (and therefore suitable as current iron abundance probe) provided by \cite{Hosek14}, direct method HII region based oxygen abundance from \cite{Pena07} and the stellar mass and SFR from \cite{Woo08}.

\subsubsection{
High redshift star forming galaxies}\label{sec, app: high z}

\cite{Steidel16} estimate the average [O/Fe] for star-forming galaxies at z$\sim$2.4 by using composite spectra of 30 main sequence objects. Photospheric and wind line features in the composite UV spectrum are fitted using both BPASS and Starburst99 models. Oxygen abundance is estimated with a number of methods (photoionization models, strong-line calibrations and the direct method including ADF=+0.24 dex offset to put the estimate on the recombination line scale).
We adopt their final estimate of (O/Fe)=4$\pm$1 (O/Fe)$_{\odot}$ as quoted in Figure 17 and the median sSFR given in Table 1 therein. 
\cite{Topping20} extend the above sample to 62 galaxies.
They construct two stacks based on the locations of their galaxies on the local BPT diagram.
The UV continuum spectra were analysed with BPASS models to determine the iron abundance. The  corresponding best fitting SPS model was used in the photoionization modelling to fit for oxygen abundances. \cite{Topping20} also show [O/Fe] estimates for several individual galaxies with the highest SNR, but their SFR and M$_{*}$ are not given. Therefore, we only show the estimates for the two stacks in Figure \ref{fig: data}.\\
\cite{Cullen21} obtain gas-phase and stellar metallicities for 4 star-forming galaxies and for two composite spectra at z$\sim$3.4.
Iron-based metallicities are determined from rest-frame UV spectra by fitting stellar-metallicity sensitive features with Starburst99 SPS models.
The gas-phase oxygen abundances are determined using the empirical calibration from \cite{Bian18}, built on the local analogues of high-redshift galaxies with direct metallicity estimates. Contrary to \cite{Steidel16} and \citet{Topping20}, \cite{Cullen21} do not use the inferred iron abundance as constraints in the oxygen abundance determination.
\\
\cite{Strom18,Sanders20,Strom22} estimate [O/Fe] using only rest-frame oxygen optical emission lines as constraints (i.e. without any UV constraints). 
\cite{Strom18} (further expanded in \citealt{Strom22}) determine abundances of both elements by simultaneously varying the metallicity of the input SPS models and gas oxygen abundance in photoionisation modelling when fitting for the observed line ratios. 
In contrast, \cite{Sanders20} first infer the oxygen abundance using direct method (including ADF=+0.24 dex offset), and use that as an input in the photoionisation modelling to infer the iron abundance.
All of those estimates used BPASS SPS models.
We only report median results from \cite{Strom18,Strom22}, in Table \ref{tab:data}, as the sSFR for individual galaxies are not given by the authors.

\subsubsection{
(Very) metal poor local dwarf galaxies}\label{sec, app: VMPDG}

\begin{figure}
\centering
\includegraphics[scale=0.5]{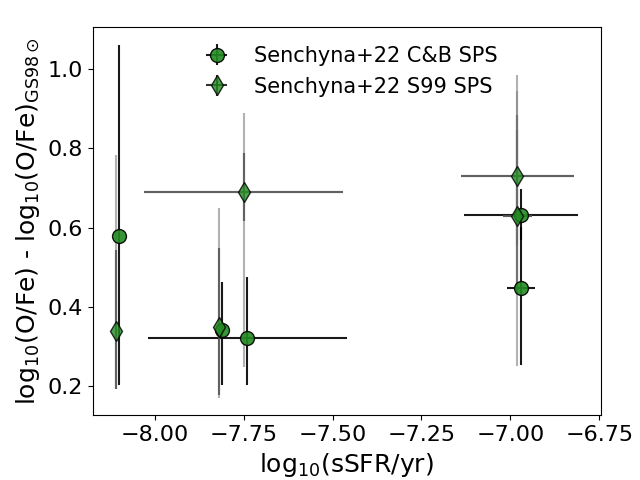}
\caption{ 
Comparison of the [O/Fe] for the sample from \cite{Senchyna22} where $Z_{\rm Fe}$ was derived with C\&B SPS models (circles, as plotted in Figure \ref{fig: data}) and S99 SPS models (diamonds, reported in Table \ref{tab:data}). We add a small offset in sSFR between the two estimates for easier comparison.
}
\label{fig: Senchyna22}
\end{figure}

\cite{Senchyna22} provide direct method HII region oxygen abundances and UV-continuum based iron abundances for 6 galaxies. 
They use different SPS models than used in the remaining studies referenced in this paper (which employ either S99 or BPASS models) and refer to the paper in preparation by Charlot \& Bruzual for their description.
They compare the $Z_{\rm Fe}$ values obtained using C\&B and S99 SPS models in table 7 therein. 
We show the differences in [O/Fe] resulting from the use of those two SPS models for their sample in Figure \ref{fig: Senchyna22}. We report the [O/Fe] obtained with S99 SPS in Table \ref{tab:data} for easier comparison with other results used in this paper.
\cite{Senchyna22}  shift their derived [O/Fe] ratios by ADF=+0.24 dex to put them on the recombination line oxygen abundance scale and include additional correction of $\sim$+0.11 dex to account for oxygen dust depletion.
We use the sSFR from \cite{Berg16} for J082555 and J104457 (no errors were given), sSFR from \cite{Senchyna19} for HS1442+4250 and sSFR from \cite{Senchyna17} for SB2 and SB82.
We do not find SFR/sSFR for J120202 and therefore, we do not include it in the analysis.
\\
The sample compiled by \cite{Kojima21} was observed as part of the Extremely Metal-Poor Representatives Explored by the Subaru Survey (EMPRESS) \citep{Kojima20} and supplemented by earlier literature results.
Their oxygen abundances have been derived with the direct method using HII region nebular oxygen lines. [Fe III]$\lambda$4658 line is also detected and the authors follow the method described in \cite{Izotov06} to estimate the iron abundance.
\cite{Kojima21} note that the dust depletion is negligible in their metal poor sample.
For the least massive object (J1142-0038) only the lower limit on the [O/Fe] could be determined.
\\
For IZw 18 we use the literature average oxygen and iron abundances listed in \cite{Lebouteiller13} (the last column of their Table 7) inferred from HII region observations
\footnote{Note that the resulting 12+log$_{10}$(O/H)=7.21
is on the high end of the 12+log$_{10}$(O/H)=7.1-7.2 suggested by the most recent direct method metallicity estimates \citep{Kehrig16}}.
\cite{Lebouteiller13} note that the dust content in IZw 18 is exceptionally low and dust depletion is expected to be insignificant. We therefore assume that its HII region-based abundance is a good measure of the iron content in the star forming material. We use the SFR and M$_{*}$ values from \cite{Zhou21}.\\
\cite{Kojima21} include a 0.2 dex systematic error in their iron abundance estimate due to varying Fe$^{2+}$ ionisation correction factors resulting from different models. We also include this systematic in the IZw18 iron abundance estimate, as it was obtained with the same method.

\begin{figure}
\centering
\includegraphics[scale=0.6]{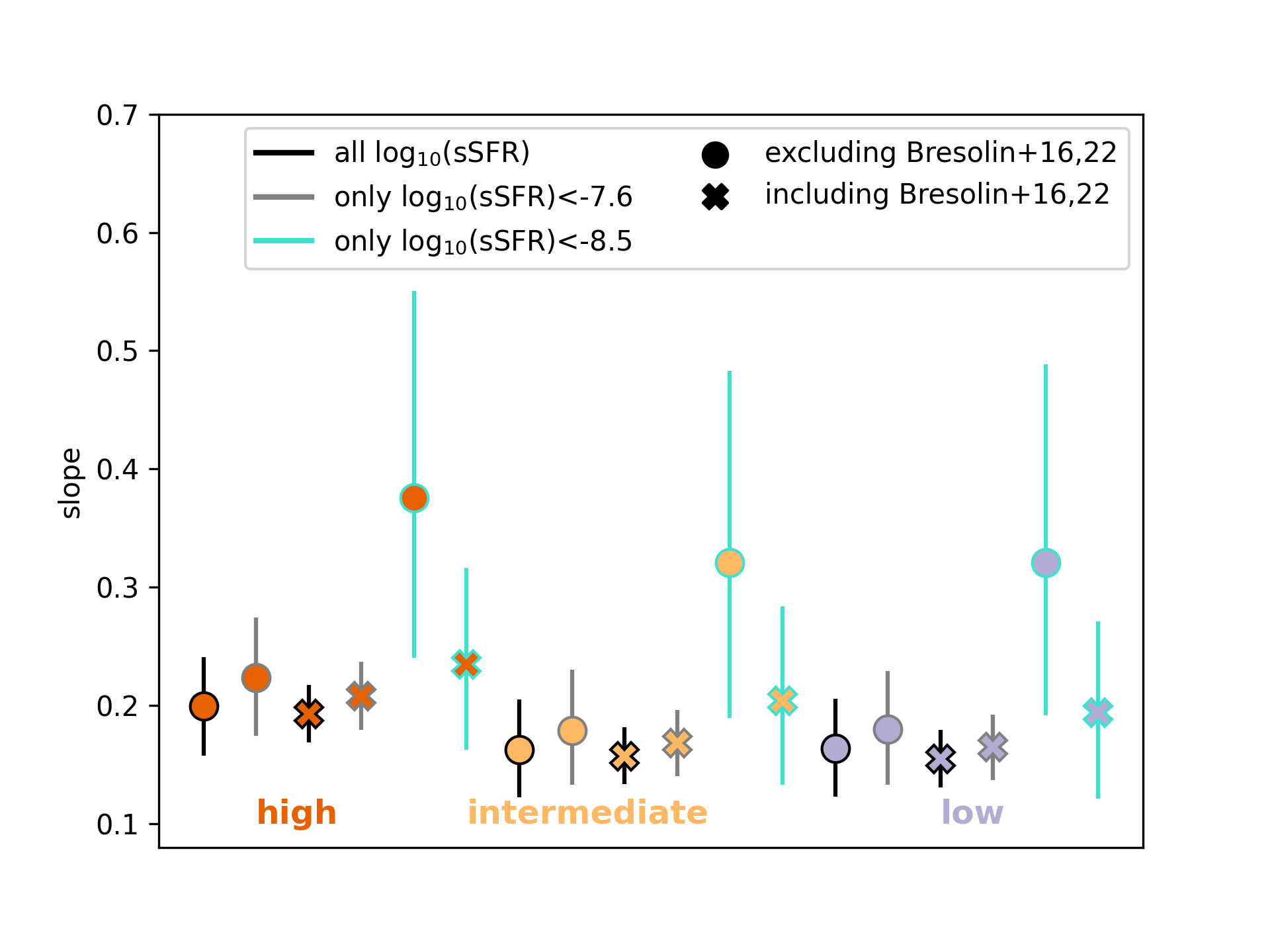}
\caption{ 
Slope of the linear fit to the [O/Fe]--log$_{10}$(sSFR) relation for different choices of the baseline (colours) and including/excluding of the sample from \cite{Bresolin16,Bresolin22} at low sSFR (crosses/circles).
We show the result obtained when fitting only the slope at log$_{10}$(sSFR)<-7.6 (symbols with grey edges), at log$_{10}$(sSFR)<-8.5 (symbols with blue edges) or when fitting the slope to the full sample (black edges).
Error bars range from 16 to 84 percentiles of the fits, the symbol indicates the median.
}
\label{fig: slopes}
\end{figure}

\section{Including the low sSFR sample with BSG based metallicity estimates}\label{sec,app: with BSG}

\begin{figure}[h!]
\centering
\includegraphics[scale=0.33]{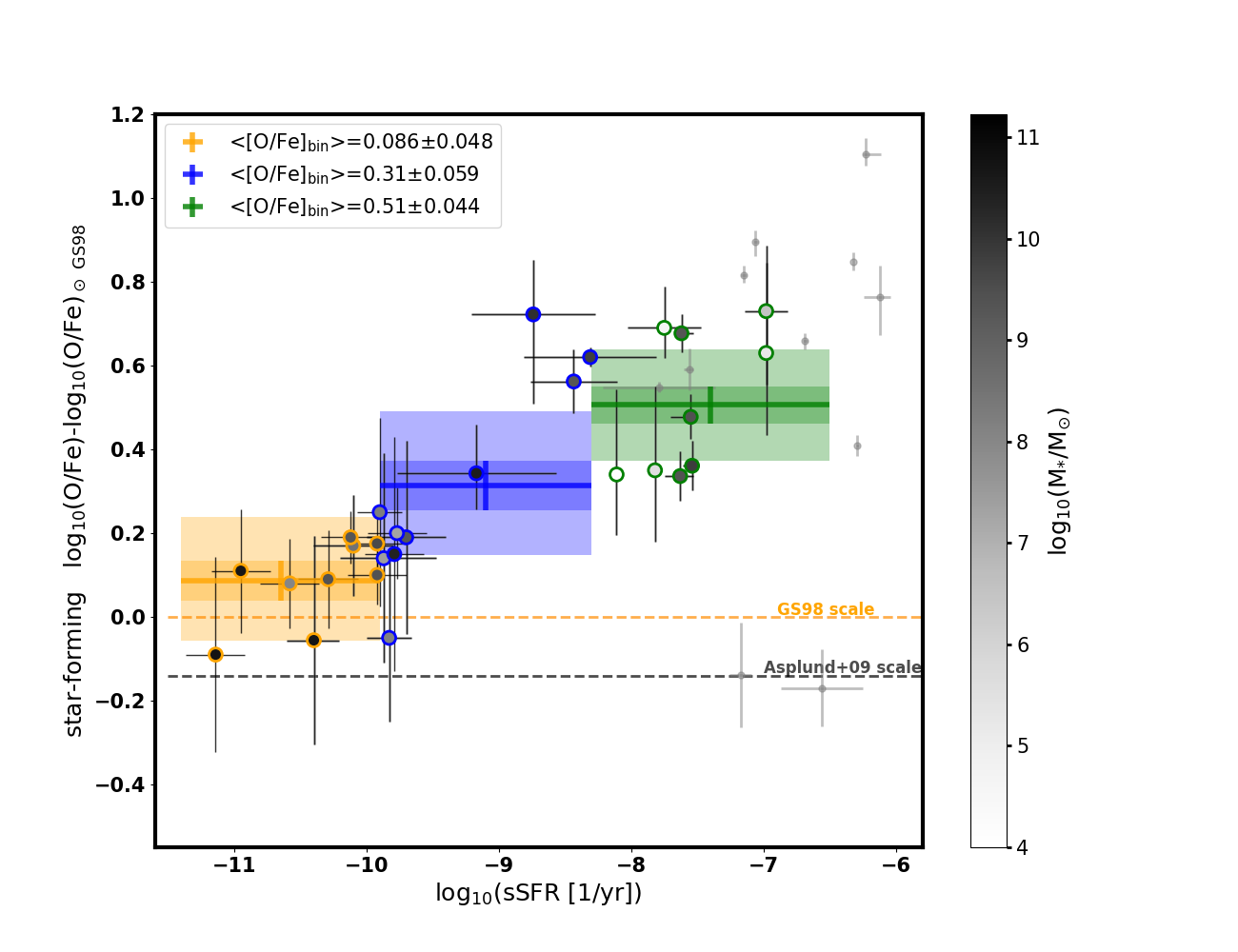}
\caption{ 
Star-forming [O/Fe] versus specific star formation rate relation including the sample from \cite{Bresolin16,Bresolin22} with BSG-based metallicities, assuming they probe Z$_{\rm Fe}$. Data points (big circles) were shifted to a common `intermediate+$\Delta_{\rm d}$' [O/Fe] baseline. Inner color of big circles indicates the stellar mass, outer color indicates the bin to which the data point belongs.
Values listed in the figure indicate median [O/Fe] in each of the log$_{10}$(sSFR) bins (see Table \ref{tab: bins with BSG} for the values found with different [O/Fe] baseline choices). 
Dark (light) area of each bin spans between 16-84 (0.13-99.87) percentiles of 10$^{5}$ draws of the average [O/Fe].
Small gray data points were not used in the analysis as it is not clear how to consistently correct them for systematic offsets. Data with indirect O or Fe estimates are not shown.
Thick horizontal lines indicate zero points for reference solar O and Fe abundance from \cite{GrevesseSauval98} (GS98) and \cite{Asplund09}.
}
\label{fig: results with BSG}
\end{figure}

Table \ref{tab: bins with BSG} summarizes the average <[O/Fe]> values found in three sSFR bins when including the low sFSR sample from \cite{Bresolin16,Bresolin22} with BSG-based metallicities and assuming those metallicities probe Z$_{\rm Fe}$.
Note that the bin edges are different than in the main analysis (Section \ref{sec: results} and Table \ref{tab: bins}) and were selected to include possibly equal number of data points in each bin.
Figure \ref{fig: results with BSG} shows the [O/Fe]--ssSFR relation with the resulting bins and data shifted to a common `intermediate + $\Delta_{\rm d}$' baseline (matching the expected high sSFR [O/Fe] level, see Section \ref{sec: results,high sSFR}).

\begin{table}[]
\centering
\resizebox{\columnwidth}{!}{%
\begin{tabular}{cccc}
\hline \\
 baseline & <[O/Fe]>  & <[O/Fe]> &  <[O/Fe]>  \\
 &  bin 1  & bin 2 & bin 3  \\ \hline
high                          & 0.085$\pm$0.15  & 0.34$\pm$0.18   & 0.62$\pm$0.12 \\
intermediate+$\Delta_{\rm d}$ & 0.087$\pm$0.15  & 0.32$\pm$0.18   & 0.51$\pm$0.13 \\
intermediate                  & -0.014$\pm$0.14 & 0.21$\pm$0.18   & 0.41$\pm$0.13 \\
low                           & -0.25$\pm$0.15  & -0.025$\pm$0.18 & 0.17$\pm$0.13 \\ \hline
\end{tabular}%
}
\caption{
Average <[O/Fe]> values in three log$_{10}$(sSFR) bins found for data selected as described in Section \ref{sec: baseline} but including the low sSFR galaxy sample with BSG based metallicities from \cite{Bresolin16,Bresolin22} and assuming that those measurements probe the iron abundance.
Column 1 indicates the choice of the common baseline. 
Bin 1 spans between log$_{10}$(sSFR [yr$^{-1}$])=-11 and -9.9 (9 data points), bin 2 between log$_{10}$(sSFR [yr$^{-1}$])=-9.9 and -8.3 (10 data points) and bin 3 between log$_{10}$(sSFR [yr$^{-1}$])=-8.3 and -6.5 (9 data points).
The errors span between 0.13 and 99.87 percentiles of the averages found when sampling the data within statistical errors (assuming split normal distribution for each data point) and the main value indicates the median.
}
\label{tab: bins with BSG}
\end{table}

\section{Comparison with theoretical expectations: additional figures}
\label{sec, app: results vs models}
\begin{figure*}
\centering
\includegraphics[width=1.0\textwidth]{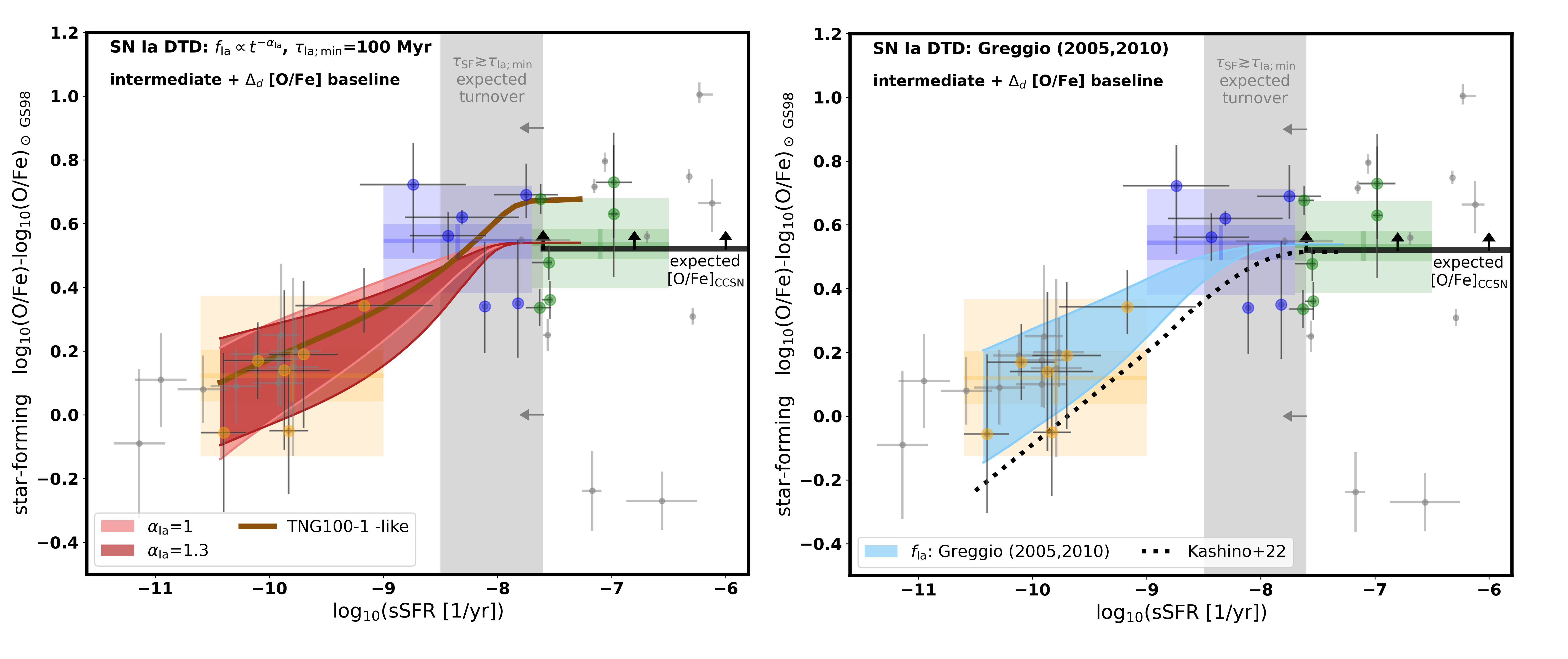}
\caption{ 
Observational [O/Fe]--sSFR relation compared with a range of model relations. Data points (big circles) were shifted to a common intermediate [O/Fe] baseline with additional offset due to possible oxygen dust depletion ($\Delta_{\rm d}$) included (see Figure \ref{fig: results_vs_models} and Section \ref{sec: results vs models}).
Dark colored ranges show the [O/Fe]--sSFR relations calculated with \ref{eq: OFe} for different SN Ia DTD and parameter choices. 
Left panel: power-law DTD with the minimum SN Ia delay time $\tau_{\rm Ia; min}$=100 Myr and different slopes $\alpha_{\rm Ia}$.
Right panel: analytic DTD for mixed single degenerate and double degenerate SN Ia scenario  from \cite{Greggio05,Greggio10}, as used by \cite{Kashino22}.
The colored ranges span between the relations calculated with C$_{\rm Ia/CC}$=0.74 (upper edges) and C$_{\rm Ia/CC}$=2.5 (bottom edges) -- see text for the details.
We also plot the relations followed by TNG100-1-like galaxies (brown line, left panel) and the relation from \cite{Kashino22} (black dotted line, right panel).
}
\label{fig: results_vs_models_extra}
\end{figure*}
Figure \ref{fig: results_vs_models_extra} compares the observational [O/Fe]--sSFR relation with model relations calculated as described in Section \ref{sec: results vs models} but for additional SN Ia DTD variations. In the left panel, we use a power-law DTD and vary the slope $\alpha_{\rm Ia}$. In the right panel, we show the relation for the analytical SN Ia DTD accounting for a mixed single-degenerate and double-degenerate progenitor SN Ia population \citep{Greggio05,Greggio10}, as used in \cite{Kashino22}.

\onecolumn
\begin{landscape}
\begin{longtable}{llllllll}
Object             & redshift    & log$_{10}$(sSFR/yr)    & {[}O/Fe{]}                  & systematics                                  & Z$_{\rm Fe}$ method & reference            & comment                        \\ \hline
Milky Way &
  0 &
  -10.4$^{+0.2}_{-0.19}$ &
  -0.36$^{+0.27}_{-0.27}$ &
  +ADF, +$\Delta_{\rm d}$ &
  young clusters &
  {[}1, 2, 3, 4{]} &
  R$_{\rm gal}$=8 kpc \\
Milky Way &
  0 &
  -10.4$^{+0.2}_{-0.19}$ &
  -0.4$^{+0.25}_{-0.25}$ &
  +ADF, +$\Delta_{\rm d}$ &
  young clusters &
  {[}1, 2, 3, 4{]} &
  R$_{\rm gal}$=6.8 kpc \\
LMC &
  0 &
  -9.7$^{+0.3}_{-0.3}$ &
  -0.15$^{+0.23}_{-0.23}$ &
  +ADF, +$\Delta_{\rm d}$ &
  stellar spectra &
  {[}5, 6, 7{]} &
   \\
SMC &
  0 &
  -10.1$^{+0.3}_{-0.3}$ &
  -0.17$^{+0.12}_{-0.12}$ &
  +ADF, +$\Delta_{\rm d}$ &
  stellar spectra &
  {[}5, 6, 7{]} &
   \\
IZw18 &
  0 &
  -7.56$^{+0.04}_{-0.03}$ &
  0.25$^{+0.05}_{-0.05}$ &
  +$\Delta_{\rm ICF,Fe}$ &
  HII region &
  {[}8, 8, 9{]} &
   \\
Sextans A &
  0 &
  -9.87$^{+0.33}_{-0.4}$ &
  -0.2$^{+0.25}_{-0.25}$ &
  +ADF, +$\Delta_{\rm d}$ &
  stellar spectra &
  {[}10, 11, 12, 13{]} &
   \\
NGC 3109 &
  0 &
  -9.83$^{+0.17}_{-0.17}$ &
  -0.39$^{+0.2}_{-0.2}$ &
  +ADF, +$\Delta_{\rm d}$ &
  BSG &
  {[}14, 15, 13{]} &
   \\
- &
  3.4 &
  -7.63$^{+0.12}_{-0.1}$ &
  -0.0038$^{+0.059}_{-0.059}$ &
  +ADF, +$\Delta_{\rm d}$, +$\Delta_{\rm SPS}$ &
  UV cont., S99 &
  {[}16{]} &
   \\
- &
  3.4 &
  -7.54$^{+0.07}_{-0.02}$ &
  0.021$^{+0.059}_{-0.059}$ &
  +ADF, +$\Delta_{\rm d}$, +$\Delta_{\rm SPS}$ &
  UV cont., S99 &
  {[}16{]} &
   \\
- &
  3.4 &
  -7.55$^{+0.15}_{-0.03}$ &
  0.14$^{+0.054}_{-0.054}$ &
  +ADF, +$\Delta_{\rm d}$, +$\Delta_{\rm SPS}$ &
  UV cont., S99 &
  {[}16{]} &
   \\
- &
  3.2 &
  -7.62$^{+0.05}_{-0.09}$ &
  0.34$^{+0.046}_{-0.046}$ &
  +ADF, +$\Delta_{\rm d}$, +$\Delta_{\rm SPS}$ &
  UV cont., S99 &
  {[}16{]} &
   \\
high M$_{*}$ stack &
  3.4 &
  -8.31$^{+0.5}_{-0.5}$ &
  0.28$^{+0.023}_{-0.023}$ &
  +ADF, +$\Delta_{\rm d}$, +$\Delta_{\rm SPS}$ &
  UV cont., S99 &
  {[}16{]} &
  stack, $\Delta_{\rm SPS}$=0.04 \\
low M$_{*}$ stack &
  3.4 &
  -7.79$^{+0.42}_{-0.42}$ &
  0.21$^{+0.013}_{-0.013}$ &
  +ADF, +$\Delta_{\rm d}$, +$\Delta_{\rm SPS}$ &
  UV cont., S99 &
  {[}16{]} &
  stack, $\Delta_{\rm SPS}$=0.6 \\
KBSS-LM1 stack &
  2.4 &
  -8.43$^{+0.33}_{-0.33}$ &
  0.46$^{+0.076}_{-0.076}$ &
  -ADF, +$\Delta_{\rm d}$ &
  UV cont., BPASS/S99 &
  {[}17{]} &
  stack \\
high stack &
  2.3 &
  -9.17$^{+0.6}_{-0.6}$ &
  0.34$^{+0.085}_{-0.12}$ &
  -ADF, +$\Delta_{\rm d}$, -$\Delta_{\rm SPS}$ &
  UV cont., BPASS &
  {[}18{]} &
  stack \\
low stack &
  2.3 &
  -8.74$^{+0.47}_{-0.47}$ &
  0.72$^{+0.21}_{-0.13}$ &
  -ADF, +$\Delta_{\rm d}$, -$\Delta_{\rm SPS}$ &
  UV cont., BPASS &
  {[}18{]} &
  stack \\
J1142-0038 &
  0.02 &
  -6.02$^{+0.03}_{-0}$ &
  0.79 &
  +$\Delta_{\rm ICF,Fe}$ &
  HII region &
  {[}19{]} &
  lower limit \\
J1429-0110 &
  0.03 &
  -6.12$^{+0.12}_{-0.08}$ &
  0.66$^{+0.091}_{-0.075}$ &
  +$\Delta_{\rm ICF,Fe}$ &
  HII region &
  {[}19{]} &
   \\
J1631+4426 &
  0.031 &
  -7.17$^{+0.09}_{-0.08}$ &
  -0.24$^{+0.12}_{-0.12}$ &
  +$\Delta_{\rm ICF,Fe}$ &
  HII region &
  {[}19, 21, 19{]} &
   \\
J0002+1715 &
  0.021 &
  -7.06$^{+0.02}_{-0.02}$ &
  0.8$^{+0.035}_{-0.027}$ &
  +$\Delta_{\rm ICF,Fe}$ &
  HII region &
  {[}19{]} &
   \\
J1642+2233 &
  0.017 &
  -6.23$^{+0.02}_{-0.12}$ &
  1$^{+0.027}_{-0.038}$ &
  +$\Delta_{\rm ICF,Fe}$ &
  HII region &
  {[}19{]} &
   \\
J2115-1734 &
  0.023 &
  -6.29$^{+0.01}_{-0.01}$ &
  0.31$^{+0.026}_{-0.026}$ &
  +$\Delta_{\rm ICF,Fe}$ &
  HII region &
  {[}19{]} &
   \\
J2253+1116 &
  0.0073 &
  -6.32$^{+0}_{-0}$ &
  0.75$^{+0.02}_{-0.022}$ &
  +$\Delta_{\rm ICF,Fe}$ &
  HII region &
  {[}19{]} &
   \\
J2310-0211 &
  0.012 &
  -7.15$^{+0.02}_{-0.02}$ &
  0.72$^{+0.019}_{-0.023}$ &
  +$\Delta_{\rm ICF,Fe}$ &
  HII region &
  {[}19{]} &
   \\
J2327-0200 &
  0.018 &
  -6.69$^{+0.01}_{-0.02}$ &
  0.56$^{+0.022}_{-0.017}$ &
  +$\Delta_{\rm ICF,Fe}$ &
  HII region &
  {[}19{]} &
   \\
J0811+4730 &
  0.044 &
  -6.56$^{+0.31}_{-0.31}$ &
  -0.27$^{+0.092}_{-0.092}$ &
  +$\Delta_{\rm ICF,Fe}$ &
  HII region &
  {[}20{]} &
   \\
J082555 &
  0.0024 &
  -8.11$^{+0}_{-0}$ &
  0.34$^{+0.15}_{-0.2}$ &
  -ADF, -$\Delta_{\rm d}$, $\Delta_{\rm SPS}^{*}$ &
  UV cont. S99/C\&B &
  {[}22, 22, 23{]} &
  {[}O/Fe{]} using S99 SPS, $\Delta_{\rm SPS}^{*}$=$^{+0.24}_{-0}$ \\
J104457 &
  0.013 &
  -7.82$^{+0}_{-0}$ &
  0.35$^{+0.17}_{-0.2}$ &
  -ADF, -$\Delta_{\rm d}$, $\Delta_{\rm SPS}^{*}$ &
  UV cont. S99/C\&B &
  {[}22, 22, 23{]} &
  {[}O/Fe{]} using S99 SPS, $\Delta_{\rm SPS}^{*}$=$^{+0.1}_{-0.0076}$ \\
HS1442+4250 &
  0.0022 &
  -7.75$^{+0.28}_{-0.28}$ &
  0.69$^{+0.072}_{-0.098}$ &
  -ADF, -$\Delta_{\rm d}$, $\Delta_{\rm SPS}^{*}$ &
  UV cont. S99/C\&B &
  {[}22, 22, 24{]} &
  {[}O/Fe{]} using S99 SPS, $\Delta_{\rm SPS}^{*}$=$^{+0.1}_{-0.37}$ \\
SB2 &
  0.0049 &
  -6.98$^{+0.04}_{-0.04}$ &
  0.63$^{+0.2}_{-0.22}$ &
  -ADF, -$\Delta_{\rm d}$, $\Delta_{\rm SPS}^{*}$ &
  UV cont. S99/C\&B &
  {[}22, 22, 25{]} &
  {[}O/Fe{]} using S99 SPS, $\Delta_{\rm SPS}^{*}$=$^{+0.1}_{-0.18}$ \\
SB82 &
  0.017 &
  -6.98$^{+0.16}_{-0.16}$ &
  0.73$^{+0.17}_{-0.16}$ &
  -ADF, -$\Delta_{\rm d}$, $\Delta_{\rm SPS}^{*}$ &
  UV cont. S99/C\&B &
  {[}22, 22, 25{]} &
  {[}O/Fe{]} using S99 SPS, $\Delta_{\rm SPS}^{*}$=$^{+0.1}_{-0.097}$ \\
- &
  2.3 &
  -8.62 &
  0.28$^{+0.2}_{-0.2}$ &
   &
  HII region/indirect &
  {[}26{]} &
  median \\
- &
  2.3 &
  -8.62 &
  0.25$^{+0.2}_{-0.2}$ &
  +$\Delta_{\rm d}$ &
  HII region/indirect &
  {[}27{]} &
  median \\
AEGIS\_11452 &
  1.7 &
  -8.25$^{+0.08}_{-0.38}$ &
  -0.13$^{+0.53}_{-2.4}$ &
  -ADF, +$\Delta_{\rm d}$ &
  HII region/indirect &
  {[}28{]} &
   \\
COSMOS\_1908 &
  3.1 &
  -7.21$^{+0.06}_{-0.37}$ &
  0.47$^{+0.16}_{-0.73}$ &
  -ADF, +$\Delta_{\rm d}$ &
  HII region/indirect &
  {[}28{]} &
   \\
C12b &
  2 &
  -9.5$^{+0.1}_{-0.12}$ &
  1.6$^{+0.032}_{-0.067}$ &
  -ADF, +$\Delta_{\rm d}$ &
  HII region/indirect &
  {[}28{]} &
   \\
Ste14a &
  2.2 &
  -7.38$^{+0.13}_{-0.12}$ &
  1.3$^{+0.29}_{-0.44}$ &
  -ADF, +$\Delta_{\rm d}$ &
  HII region/indirect &
  {[}28{]} &
   \\
C12c &
  3.5 &
  -8.61$^{+0.17}_{-0.6}$ &
  0.38$^{+0.14}_{-2}$ &
  -ADF, +$\Delta_{\rm d}$ &
  HII region/indirect &
  {[}28{]} &
   \\
KBSS-LM1\_stack &
  2.4 &
  -8.35$^{+0.3}_{-0.3}$ &
  1.6$^{+0.13}_{-0.21}$ &
  -ADF, +$\Delta_{\rm d}$ &
  HII region/indirect &
  {[}28{]} &
   \\
GOODS-S\_41547 &
  2.5 &
  -7.42$^{+0.16}_{-0.14}$ &
  -0.6 &
  -ADF, +$\Delta_{\rm d}$ &
  HII region/indirect &
  {[}28{]} &
  lower limit \\
Ste14b &
  2.3 &
  -7.45$^{+0.08}_{-0.12}$ &
  0.07 &
  -ADF, +$\Delta_{\rm d}$ &
  HII region/indirect &
  {[}28{]} &
  lower limit \\
S13 &
  1.4 &
  -7.52$^{+0.09}_{-0.18}$ &
  -1.6 &
  -ADF, +$\Delta_{\rm d}$ &
  HII region/indirect &
  {[}28{]} &
  lower limit \\
B14 &
  3.6 &
  -8.07$^{+0.38}_{-0.4}$ &
  1.8 &
  -ADF, +$\Delta_{\rm d}$ &
  HII region/indirect &
  {[}28{]} &
  lower limit \\
COSMOS\_23895 &
  3.6 &
  -7.93$^{+0.14}_{-0.08}$ &
  0.96 &
  -ADF, +$\Delta_{\rm d}$ &
  HII region/indirect &
  {[}28{]} &
  lower limit \\
MOSDEF\_composite &
  2.8 &
  -7.75$^{+0.11}_{-0.09}$ &
  0.29 &
  -ADF, +$\Delta_{\rm d}$ &
  HII region/indirect &
  {[}28{]} &
  lower limit \\
WLM &
  0 &
  -9.77$^{+0.22}_{-0.22}$ &
  -0.14$^{+0.11}_{-0.11}$ &
  +ADF, +$\Delta_{\rm d}$ &
  BSG &
  {[}29, 29, 30{]} &
   \\
IC1613 &
  0 &
  -10.6$^{+0.22}_{-0.22}$ &
  -0.26$^{+0.11}_{-0.11}$ &
  +ADF, +$\Delta_{\rm d}$ &
  BSG &
  {[}29, 29, 30{]} &
   \\
NGC6822 &
  0 &
  -9.9$^{+0.17}_{-0.17}$ &
  -0.08$^{+0.22}_{-0.22}$ &
  +ADF, +$\Delta_{\rm d}$ &
  BSG &
  {[}29, 29, 13{]} &
  ADF=0.23 \\
NGC55 &
  0 &
  -10.3$^{+0.22}_{-0.22}$ &
  -0.25$^{+0.12}_{-0.12}$ &
  +ADF, +$\Delta_{\rm d}$ &
  BSG &
  {[}29, 29, 30{]} &
   \\
NGC300 &
  0 &
  -10.1$^{+0.22}_{-0.22}$ &
  -0.13$^{+0.063}_{-0.063}$ &
  +ADF, +$\Delta_{\rm d}$ &
  BSG &
  {[}29, 29, 30{]} &
  ADF=0.22 \\
M33 &
  0 &
  -9.92$^{+0.22}_{-0.22}$ &
  -0.27$^{+0.071}_{-0.071}$ &
  +ADF, +$\Delta_{\rm d}$ &
  BSG &
  {[}29, 29, 30{]} &
  ADF=0.27 \\
M31 &
  0 &
  -11.1$^{+0.22}_{-0.22}$ &
  -0.37$^{+0.23}_{-0.23}$ &
  +ADF, +$\Delta_{\rm d}$ &
  BSG &
  {[}29, 29, 30{]} &
  ADF=0.18 \\
M81 &
  0 &
  -10.9$^{+0.22}_{-0.22}$ &
  -0.23$^{+0.15}_{-0.15}$ &
  +ADF, +$\Delta_{\rm d}$ &
  BSG &
  {[}29, 29, 30{]} &
   \\
M83 &
  0 &
  -9.79$^{+0.22}_{-0.22}$ &
  -0.19$^{+0.28}_{-0.28}$ &
  +ADF, +$\Delta_{\rm d}$ &
  BSG &
  {[}29, 29, 30{]} &
   \\
NGC2403 &
  0 &
  -9.92$^{+0.22}_{-0.22}$ &
  -0.17$^{+0.071}_{-0.071}$ &
  +ADF, +$\Delta_{\rm d}$ &
  BSG &
  {[}31, 31, 30{]} & 
  \\ 
  - &
  2.2 &
  -8.75 &
  0.33$^{+0.12}_{-0.12}$ &
  +ADF, +$\Delta_{\rm d}$, -$\Delta_{\rm SPS}$ &
  UV cont., BPASS &
  {[}32{]} & stack, Z$_{\rm O}$ assigned from MZR
   \\  
  \hline
\caption{ 
Observational sSFR (column 3) and [O/Fe] (column 4) estimates used in this study. The latter assume log$_{10}$(O/Fe)$_{\odot}$=1.33 dex \cite{GrevesseSauval98}.
Column 5 lists sources of systematic [O/Fe] offsets between the studies. Unless a different value is specified in the last column, we assume:
ADF=0.24 dex (offset between direct method and recombination line Z$_{\rm O/H}$ scales), $\Delta_{d}$=0.1 dex (oxygen dust depletion correction), $\Delta_{\rm ICF, Fe}$=0.2 dex (Z$_{\rm Fe}$ ionisation correction factor), $\Delta_{\rm SPS}$=0.1 dex (Z$_{\rm Fe}$ offset between BPASS and S99 SPS models).
The first reference in column 7 points to Z$_{\rm Fe}$, the second to Z$_{\rm O/H}$, the third to sSFR or SFR and the last (if applicable) to to M$_{*}$. If only one reference is given, all properties come from the same source.
[1] \cite{Spina22}, [2] \cite{Arellano-Cordova20}, [3] \cite{Elia_2022}, [4] \cite{Bland-Hawthorn16}, [5] \cite{Song21}, [6] \cite{Dominguez-Guzman22}, [7] \cite{Skibba12}, 
[8] \cite{Lebouteiller13}, [9] \cite{Zhou21},
[10] \cite{Kaufer04}, [11] \cite{Magrini05}, [12] \cite{Hunter10}, [13] \cite{Woo08},
[14] \cite{Hosek14} , [15] \cite{Pena07},
[16] \cite{Cullen21}, [17] \cite{Steidel16}, [18]
 \cite{Topping20}, [19] \cite{Kojima21}, [20] \cite{Izotov18}, [21] \cite{Thuan22}, [22] \cite{Senchyna22}, [23] \cite{Berg16}, [24] \cite{Senchyna19}, [25] \cite{Senchyna17}, [26] \cite{Strom18}, [27] \cite{Strom22}, [28] \cite{Sanders20}, [29] \cite{Bresolin16}, [30] \cite{Leroy19}, [31] \cite{Bresolin22}, [32] \cite{Kashino22}
 }
\label{tab:data}
\end{longtable}
\end{landscape}
 \twocolumn
\begin{acknowledgements} 
MCh is grateful for many insightful discussions on various aspects of this work, and would like to thank in particular to Allison Strom, Claudio Dalla Vecchia, Jorge S{\'a}nchez Almeida, Gijs Nelemans, Svea Hernandez, Sabyasachi Goswami, Valeriya Korol, Alexandre Vazdekis, Robert Grand, Andre Sieverding, Jakub Klencki, S{\o}ren Larsen and Scott Trager.
TM acknowledges support by a Spinoza Grant from the Dutch Research Council (NWO).
\end{acknowledgements}

\bibliographystyle{aa}
\bibliography{bibliography.bib}

\begin{thebibliography}{212}
\expandafter\ifx\csname natexlab\endcsname\relax\def\natexlab#1{#1}\fi

\bibitem[{{Abbott}(1982)}]{Abbott82}
{Abbott}, D.~C. 1982, \apj, 259, 282

\bibitem[{{Afsariardchi} {et~al.}(2021){Afsariardchi}, {Drout}, {Khatami},
  {Matzner}, {Moon}, \& {Ni}}]{Afsariardchi21}
{Afsariardchi}, N., {Drout}, M.~R., {Khatami}, D.~K., {et~al.} 2021, \apj, 918,
  89

\bibitem[{{Amarsi} {et~al.}(2020){Amarsi}, {Lind}, {Osorio}, {Nordlander},
  {Bergemann}, {Reggiani}, {Wang}, {Buder}, {Asplund}, {Barklem}, {Wehrhahn},
  {Sk{\'u}lad{\'o}ttir}, {Kobayashi}, {Karakas}, {Gao}, {Bland-Hawthorn}, {de
  Silva}, {Kos}, {Lewis}, {Martell}, {Sharma}, {Simpson}, {Zucker},
  {{\v{C}}otar}, {Horner}, \& {GALAH Collaboration}}]{Amarsi20}
{Amarsi}, A.~M., {Lind}, K., {Osorio}, Y., {et~al.} 2020, \aap, 642, A62

\bibitem[{{Amarsi} {et~al.}(2019){Amarsi}, {Nissen}, \&
  {Sk{\'u}lad{\'o}ttir}}]{Amarsi19}
{Amarsi}, A.~M., {Nissen}, P.~E., \& {Sk{\'u}lad{\'o}ttir}, {\'A}. 2019, \aap,
  630, A104

\bibitem[{{Anderson}(2019)}]{Anderson19}
{Anderson}, J.~P. 2019, \aap, 628, A7

\bibitem[{{Arellano-C{\'o}rdova} {et~al.}(2022){Arellano-C{\'o}rdova}, {Berg},
  {Chisholm}, {Arrabal Haro}, {Dickinson}, {Finkelstein}, {Leclercq}, {Rogers},
  {Simons}, {Skillman}, {Trump}, \& {Kartaltepe}}]{Arellano-Cordova22}
{Arellano-C{\'o}rdova}, K.~Z., {Berg}, D.~A., {Chisholm}, J., {et~al.} 2022,
  \apjl, 940, L23

\bibitem[{{Arellano-C{\'o}rdova} {et~al.}(2020){Arellano-C{\'o}rdova},
  {Esteban}, {Garc{\'\i}a-Rojas}, \& {M{\'e}ndez-Delgado}}]{Arellano-Cordova20}
{Arellano-C{\'o}rdova}, K.~Z., {Esteban}, C., {Garc{\'\i}a-Rojas}, J., \&
  {M{\'e}ndez-Delgado}, J.~E. 2020, \mnras, 496, 1051

\bibitem[{{Asplund} {et~al.}(2009){Asplund}, {Grevesse}, {Sauval}, \&
  {Scott}}]{Asplund09}
{Asplund}, M., {Grevesse}, N., {Sauval}, A.~J., \& {Scott}, P. 2009, \araa, 47,
  481

\bibitem[{{Aumer} \& {Binney}(2009)}]{Aumer09}
{Aumer}, M. \& {Binney}, J.~J. 2009, \mnras, 397, 1286

\bibitem[{{Baratella} {et~al.}(2020){Baratella}, {D'Orazi}, {Carraro},
  {Desidera}, {Randich}, {Magrini}, {Adibekyan}, {Smiljanic}, {Spina},
  {Tsantaki}, {Tautvai{\v{s}}ien{\.{e}}}, {Sousa}, {Jofr{\'e}},
  {Jim{\'e}nez-Esteban}, {Delgado-Mena}, {Martell}, {Van der Swaelmen},
  {Roccatagliata}, {Gilmore}, {Alfaro}, {Bayo}, {Bensby}, {Bragaglia},
  {Franciosini}, {Gonneau}, {Heiter}, {Hourihane}, {Jeffries}, {Koposov},
  {Morbidelli}, {Prisinzano}, {Sacco}, {Sbordone}, {Worley}, {Zaggia}, \&
  {Lewis}}]{Baratella20}
{Baratella}, M., {D'Orazi}, V., {Carraro}, G., {et~al.} 2020, \aap, 634, A34

\bibitem[{{Berg} {et~al.}(2016){Berg}, {Skillman}, {Henry}, {Erb}, \&
  {Carigi}}]{Berg16}
{Berg}, D.~A., {Skillman}, E.~D., {Henry}, R. B.~C., {Erb}, D.~K., \& {Carigi},
  L. 2016, \apj, 827, 126

\bibitem[{{Bian} {et~al.}(2018){Bian}, {Kewley}, \& {Dopita}}]{Bian18}
{Bian}, F., {Kewley}, L.~J., \& {Dopita}, M.~A. 2018, \apj, 859, 175

\bibitem[{{Bland-Hawthorn} \& {Gerhard}(2016)}]{Bland-Hawthorn16}
{Bland-Hawthorn}, J. \& {Gerhard}, O. 2016, \araa, 54, 529

\bibitem[{{Bonaca} {et~al.}(2020){Bonaca}, {Conroy}, {Cargile}, {Naidu},
  {Johnson}, {Zaritsky}, {Ting}, {Caldwell}, {Han}, \& {van Dokkum}}]{Bonaca20}
{Bonaca}, A., {Conroy}, C., {Cargile}, P.~A., {et~al.} 2020, \apjl, 897, L18

\bibitem[{{Boogaard} {et~al.}(2018){Boogaard}, {Brinchmann}, {Bouch{\'e}},
  {Paalvast}, {Bacon}, {Bouwens}, {Contini}, {Gunawardhana}, {Inami}, {Marino},
  {Maseda}, {Mitchell}, {Nanayakkara}, {Richard}, {Schaye}, {Schreiber},
  {Tacchella}, {Wisotzki}, \& {Zabl}}]{Boogaard18}
{Boogaard}, L.~A., {Brinchmann}, J., {Bouch{\'e}}, N., {et~al.} 2018, \aap,
  619, A27

\bibitem[{{Bresolin} {et~al.}(2022){Bresolin}, {Kudritzki}, \&
  {Urbaneja}}]{Bresolin22}
{Bresolin}, F., {Kudritzki}, R.-P., \& {Urbaneja}, M.~A. 2022, \apj, 940, 32

\bibitem[{{Bresolin} {et~al.}(2016){Bresolin}, {Kudritzki}, {Urbaneja},
  {Gieren}, {Ho}, \& {Pietrzy{\'n}ski}}]{Bresolin16}
{Bresolin}, F., {Kudritzki}, R.-P., {Urbaneja}, M.~A., {et~al.} 2016, \apj,
  830, 64

\bibitem[{{Bresolin} {et~al.}(2007){Bresolin}, {Urbaneja}, {Gieren},
  {Pietrzy{\'n}ski}, \& {Kudritzki}}]{Bresolin07}
{Bresolin}, F., {Urbaneja}, M.~A., {Gieren}, W., {Pietrzy{\'n}ski}, G., \&
  {Kudritzki}, R.-P. 2007, \apj, 671, 2028

\bibitem[{{Brinchmann} {et~al.}(2004){Brinchmann}, {Charlot}, {White},
  {Tremonti}, {Kauffmann}, {Heckman}, \& {Brinkmann}}]{Brinchmann04}
{Brinchmann}, J., {Charlot}, S., {White}, S.~D.~M., {et~al.} 2004, \mnras, 351,
  1151

\bibitem[{{Bromm} \& {Loeb}(2003)}]{Bromm03}
{Bromm}, V. \& {Loeb}, A. 2003, \nat, 425, 812

\bibitem[{{Brott} {et~al.}(2011){Brott}, {de Mink}, {Cantiello}, {Langer}, {de
  Koter}, {Evans}, {Hunter}, {Trundle}, \& {Vink}}]{Brott11I}
{Brott}, I., {de Mink}, S.~E., {Cantiello}, M., {et~al.} 2011, \aap, 530, A115

\bibitem[{{Buder} {et~al.}(2021){Buder}, {Sharma}, {Kos}, {Amarsi},
  {Nordlander}, {Lind}, {Martell}, {Asplund}, {Bland-Hawthorn}, {Casey}, {de
  Silva}, {D'Orazi}, {Freeman}, {Hayden}, {Lewis}, {Lin}, {Schlesinger},
  {Simpson}, {Stello}, {Zucker}, {Zwitter}, {Beeson}, {Buck}, {Casagrande},
  {Clark}, {{\v{C}}otar}, {da Costa}, {de Grijs}, {Feuillet}, {Horner},
  {Kafle}, {Khanna}, {Kobayashi}, {Liu}, {Montet}, {Nandakumar}, {Nataf},
  {Ness}, {Spina}, {Tepper-Garc{\'\i}a}, {Ting}, {Traven},
  {Vogrin{\v{c}}i{\v{c}}}, {Wittenmyer}, {Wyse}, {{\v{Z}}erjal}, \& {Galah
  Collaboration}}]{Buder21}
{Buder}, S., {Sharma}, S., {Kos}, J., {et~al.} 2021, \mnras, 506, 150

\bibitem[{{Calabr{\`o}} {et~al.}(2021){Calabr{\`o}}, {Castellano},
  {Pentericci}, {Fontanot}, {Menci}, {Cullen}, {McLure}, {Bolzonella},
  {Cimatti}, {Marchi}, {Talia}, {Amor{\'\i}n}, {Cresci}, {De Lucia}, {Fynbo},
  {Fontana}, {Franco}, {Hathi}, {Hibon}, {Hirschmann}, {Mannucci}, {Santini},
  {Saxena}, {Schaerer}, {Xie}, \& {Zamorani}}]{Calabro21}
{Calabr{\`o}}, A., {Castellano}, M., {Pentericci}, L., {et~al.} 2021, \aap,
  646, A39

\bibitem[{{Carton} {et~al.}(2018){Carton}, {Brinchmann}, {Contini}, {Epinat},
  {Finley}, {Richard}, {Patr{\'\i}cio}, {Schaye}, {Nanayakkara}, {Weilbacher},
  \& {Wisotzki}}]{Carton18}
{Carton}, D., {Brinchmann}, J., {Contini}, T., {et~al.} 2018, \mnras, 478, 4293

\bibitem[{{Chen} {et~al.}(2023){Chen}, {Jones}, {Sanders}, {Fadda}, {Sutter},
  {Minchin}, {Huntzinger}, {Senchyna}, {Stark}, {Spilker}, {Weiner}, \&
  {Roberts-Borsani}}]{Chen23}
{Chen}, Y., {Jones}, T., {Sanders}, R., {et~al.} 2023, Nature Astronomy

\bibitem[{{Chisholm} {et~al.}(2019){Chisholm}, {Rigby}, {Bayliss}, {Berg},
  {Dahle}, {Gladders}, \& {Sharon}}]{Chisholm19}
{Chisholm}, J., {Rigby}, J.~R., {Bayliss}, M., {et~al.} 2019, \apj, 882, 182

\bibitem[{{Chru{\'s}li{\'n}ska}(2022)}]{Chruslinska22}
{Chru{\'s}li{\'n}ska}, M. 2022, arXiv e-prints, arXiv:2206.10622

\bibitem[{{Chru{\'s}li{\'n}ska} {et~al.}(2021){Chru{\'s}li{\'n}ska},
  {Nelemans}, {Boco}, \& {Lapi}}]{Chruslinska21}
{Chru{\'s}li{\'n}ska}, M., {Nelemans}, G., {Boco}, L., \& {Lapi}, A. 2021,
  \mnras, 508, 4994

\bibitem[{{Conroy} {et~al.}(2014){Conroy}, {Graves}, \& {van
  Dokkum}}]{Conroy14}
{Conroy}, C., {Graves}, G.~J., \& {van Dokkum}, P.~G. 2014, \apj, 780, 33

\bibitem[{{Cooper} {et~al.}(2009){Cooper}, {Newman}, \& {Yan}}]{Cooper09}
{Cooper}, M.~C., {Newman}, J.~A., \& {Yan}, R. 2009, \apj, 704, 687

\bibitem[{{Crain} {et~al.}(2015){Crain}, {Schaye}, {Bower}, {Furlong},
  {Schaller}, {Theuns}, {Dalla Vecchia}, {Frenk}, {McCarthy}, {Helly},
  {Jenkins}, {Rosas-Guevara}, {White}, \& {Trayford}}]{Crain15}
{Crain}, R.~A., {Schaye}, J., {Bower}, R.~G., {et~al.} 2015, \mnras, 450, 1937

\bibitem[{{Crowther} {et~al.}(2006){Crowther}, {Prinja}, {Pettini}, \&
  {Steidel}}]{Crowther06}
{Crowther}, P.~A., {Prinja}, R.~K., {Pettini}, M., \& {Steidel}, C.~C. 2006,
  \mnras, 368, 895

\bibitem[{{Cullen} {et~al.}(2019){Cullen}, {McLure}, {Dunlop}, {Khochfar},
  {Dav{\'e}}, {Amor{\'\i}n}, {Bolzonella}, {Carnall}, {Castellano}, {Cimatti},
  {Cirasuolo}, {Cresci}, {Fynbo}, {Fontanot}, {Gargiulo}, {Garilli}, {Guaita},
  {Hathi}, {Hibon}, {Mannucci}, {Marchi}, {McLeod}, {Pentericci}, {Pozzetti},
  {Shapley}, {Talia}, \& {Zamorani}}]{Cullen19}
{Cullen}, F., {McLure}, R.~J., {Dunlop}, J.~S., {et~al.} 2019, \mnras, 487,
  2038

\bibitem[{{Cullen} {et~al.}(2021){Cullen}, {Shapley}, {McLure}, {Dunlop},
  {Sanders}, {Topping}, {Reddy}, {Amor{\'\i}n}, {Begley}, {Bolzonella},
  {Calabr{\`o}}, {Carnall}, {Castellano}, {Cimatti}, {Cirasuolo}, {Cresci},
  {Fontana}, {Fontanot}, {Garilli}, {Guaita}, {Hamadouche}, {Hathi},
  {Mannucci}, {McLeod}, {Pentericci}, {Saxena}, {Talia}, \&
  {Zamorani}}]{Cullen21}
{Cullen}, F., {Shapley}, A.~E., {McLure}, R.~J., {et~al.} 2021, \mnras, 505,
  903

\bibitem[{{Curti} {et~al.}(2023){Curti}, {D'Eugenio}, {Carniani}, {Maiolino},
  {Sandles}, {Witstok}, {Baker}, {Bennett}, {Piotrowska}, {Tacchella},
  {Charlot}, {Nakajima}, {Maheson}, {Mannucci}, {Amiri}, {Arribas}, {Belfiore},
  {Bonaventura}, {Bunker}, {Chevallard}, {Cresci}, {Curtis-Lake},
  {Hayden-Pawson}, {Jones}, {Kumari}, {Laseter}, {Looser}, {Marconi}, {Maseda},
  {Scholtz}, {Smit}, {{\"U}bler}, \& {Wallace}}]{Curti23}
{Curti}, M., {D'Eugenio}, F., {Carniani}, S., {et~al.} 2023, \mnras, 518, 425

\bibitem[{{Curtis} {et~al.}(2019){Curtis}, {Ebinger}, {Fr{\"o}hlich}, {Hempel},
  {Perego}, {Liebend{\"o}rfer}, \& {Thielemann}}]{Curtis19}
{Curtis}, S., {Ebinger}, K., {Fr{\"o}hlich}, C., {et~al.} 2019, \apj, 870, 2

\bibitem[{Curtis {et~al.}(2018)Curtis, Ebinger, Fröhlich, Hempel, Perego,
  Liebendörfer, \& Thielemann}]{Curtis_2019}
Curtis, S., Ebinger, K., Fröhlich, C., {et~al.} 2018, 870, 2

\bibitem[{{Davies} {et~al.}(2017){Davies}, {Kudritzki}, {Lardo}, {Bergemann},
  {Beasor}, {Plez}, {Evans}, {Bastian}, \& {Patrick}}]{Davies17}
{Davies}, B., {Kudritzki}, R.-P., {Lardo}, C., {et~al.} 2017, \apj, 847, 112

\bibitem[{{de los Reyes} \& {Kennicutt}(2019)}]{delosReyes19}
{de los Reyes}, M. A.~C. \& {Kennicutt}, Robert~C., J. 2019, \apj, 872, 16

\bibitem[{{Dom{\'\i}nguez-Guzm{\'a}n}
  {et~al.}(2022){Dom{\'\i}nguez-Guzm{\'a}n}, {Rodr{\'\i}guez},
  {Garc{\'\i}a-Rojas}, {Esteban}, \& {Toribio San
  Cipriano}}]{Dominguez-Guzman22}
{Dom{\'\i}nguez-Guzm{\'a}n}, G., {Rodr{\'\i}guez}, M., {Garc{\'\i}a-Rojas}, J.,
  {Esteban}, C., \& {Toribio San Cipriano}, L. 2022, \mnras, 517, 4497

\bibitem[{{Donnari} {et~al.}(2019){Donnari}, {Pillepich}, {Nelson},
  {Vogelsberger}, {Genel}, {Weinberger}, {Marinacci}, {Springel}, \&
  {Hernquist}}]{Donnari19}
{Donnari}, M., {Pillepich}, A., {Nelson}, D., {et~al.} 2019, \mnras, 485, 4817

\bibitem[{{Ebinger} {et~al.}(2020){Ebinger}, {Curtis}, {Ghosh}, {Fr{\"o}hlich},
  {Hempel}, {Perego}, {Liebend{\"o}rfer}, \& {Thielemann}}]{Ebinger20}
{Ebinger}, K., {Curtis}, S., {Ghosh}, S., {et~al.} 2020, \apj, 888, 91

\bibitem[{Ebinger {et~al.}(2020)Ebinger, Curtis, Ghosh, Fröhlich, Hempel,
  Perego, Liebendörfer, \& Thielemann}]{Ebinger_2020}
Ebinger, K., Curtis, S., Ghosh, S., {et~al.} 2020, The Astrophysical Journal,
  888, 91

\bibitem[{{El Eid} \& {Langer}(1986)}]{ElEid86}
{El Eid}, M.~F. \& {Langer}, N. 1986, \aap, 167, 274

\bibitem[{{Eldridge} \& {Stanway}(2022)}]{Eldridge22}
{Eldridge}, J.~J. \& {Stanway}, E.~R. 2022, \araa, 60, 455

\bibitem[{{Eldridge} {et~al.}(2017){Eldridge}, {Stanway}, {Xiao}, {McClelland},
  {Taylor}, {Ng}, {Greis}, \& {Bray}}]{Eldridge17}
{Eldridge}, J.~J., {Stanway}, E.~R., {Xiao}, L., {et~al.} 2017, \pasa, 34, e058

\bibitem[{Elia {et~al.}(2022)Elia, Molinari, Schisano, Soler, Merello, Russeil,
  Veneziani, Zavagno, Noriega-Crespo, Olmi, Benedettini, Hennebelle, Klessen,
  Leurini, Paladini, Pezzuto, Traficante, Eden, Martin, Sormani, Coletta,
  Colman, Plume, Maruccia, Mininni, \& Liu}]{Elia_2022}
Elia, D., Molinari, S., Schisano, E., {et~al.} 2022, The Astrophysical Journal,
  941, 162

\bibitem[{{Ertl} {et~al.}(2020){Ertl}, {Woosley}, {Sukhbold}, \&
  {Janka}}]{Ertl20}
{Ertl}, T., {Woosley}, S.~E., {Sukhbold}, T., \& {Janka}, H.~T. 2020, \apj,
  890, 51

\bibitem[{{Esteban} {et~al.}(2014){Esteban}, {Garc{\'\i}a-Rojas}, {Carigi},
  {Peimbert}, {Bresolin}, {L{\'o}pez-S{\'a}nchez}, \&
  {Mesa-Delgado}}]{Esteban14}
{Esteban}, C., {Garc{\'\i}a-Rojas}, J., {Carigi}, L., {et~al.} 2014, \mnras,
  443, 624

\bibitem[{{Fantin} {et~al.}(2019){Fantin}, {C{\^o}t{\'e}}, {McConnachie},
  {Bergeron}, {Cuillandre}, {Gwyn}, {Ibata}, {Thomas}, {Carlberg}, {Fabbro},
  {Haywood}, {Lan{\c{c}}on}, {Lewis}, {Malhan}, {Martin}, {Navarro}, {Scott},
  \& {Starkenburg}}]{Fantin19}
{Fantin}, N.~J., {C{\^o}t{\'e}}, P., {McConnachie}, A.~W., {et~al.} 2019, \apj,
  887, 148

\bibitem[{{Fryer}(1999)}]{Fryer99}
{Fryer}, C.~L. 1999, \apj, 522, 413

\bibitem[{{Fryer} {et~al.}(2006){Fryer}, {Young}, \& {Hungerford}}]{Fryer06}
{Fryer}, C.~L., {Young}, P.~A., \& {Hungerford}, A.~L. 2006, \apj, 650, 1028

\bibitem[{{Garcia} {et~al.}(2021){Garcia}, {Evans}, {Bestenlehner}, {Bouret},
  {Castro}, {Cervi{\~n}o}, {Fullerton}, {Gieles}, {Herrero}, {de Koter},
  {Lennon}, {van Loon}, {Martins}, {de Mink}, {Najarro}, {Negueruela}, {Sana},
  {Sim{\'o}n-D{\'\i}az}, {Sz{\'e}csi}, {Tramper}, {Vink}, \&
  {Wofford}}]{Garcia21}
{Garcia}, M., {Evans}, C.~J., {Bestenlehner}, J.~M., {et~al.} 2021,
  Experimental Astronomy, 51, 887

\bibitem[{{Garcia} {et~al.}(2014){Garcia}, {Herrero}, {Najarro}, {Lennon}, \&
  {Alejandro Urbaneja}}]{Garcia14}
{Garcia}, M., {Herrero}, A., {Najarro}, F., {Lennon}, D.~J., \& {Alejandro
  Urbaneja}, M. 2014, \apj, 788, 64

\bibitem[{{Goswami} {et~al.}(2022){Goswami}, {Silva}, {Bressan}, {Grisoni},
  {Costa}, {Marigo}, {Granato}, {Lapi}, \& {Spera}}]{Goswami22}
{Goswami}, S., {Silva}, L., {Bressan}, A., {et~al.} 2022, \aap, 663, A1

\bibitem[{{G{\"o}tberg} {et~al.}(2019){G{\"o}tberg}, {de Mink}, {Groh},
  {Leitherer}, \& {Norman}}]{Gotberg19}
{G{\"o}tberg}, Y., {de Mink}, S.~E., {Groh}, J.~H., {Leitherer}, C., \&
  {Norman}, C. 2019, \aap, 629, A134

\bibitem[{{G{\"o}tberg} {et~al.}(2020){G{\"o}tberg}, {de Mink}, {McQuinn},
  {Zapartas}, {Groh}, \& {Norman}}]{Gotberg20}
{G{\"o}tberg}, Y., {de Mink}, S.~E., {McQuinn}, M., {et~al.} 2020, \aap, 634,
  A134

\bibitem[{{Gratton} {et~al.}(2000){Gratton}, {Carretta}, {Matteucci}, \&
  {Sneden}}]{Gratton00}
{Gratton}, R.~G., {Carretta}, E., {Matteucci}, F., \& {Sneden}, C. 2000, \aap,
  358, 671

\bibitem[{{Greggio}(2005)}]{Greggio05}
{Greggio}, L. 2005, \aap, 441, 1055

\bibitem[{{Greggio}(2010)}]{Greggio10}
{Greggio}, L. 2010, \mnras, 406, 22

\bibitem[{{Grevesse} \& {Sauval}(1998)}]{GrevesseSauval98}
{Grevesse}, N. \& {Sauval}, A.~J. 1998, \ssr, 85, 161

\bibitem[{{Grimmett} {et~al.}(2018){Grimmett}, {Heger}, {Karakas}, \&
  {M{\"u}ller}}]{Grimmett18}
{Grimmett}, J.~J., {Heger}, A., {Karakas}, A.~I., \& {M{\"u}ller}, B. 2018,
  \mnras, 479, 495

\bibitem[{{Grimmett} {et~al.}(2021){Grimmett}, {M{\"u}ller}, {Heger},
  {Banerjee}, \& {Obergaulinger}}]{Grimmet21}
{Grimmett}, J.~J., {M{\"u}ller}, B., {Heger}, A., {Banerjee}, P., \&
  {Obergaulinger}, M. 2021, \mnras, 501, 2764

\bibitem[{{Gvozdenko} {et~al.}(2022){Gvozdenko}, {Larsen}, {Beasley}, \&
  {Brodie}}]{Gvozdenko22}
{Gvozdenko}, A., {Larsen}, S.~S., {Beasley}, M.~A., \& {Brodie}, J. 2022, \aap,
  666, A159

\bibitem[{{Hayden} {et~al.}(2022){Hayden}, {Sharma}, {Bland-Hawthorn}, {Spina},
  {Buder}, {Ciuc{\u{a}}}, {Asplund}, {Casey}, {De Silva}, {D'Orazi}, {Freeman},
  {Kos}, {Lewis}, {Lin}, {Lind}, {Martell}, {Schlesinger}, {Simpson}, {Zucker},
  {Zwitter}, {Chen}, {{\v{C}}otar}, {Feuillet}, {Horner}, {Joyce},
  {Nordlander}, {Stello}, {Tepper-Garcia}, {Ting}, {Wang}, {Wittenmyer}, \&
  {Wyse}}]{Hayden22}
{Hayden}, M.~R., {Sharma}, S., {Bland-Hawthorn}, J., {et~al.} 2022, \mnras,
  517, 5325

\bibitem[{{Heckman} {et~al.}(1998){Heckman}, {Robert}, {Leitherer}, {Garnett},
  \& {van der Rydt}}]{Heckman98}
{Heckman}, T.~M., {Robert}, C., {Leitherer}, C., {Garnett}, D.~R., \& {van der
  Rydt}, F. 1998, \apj, 503, 646

\bibitem[{{Heger} {et~al.}(2003){Heger}, {Fryer}, {Woosley}, {Langer}, \&
  {Hartmann}}]{Heger03}
{Heger}, A., {Fryer}, C.~L., {Woosley}, S.~E., {Langer}, N., \& {Hartmann},
  D.~H. 2003, \apj, 591, 288

\bibitem[{{Heger} \& {Woosley}(2002)}]{HegerWoosley02}
{Heger}, A. \& {Woosley}, S.~E. 2002, \apj, 567, 532

\bibitem[{{Heger} \& {Woosley}(2010)}]{HegerWoosley10}
{Heger}, A. \& {Woosley}, S.~E. 2010, \apj, 724, 341

\bibitem[{{Heringer} {et~al.}(2019){Heringer}, {Pritchet}, \& {van
  Kerkwijk}}]{Heringer19}
{Heringer}, E., {Pritchet}, C., \& {van Kerkwijk}, M.~H. 2019, \apj, 882, 52

\bibitem[{{Hernandez} {et~al.}(2019){Hernandez}, {Larsen}, {Aloisi}, {Berg},
  {Blair}, {Fox}, {Heckman}, {James}, {Long}, {Skillman}, \&
  {Whitmore}}]{Hernandez19}
{Hernandez}, S., {Larsen}, S., {Aloisi}, A., {et~al.} 2019, \apj, 872, 116

\bibitem[{{Hernandez} {et~al.}(2017){Hernandez}, {Larsen}, {Trager}, {Groot},
  \& {Kaper}}]{Hernandez17}
{Hernandez}, S., {Larsen}, S., {Trager}, S., {Groot}, P., \& {Kaper}, L. 2017,
  \aap, 603, A119

\bibitem[{{Hosek} {et~al.}(2014){Hosek}, {Kudritzki}, {Bresolin}, {Urbaneja},
  {Evans}, {Pietrzy{\'n}ski}, {Gieren}, {Przybilla}, \& {Carraro}}]{Hosek14}
{Hosek}, Matthew~W., J., {Kudritzki}, R.-P., {Bresolin}, F., {et~al.} 2014,
  \apj, 785, 151

\bibitem[{{Hunter} {et~al.}(2010){Hunter}, {Elmegreen}, \& {Ludka}}]{Hunter10}
{Hunter}, D.~A., {Elmegreen}, B.~G., \& {Ludka}, B.~C. 2010, \aj, 139, 447

\bibitem[{{Iben} \& {Tutukov}(1984)}]{IbenTutukov84}
{Iben}, I., J. \& {Tutukov}, A.~V. 1984, \apjs, 54, 335

\bibitem[{{Imasheva} {et~al.}(2023){Imasheva}, {Janka}, \&
  {Weiss}}]{Imasheva23}
{Imasheva}, L., {Janka}, H.-T., \& {Weiss}, A. 2023, \mnras, 518, 1818

\bibitem[{{Isobe} {et~al.}(2022){Isobe}, {Ouchi}, {Suzuki}, {Moriya},
  {Nakajima}, {Nomoto}, {Rauch}, {Harikane}, {Kojima}, {Ono}, {Fujimoto},
  {Inoue}, {Kim}, {Komiyama}, {Kusakabe}, {Lee}, {Maseda}, {Matthee},
  {Michel-Dansac}, {Nagao}, {Nanayakkara}, {Nishigaki}, {Onodera}, {Sugahara},
  \& {Xu}}]{Isobe22}
{Isobe}, Y., {Ouchi}, M., {Suzuki}, A., {et~al.} 2022, \apj, 925, 111

\bibitem[{{Iwamoto} {et~al.}(1999){Iwamoto}, {Brachwitz}, {Nomoto},
  {Kishimoto}, {Umeda}, {Hix}, \& {Thielemann}}]{Iwamoto99}
{Iwamoto}, K., {Brachwitz}, F., {Nomoto}, K., {et~al.} 1999, \apjs, 125, 439

\bibitem[{{Izotov} {et~al.}(2006){Izotov}, {Stasi{\'n}ska}, {Meynet}, {Guseva},
  \& {Thuan}}]{Izotov06}
{Izotov}, Y.~I., {Stasi{\'n}ska}, G., {Meynet}, G., {Guseva}, N.~G., \&
  {Thuan}, T.~X. 2006, \aap, 448, 955

\bibitem[{{Izotov} \& {Thuan}(1999)}]{Izotov99}
{Izotov}, Y.~I. \& {Thuan}, T.~X. 1999, \apj, 511, 639

\bibitem[{{Izotov} {et~al.}(2018){Izotov}, {Thuan}, {Guseva}, \&
  {Liss}}]{Izotov18}
{Izotov}, Y.~I., {Thuan}, T.~X., {Guseva}, N.~G., \& {Liss}, S.~E. 2018,
  \mnras, 473, 1956

\bibitem[{{Janka} {et~al.}(2007){Janka}, {Langanke}, {Marek},
  {Mart{\'\i}nez-Pinedo}, \& {M{\"u}ller}}]{Janka07}
{Janka}, H.~T., {Langanke}, K., {Marek}, A., {Mart{\'\i}nez-Pinedo}, G., \&
  {M{\"u}ller}, B. 2007, \physrep, 442, 38

\bibitem[{{Je{\v{r}}{\'a}bkov{\'a}} {et~al.}(2018){Je{\v{r}}{\'a}bkov{\'a}},
  {Hasani Zonoozi}, {Kroupa}, {Beccari}, {Yan}, {Vazdekis}, \&
  {Zhang}}]{Jerabkova18}
{Je{\v{r}}{\'a}bkov{\'a}}, T., {Hasani Zonoozi}, A., {Kroupa}, P., {et~al.}
  2018, \aap, 620, A39

\bibitem[{{Jones} {et~al.}(2020){Jones}, {Sanders}, {Roberts-Borsani}, {Ellis},
  {Laporte}, {Treu}, \& {Harikane}}]{Jones20}
{Jones}, T., {Sanders}, R., {Roberts-Borsani}, G., {et~al.} 2020, \apj, 903,
  150

\bibitem[{{Kashino} {et~al.}(2022){Kashino}, {Lilly}, {Renzini}, {Daddi},
  {Zamorani}, {Silverman}, {Ilbert}, {Peng}, {Mainieri}, {Bardelli}, {Zucca},
  {Kartaltepe}, \& {Sanders}}]{Kashino22}
{Kashino}, D., {Lilly}, S.~J., {Renzini}, A., {et~al.} 2022, \apj, 925, 82

\bibitem[{{Katz} {et~al.}(2023){Katz}, {Saxena}, {Cameron}, {Carniani},
  {Bunker}, {Arribas}, {Bhatawdekar}, {Bowler}, {Boyett}, {Cresci},
  {Curtis-Lake}, {D'Eugenio}, {Kumari}, {Looser}, {Maiolino}, {{\"U}bler},
  {Willott}, \& {Witstok}}]{Katz23}
{Katz}, H., {Saxena}, A., {Cameron}, A.~J., {et~al.} 2023, \mnras, 518, 592

\bibitem[{{Kaufer} {et~al.}(2004){Kaufer}, {Venn}, {Tolstoy}, {Pinte}, \&
  {Kudritzki}}]{Kaufer04}
{Kaufer}, A., {Venn}, K.~A., {Tolstoy}, E., {Pinte}, C., \& {Kudritzki}, R.-P.
  2004, \aj, 127, 2723

\bibitem[{{Kehrig} {et~al.}(2016){Kehrig}, {V{\'\i}lchez}, {P{\'e}rez-Montero},
  {Iglesias-P{\'a}ramo}, {Hern{\'a}ndez-Fern{\'a}ndez}, {Duarte Puertas},
  {Brinchmann}, {Durret}, \& {Kunth}}]{Kehrig16}
{Kehrig}, C., {V{\'\i}lchez}, J.~M., {P{\'e}rez-Montero}, E., {et~al.} 2016,
  \mnras, 459, 2992

\bibitem[{{Kewley} \& {Ellison}(2008)}]{KewleyEllison08}
{Kewley}, L.~J. \& {Ellison}, S.~L. 2008, \apj, 681, 1183

\bibitem[{{Kewley} {et~al.}(2019){Kewley}, {Nicholls}, \&
  {Sutherland}}]{Kewley19}
{Kewley}, L.~J., {Nicholls}, D.~C., \& {Sutherland}, R.~S. 2019, \araa, 57, 511

\bibitem[{{Kobayashi} {et~al.}(2020{\natexlab{a}}){Kobayashi}, {Karakas}, \&
  {Lugaro}}]{KobayashiKarakasLugaro20}
{Kobayashi}, C., {Karakas}, A.~I., \& {Lugaro}, M. 2020{\natexlab{a}}, \apj,
  900, 179

\bibitem[{{Kobayashi} {et~al.}(2020{\natexlab{b}}){Kobayashi}, {Leung}, \&
  {Nomoto}}]{Kobayashi20}
{Kobayashi}, C., {Leung}, S.-C., \& {Nomoto}, K. 2020{\natexlab{b}}, \apj, 895,
  138

\bibitem[{{Kojima} {et~al.}(2020){Kojima}, {Ouchi}, {Rauch}, {Ono}, {Nakajima},
  {Isobe}, {Fujimoto}, {Harikane}, {Hashimoto}, {Hayashi}, {Komiyama},
  {Kusakabe}, {Kim}, {Lee}, {Mukae}, {Nagao}, {Onodera}, {Shibuya}, {Sugahara},
  {Umemura}, \& {Yabe}}]{Kojima20}
{Kojima}, T., {Ouchi}, M., {Rauch}, M., {et~al.} 2020, \apj, 898, 142

\bibitem[{{Kojima} {et~al.}(2021){Kojima}, {Ouchi}, {Rauch}, {Ono}, {Nakajima},
  {Isobe}, {Fujimoto}, {Harikane}, {Hashimoto}, {Hayashi}, {Komiyama},
  {Kusakabe}, {Kim}, {Lee}, {Mukae}, {Nagao}, {Onodera}, {Shibuya}, {Sugahara},
  {Umemura}, \& {Yabe}}]{Kojima21}
{Kojima}, T., {Ouchi}, M., {Rauch}, M., {et~al.} 2021, \apj, 913, 22

\bibitem[{{Kudritzki}(2002)}]{Kudritzki02}
{Kudritzki}, R.~P. 2002, \apj, 577, 389

\bibitem[{{Kudritzki} {et~al.}(2016){Kudritzki}, {Castro}, {Urbaneja}, {Ho},
  {Bresolin}, {Gieren}, {Pietrzy{\'n}ski}, \& {Przybilla}}]{Kudritzki16}
{Kudritzki}, R.~P., {Castro}, N., {Urbaneja}, M.~A., {et~al.} 2016, \apj, 829,
  70

\bibitem[{{Kudritzki} {et~al.}(1987){Kudritzki}, {Pauldrach}, \&
  {Puls}}]{Kudritzki87}
{Kudritzki}, R.~P., {Pauldrach}, A., \& {Puls}, J. 1987, \aap, 173, 293

\bibitem[{{Kudritzki} {et~al.}(2008){Kudritzki}, {Urbaneja}, {Bresolin},
  {Przybilla}, {Gieren}, \& {Pietrzy{\'n}ski}}]{Kudritzki08}
{Kudritzki}, R.-P., {Urbaneja}, M.~A., {Bresolin}, F., {et~al.} 2008, \apj,
  681, 269

\bibitem[{{Lach} {et~al.}(2020){Lach}, {R{\"o}pke}, {Seitenzahl}, {Cot{\'e}},
  {Gronow}, \& {Ruiter}}]{Lach20}
{Lach}, F., {R{\"o}pke}, F.~K., {Seitenzahl}, I.~R., {et~al.} 2020, \aap, 644,
  A118

\bibitem[{{Langer} {et~al.}(2007){Langer}, {Norman}, {de Koter}, {Vink},
  {Cantiello}, \& {Yoon}}]{Langer07}
{Langer}, N., {Norman}, C.~A., {de Koter}, A., {et~al.} 2007, \aap, 475, L19

\bibitem[{{Lebouteiller} {et~al.}(2013){Lebouteiller}, {Heap}, {Hubeny}, \&
  {Kunth}}]{Lebouteiller13}
{Lebouteiller}, V., {Heap}, S., {Hubeny}, I., \& {Kunth}, D. 2013, \aap, 553,
  A16

\bibitem[{{Lee} {et~al.}(2006){Lee}, {Skillman}, {Cannon}, {Jackson}, {Gehrz},
  {Polomski}, \& {Woodward}}]{Lee06}
{Lee}, H., {Skillman}, E.~D., {Cannon}, J.~M., {et~al.} 2006, \apj, 647, 970

\bibitem[{{Lee} {et~al.}(2002){Lee}, {Salzer}, {Impey}, {Thuan}, \&
  {Gronwall}}]{Lee02}
{Lee}, J.~C., {Salzer}, J.~J., {Impey}, C., {Thuan}, T.~X., \& {Gronwall}, C.
  2002, \aj, 124, 3088

\bibitem[{{Leitherer} {et~al.}(2014){Leitherer}, {Ekstr{\"o}m}, {Meynet},
  {Schaerer}, {Agienko}, \& {Levesque}}]{Leitherer14}
{Leitherer}, C., {Ekstr{\"o}m}, S., {Meynet}, G., {et~al.} 2014, \apjs, 212, 14

\bibitem[{{Lemasle} {et~al.}(2017){Lemasle}, {Groenewegen}, {Grebel}, {Bono},
  {Fiorentino}, {Fran{\c{c}}ois}, {Inno}, {Kovtyukh}, {Matsunaga}, {Pedicelli},
  {Primas}, {Pritchard}, {Romaniello}, \& {da Silva}}]{Lemasle17}
{Lemasle}, B., {Groenewegen}, M.~A.~T., {Grebel}, E.~K., {et~al.} 2017, \aap,
  608, A85

\bibitem[{{Leroy} {et~al.}(2019){Leroy}, {Sandstrom}, {Lang}, {Lewis}, {Salim},
  {Behrens}, {Chastenet}, {Chiang}, {Gallagher}, {Kessler}, \&
  {Utomo}}]{Leroy19}
{Leroy}, A.~K., {Sandstrom}, K.~M., {Lang}, D., {et~al.} 2019, \apjs, 244, 24

\bibitem[{{Lilly} {et~al.}(2013){Lilly}, {Carollo}, {Pipino}, {Renzini}, \&
  {Peng}}]{Lilly13}
{Lilly}, S.~J., {Carollo}, C.~M., {Pipino}, A., {Renzini}, A., \& {Peng}, Y.
  2013, \apj, 772, 119

\bibitem[{{Limongi} \& {Chieffi}(2018)}]{LimongiChieffi18}
{Limongi}, M. \& {Chieffi}, A. 2018, \apjs, 237, 13

\bibitem[{{Liu} {et~al.}(2022){Liu}, {Kudritzki}, {Zhao}, {Urbaneja}, {Huang},
  {Zhang}, \& {Zhao}}]{Liu22}
{Liu}, C., {Kudritzki}, R.-P., {Zhao}, G., {et~al.} 2022, \apj, 932, 29

\bibitem[{{Livio} \& {Mazzali}(2018)}]{LivioMazzali18}
{Livio}, M. \& {Mazzali}, P. 2018, \physrep, 736, 1

\bibitem[{{Madau} \& {Dickinson}(2014)}]{MadauDickinson14}
{Madau}, P. \& {Dickinson}, M. 2014, \araa, 52, 415

\bibitem[{{Maeder} {et~al.}(2014){Maeder}, {Przybilla}, {Nieva}, {Georgy},
  {Meynet}, {Ekstr{\"o}m}, \& {Eggenberger}}]{Maeder14}
{Maeder}, A., {Przybilla}, N., {Nieva}, M.-F., {et~al.} 2014, \aap, 565, A39

\bibitem[{{Magrini} {et~al.}(2005){Magrini}, {Leisy}, {Corradi}, {Perinotto},
  {Mampaso}, \& {V{\'\i}lchez}}]{Magrini05}
{Magrini}, L., {Leisy}, P., {Corradi}, R.~L.~M., {et~al.} 2005, \aap, 443, 115

\bibitem[{{Maiolino} {et~al.}(2020){Maiolino}, {Cirasuolo}, {Afonso}, {Bauer},
  {Bowler}, {Cucciati}, {Daddi}, {De Lucia}, {Evans}, {Flores}, {Gargiulo},
  {Garilli}, {Jablonka}, {Jarvis}, {Kneib}, {Lilly}, {Looser}, {Magliocchetti},
  {Man}, {Mannucci}, {Maurogordato}, {McLure}, {Norberg}, {Oesch}, {Oliva},
  {Paltani}, {Pappalardo}, {Peng}, {Pentericci}, {Pozzetti}, {Renzini},
  {Rodrigues}, {Royer}, {Serjeant}, {Vanzi}, {Wild}, \&
  {Zamorani}}]{Maiolino20}
{Maiolino}, R., {Cirasuolo}, M., {Afonso}, J., {et~al.} 2020, The Messenger,
  180, 24

\bibitem[{{Maiolino} \& {Mannucci}(2019)}]{MaiolinoMannucci19}
{Maiolino}, R. \& {Mannucci}, F. 2019, \aapr, 27, 3

\bibitem[{{Maoz} \& {Graur}(2017)}]{MaozGraur17}
{Maoz}, D. \& {Graur}, O. 2017, \apj, 848, 25

\bibitem[{{Maoz} \& {Mannucci}(2012)}]{MaozMannucci12}
{Maoz}, D. \& {Mannucci}, F. 2012, \pasa, 29, 447

\bibitem[{{Maoz} {et~al.}(2014){Maoz}, {Mannucci}, \&
  {Nelemans}}]{MaozMannucciNelemans14}
{Maoz}, D., {Mannucci}, F., \& {Nelemans}, G. 2014, \araa, 52, 107

\bibitem[{{Marks} {et~al.}(2012){Marks}, {Kroupa}, {Dabringhausen}, \&
  {Pawlowski}}]{Marks12}
{Marks}, M., {Kroupa}, P., {Dabringhausen}, J., \& {Pawlowski}, M.~S. 2012,
  \mnras, 422, 2246

\bibitem[{{Martinez} {et~al.}(2022){Martinez}, {Bersten}, {Anderson}, {Hamuy},
  {Gonz{\'a}lez-Gait{\'a}n}, {F{\"o}rster}, {Orellana}, {Stritzinger},
  {Phillips}, {Guti{\'e}rrez}, {Burns}, {Contreras}, {de Jaeger}, {Ertini},
  {Folatelli}, {Galbany}, {Hoeflich}, {Hsiao}, {Morrell}, {Pessi}, \&
  {Suntzeff}}]{Martinez22}
{Martinez}, L., {Bersten}, M.~C., {Anderson}, J.~P., {et~al.} 2022, \aap, 660,
  A41

\bibitem[{{Matteucci} \& {Greggio}(1986)}]{MatteucciGreggio86}
{Matteucci}, F. \& {Greggio}, L. 1986, \aap, 154, 279

\bibitem[{{Matthee} {et~al.}(2022){Matthee}, {Feltre}, {Maseda}, {Nanayakkara},
  {Boogaard}, {Bacon}, {Verhamme}, {Leclercq}, {Kusakabe}, {Urrutia}, \&
  {Wisotzki}}]{Matthee22}
{Matthee}, J., {Feltre}, A., {Maseda}, M., {et~al.} 2022, \aap, 660, A10

\bibitem[{{Matthee} \& {Schaye}(2018)}]{MattheeSchaye18}
{Matthee}, J. \& {Schaye}, J. 2018, \mnras, 479, L34

\bibitem[{{Matthee} \& {Schaye}(2019)}]{MattheeSchaye19}
{Matthee}, J. \& {Schaye}, J. 2019, \mnras, 484, 915

\bibitem[{{Mazzali} {et~al.}(2007){Mazzali}, {R{\"o}pke}, {Benetti}, \&
  {Hillebrandt}}]{Mazzali07}
{Mazzali}, P.~A., {R{\"o}pke}, F.~K., {Benetti}, S., \& {Hillebrandt}, W. 2007,
  Science, 315, 825

\bibitem[{{McAlpine} {et~al.}(2016){McAlpine}, {Helly}, {Schaller}, {Trayford},
  {Qu}, {Furlong}, {Bower}, {Crain}, {Schaye}, {Theuns}, {Dalla Vecchia},
  {Frenk}, {McCarthy}, {Jenkins}, {Rosas-Guevara}, {White}, {Baes}, {Camps}, \&
  {Lemson}}]{McAlpine16}
{McAlpine}, S., {Helly}, J.~C., {Schaller}, M., {et~al.} 2016, Astronomy and
  Computing, 15, 72

\bibitem[{{Miglio} {et~al.}(2021){Miglio}, {Chiappini}, {Mackereth}, {Davies},
  {Brogaard}, {Casagrande}, {Chaplin}, {Girardi}, {Kawata}, {Khan}, {Izzard},
  {Montalb{\'a}n}, {Mosser}, {Vincenzo}, {Bossini}, {Noels}, {Rodrigues},
  {Valentini}, \& {Mandel}}]{Miglio18}
{Miglio}, A., {Chiappini}, C., {Mackereth}, J.~T., {et~al.} 2021, \aap, 645,
  A85

\bibitem[{{M{\"u}ller} {et~al.}(2017){M{\"u}ller}, {Prieto}, {Pejcha}, \&
  {Clocchiatti}}]{Muller17}
{M{\"u}ller}, T., {Prieto}, J.~L., {Pejcha}, O., \& {Clocchiatti}, A. 2017,
  \apj, 841, 127

\bibitem[{{Naiman} {et~al.}(2018){Naiman}, {Pillepich}, {Springel},
  {Ramirez-Ruiz}, {Torrey}, {Vogelsberger}, {Pakmor}, {Nelson}, {Marinacci},
  {Hernquist}, {Weinberger}, \& {Genel}}]{Naiman18}
{Naiman}, J.~P., {Pillepich}, A., {Springel}, V., {et~al.} 2018, \mnras, 477,
  1206

\bibitem[{{Nakajima} {et~al.}(2023){Nakajima}, {Ouchi}, {Isobe}, {Harikane},
  {Zhang}, {Ono}, {Umeda}, \& {Oguri}}]{Nakajima23}
{Nakajima}, K., {Ouchi}, M., {Isobe}, Y., {et~al.} 2023, arXiv e-prints,
  arXiv:2301.12825

\bibitem[{{Nanayakkara} {et~al.}(2019){Nanayakkara}, {Brinchmann}, {Boogaard},
  {Bouwens}, {Cantalupo}, {Feltre}, {Kollatschny}, {Marino}, {Maseda},
  {Matthee}, {Paalvast}, {Richard}, \& {Verhamme}}]{Nanayakkara19}
{Nanayakkara}, T., {Brinchmann}, J., {Boogaard}, L., {et~al.} 2019, \aap, 624,
  A89

\bibitem[{{Nelemans} {et~al.}(2013){Nelemans}, {Toonen}, \&
  {Bours}}]{Nelemans13}
{Nelemans}, G., {Toonen}, S., \& {Bours}, M. 2013, in Binary Paths to Type Ia
  Supernovae Explosions, ed. R.~{Di Stefano}, M.~{Orio}, \& M.~{Moe}, Vol. 281,
  225--231

\bibitem[{{Nelson} {et~al.}(2019){Nelson}, {Springel}, {Pillepich},
  {Rodriguez-Gomez}, {Torrey}, {Genel}, {Vogelsberger}, {Pakmor}, {Marinacci},
  {Weinberger}, {Kelley}, {Lovell}, {Diemer}, \& {Hernquist}}]{Nelson19_TNG}
{Nelson}, D., {Springel}, V., {Pillepich}, A., {et~al.} 2019, Computational
  Astrophysics and Cosmology, 6, 2

\bibitem[{{Nomoto}(1982)}]{Nomoto82}
{Nomoto}, K. 1982, \apj, 253, 798

\bibitem[{{Nomoto} {et~al.}(1997){Nomoto}, {Iwamoto}, {Nakasato}, {Thielemann},
  {Brachwitz}, {Tsujimoto}, {Kubo}, \& {Kishimoto}}]{Nomoto97}
{Nomoto}, K., {Iwamoto}, K., {Nakasato}, N., {et~al.} 1997, \nphysa, 621, 467

\bibitem[{{Nomoto} {et~al.}(2013){Nomoto}, {Kobayashi}, \&
  {Tominaga}}]{Nomoto13}
{Nomoto}, K., {Kobayashi}, C., \& {Tominaga}, N. 2013, \araa, 51, 457

\bibitem[{{Nomoto} {et~al.}(2005){Nomoto}, {Tominaga}, {Umeda}, {Maeda},
  {Ohkubo}, {Deng}, \& {Mazzali}}]{Nomoto05}
{Nomoto}, K., {Tominaga}, N., {Umeda}, H., {et~al.} 2005, in Astronomical
  Society of the Pacific Conference Series, Vol. 332, The Fate of the Most
  Massive Stars, ed. R.~{Humphreys} \& K.~{Stanek}, 384

\bibitem[{{Pagel} \& {Tautvaisiene}(1998)}]{Pagel98}
{Pagel}, B.~E.~J. \& {Tautvaisiene}, G. 1998, \mnras, 299, 535

\bibitem[{{Pauldrach} {et~al.}(1993){Pauldrach}, {Feldmeier}, {Puls}, \&
  {Kudritzki}}]{Pauldrach93}
{Pauldrach}, A.~W.~A., {Feldmeier}, A., {Puls}, J., \& {Kudritzki}, R.~P. 1993,
  \ssr, 66, 105

\bibitem[{{Peimbert}(1967)}]{Peimbert67}
{Peimbert}, M. 1967, \apj, 150, 825

\bibitem[{{Pejcha} \& {Thompson}(2015)}]{PejchaThompson15}
{Pejcha}, O. \& {Thompson}, T.~A. 2015, \apj, 801, 90

\bibitem[{Pe{\~{n} }a {et~al.}(2007)Pe{\~{n} }a, Stasi{\'{n}}ska, \&
  Richer}]{Pena07}
Pe{\~{n} }a, M., Stasi{\'{n}}ska, G., \& Richer, M.~G. 2007, \aap, 476, 745

\bibitem[{{Pillepich} {et~al.}(2018){Pillepich}, {Springel}, {Nelson}, {Genel},
  {Naiman}, {Pakmor}, {Hernquist}, {Torrey}, {Vogelsberger}, {Weinberger}, \&
  {Marinacci}}]{Pillepich18}
{Pillepich}, A., {Springel}, V., {Nelson}, D., {et~al.} 2018, \mnras, 473, 4077

\bibitem[{{Popesso} {et~al.}(2023){Popesso}, {Concas}, {Cresci}, {Belli},
  {Rodighiero}, {Inami}, {Dickinson}, {Ilbert}, {Pannella}, \&
  {Elbaz}}]{Popesso23}
{Popesso}, P., {Concas}, A., {Cresci}, G., {et~al.} 2023, \mnras, 519, 1526

\bibitem[{{Przybilla} {et~al.}(2006){Przybilla}, {Butler}, {Becker}, \&
  {Kudritzki}}]{Przybilla06}
{Przybilla}, N., {Butler}, K., {Becker}, S.~R., \& {Kudritzki}, R.~P. 2006,
  \aap, 445, 1099

\bibitem[{{Rajamuthukumar} {et~al.}(2023){Rajamuthukumar}, {Hamers},
  {Neunteufel}, {Pakmor}, \& {de Mink}}]{Rajamuthukumar23}
{Rajamuthukumar}, A.~S., {Hamers}, A.~S., {Neunteufel}, P., {Pakmor}, R., \&
  {de Mink}, S.~E. 2023, \apj, 950, 9

\bibitem[{{Rix} {et~al.}(2004){Rix}, {Pettini}, {Leitherer}, {Bresolin},
  {Kudritzki}, \& {Steidel}}]{Rix04}
{Rix}, S.~A., {Pettini}, M., {Leitherer}, C., {et~al.} 2004, \apj, 615, 98

\bibitem[{{Rodr{\'\i}guez} \& {Rubin}(2005)}]{Rodriguez05}
{Rodr{\'\i}guez}, M. \& {Rubin}, R.~H. 2005, \apj, 626, 900

\bibitem[{{Rodr{\'\i}guez} {et~al.}(2022){Rodr{\'\i}guez}, {Maoz}, \&
  {Nakar}}]{Rodriguez22}
{Rodr{\'\i}guez}, {\'O}., {Maoz}, D., \& {Nakar}, E. 2022, arXiv e-prints,
  arXiv:2209.05552

\bibitem[{{Rodr{\'\i}guez} {et~al.}(2021){Rodr{\'\i}guez}, {Meza},
  {Pineda-Garc{\'\i}a}, \& {Ramirez}}]{Rodriguez21}
{Rodr{\'\i}guez}, {\'O}., {Meza}, N., {Pineda-Garc{\'\i}a}, J., \& {Ramirez},
  M. 2021, \mnras, 505, 1742

\bibitem[{{Rolleston} {et~al.}(2002){Rolleston}, {Trundle}, \&
  {Dufton}}]{Rolleston02}
{Rolleston}, W.~R.~J., {Trundle}, C., \& {Dufton}, P.~L. 2002, \aap, 396, 53

\bibitem[{Roman-Duval {et~al.}(2021)Roman-Duval, Jenkins, Tchernyshyov,
  Williams, Clark, Gordon, Meixner, Hagen, Peek, Sandstrom, Werk, \&
  Merica-Jones}]{Roman-Duval2021}
Roman-Duval, J., Jenkins, E.~B., Tchernyshyov, K., {et~al.} 2021, The
  Astrophysical Journal, 910, 95

\bibitem[{{Runco} {et~al.}(2021){Runco}, {Shapley}, {Sanders}, {Topping},
  {Kriek}, {Reddy}, {Coil}, {Mobasher}, {Siana}, {Freeman}, {Shivaei}, {Azadi},
  {Price}, {Leung}, {Fetherolf}, {de Groot}, {Zick}, {Fornasini}, \&
  {Barro}}]{Runco21}
{Runco}, J.~N., {Shapley}, A.~E., {Sanders}, R.~L., {et~al.} 2021, \mnras, 502,
  2600

\bibitem[{{Russell} \& {Dopita}(1992)}]{RusselDopita92}
{Russell}, S.~C. \& {Dopita}, M.~A. 1992, \apj, 384, 508

\bibitem[{{Salim} {et~al.}(2007){Salim}, {Rich}, {Charlot}, {Brinchmann},
  {Johnson}, {Schiminovich}, {Seibert}, {Mallery}, {Heckman}, {Forster},
  {Friedman}, {Martin}, {Morrissey}, {Neff}, {Small}, {Wyder}, {Bianchi},
  {Donas}, {Lee}, {Madore}, {Milliard}, {Szalay}, {Welsh}, \& {Yi}}]{Salim07}
{Salim}, S., {Rich}, R.~M., {Charlot}, S., {et~al.} 2007, \apjs, 173, 267

\bibitem[{{S{\'a}nchez} {et~al.}(2014){S{\'a}nchez}, {Rosales-Ortega},
  {Iglesias-P{\'a}ramo}, {Moll{\'a}}, {Barrera-Ballesteros}, {Marino},
  {P{\'e}rez}, {S{\'a}nchez-Blazquez}, {Gonz{\'a}lez Delgado}, {Cid Fernandes},
  {de Lorenzo-C{\'a}ceres}, {Mendez-Abreu}, {Galbany}, {Falcon-Barroso},
  {Miralles-Caballero}, {Husemann}, {Garc{\'\i}a-Benito}, {Mast}, {Walcher},
  {Gil de Paz}, {Garc{\'\i}a-Lorenzo}, {Jungwiert}, {V{\'\i}lchez},
  {J{\'\i}lkov{\'a}}, {Lyubenova}, {Cortijo-Ferrero}, {D{\'\i}az}, {Wisotzki},
  {M{\'a}rquez}, {Bland-Hawthorn}, {Ellis}, {van de Ven}, {Jahnke},
  {Papaderos}, {Gomes}, {Mendoza}, \& {L{\'o}pez-S{\'a}nchez}}]{Sanchez14}
{S{\'a}nchez}, S.~F., {Rosales-Ortega}, F.~F., {Iglesias-P{\'a}ramo}, J.,
  {et~al.} 2014, \aap, 563, A49

\bibitem[{{Sander} \& {Vink}(2020)}]{Sander20}
{Sander}, A. A.~C. \& {Vink}, J.~S. 2020, \mnras, 499, 873

\bibitem[{{Sanders} {et~al.}(2020){Sanders}, {Shapley}, {Reddy}, {Kriek},
  {Siana}, {Coil}, {Mobasher}, {Shivaei}, {Freeman}, {Azadi}, {Price}, {Leung},
  {Fetherolf}, {de Groot}, {Zick}, {Fornasini}, \& {Barro}}]{Sanders20}
{Sanders}, R.~L., {Shapley}, A.~E., {Reddy}, N.~A., {et~al.} 2020, \mnras, 491,
  1427

\bibitem[{{Sawada} \& {Suwa}(2023)}]{Sawada23}
{Sawada}, R. \& {Suwa}, Y. 2023, arXiv e-prints, arXiv:2301.03610

\bibitem[{{Schaye} {et~al.}(2015){Schaye}, {Crain}, {Bower}, {Furlong},
  {Schaller}, {Theuns}, {Dalla Vecchia}, {Frenk}, {McCarthy}, {Helly},
  {Jenkins}, {Rosas-Guevara}, {White}, {Baes}, {Booth}, {Camps}, {Navarro},
  {Qu}, {Rahmati}, {Sawala}, {Thomas}, \& {Trayford}}]{Schaye15}
{Schaye}, J., {Crain}, R.~A., {Bower}, R.~G., {et~al.} 2015, \mnras, 446, 521

\bibitem[{{Schneider} {et~al.}(2021){Schneider}, {Podsiadlowski}, \&
  {M{\"u}ller}}]{Schneider21}
{Schneider}, F.~R.~N., {Podsiadlowski}, P., \& {M{\"u}ller}, B. 2021, \aap,
  645, A5

\bibitem[{{Sch{\"o}nrich} \& {Binney}(2009)}]{SchonrichBinney09}
{Sch{\"o}nrich}, R. \& {Binney}, J. 2009, \mnras, 396, 203

\bibitem[{{Senchyna} {et~al.}(2022){Senchyna}, {Stark}, {Charlot}, {Plat},
  {Chevallard}, {Chen}, {Jones}, {Sanders}, {Rudie}, {Cooper}, \&
  {Bruzual}}]{Senchyna22}
{Senchyna}, P., {Stark}, D.~P., {Charlot}, S., {et~al.} 2022, \apj, 930, 105

\bibitem[{{Senchyna} {et~al.}(2019){Senchyna}, {Stark}, {Chevallard},
  {Charlot}, {Jones}, \& {Vidal-Garc{\'\i}a}}]{Senchyna19}
{Senchyna}, P., {Stark}, D.~P., {Chevallard}, J., {et~al.} 2019, \mnras, 488,
  3492

\bibitem[{{Senchyna} {et~al.}(2017){Senchyna}, {Stark}, {Vidal-Garc{\'\i}a},
  {Chevallard}, {Charlot}, {Mainali}, {Jones}, {Wofford}, {Feltre}, \&
  {Gutkin}}]{Senchyna17}
{Senchyna}, P., {Stark}, D.~P., {Vidal-Garc{\'\i}a}, A., {et~al.} 2017, \mnras,
  472, 2608

\bibitem[{{Sharma} {et~al.}(2018){Sharma}, {Stello}, {Buder}, {Kos},
  {Bland-Hawthorn}, {Asplund}, {Duong}, {Lin}, {Lind}, {Ness}, {Huber},
  {Zwitter}, {Traven}, {Hon}, {Kafle}, {Khanna}, {Saddon}, {Anguiano}, {Casey},
  {Freeman}, {Martell}, {De Silva}, {Simpson}, {Wittenmyer}, \&
  {Zucker}}]{Sharma18}
{Sharma}, S., {Stello}, D., {Buder}, S., {et~al.} 2018, \mnras, 473, 2004

\bibitem[{{Skibba} {et~al.}(2012){Skibba}, {Engelbracht}, {Aniano}, {Babler},
  {Bernard}, {Bot}, {Carlson}, {Galametz}, {Galliano}, {Gordon}, {Hony},
  {Israel}, {Lebouteiller}, {Li}, {Madden}, {Meixner}, {Misselt}, {Montiel},
  {Okumura}, {Panuzzo}, {Paradis}, {Roman-Duval}, {Rubio}, {Sauvage}, {Seale},
  {Srinivasan}, \& {van Loon}}]{Skibba12}
{Skibba}, R.~A., {Engelbracht}, C.~W., {Aniano}, G., {et~al.} 2012, \apj, 761,
  42

\bibitem[{{Sollima} {et~al.}(2022){Sollima}, {D'Orazi}, {Gratton}, {Carini},
  {Carretta}, {Bragaglia}, \& {Lucatello}}]{Sollima22}
{Sollima}, A., {D'Orazi}, V., {Gratton}, R., {et~al.} 2022, \aap, 661, A69

\bibitem[{{Sommariva} {et~al.}(2012){Sommariva}, {Mannucci}, {Cresci},
  {Maiolino}, {Marconi}, {Nagao}, {Baroni}, \& {Grazian}}]{Sommariva12}
{Sommariva}, V., {Mannucci}, F., {Cresci}, G., {et~al.} 2012, \aap, 539, A136

\bibitem[{{Song} {et~al.}(2021){Song}, {Mateo}, {Bailey}, {Walker}, {Roederer},
  {Olszewski}, {Reiter}, \& {Kremin}}]{Song21}
{Song}, Y.-Y., {Mateo}, M., {Bailey}, J.~I., {et~al.} 2021, \mnras, 504, 4160

\bibitem[{{Speagle} {et~al.}(2014){Speagle}, {Steinhardt}, {Capak}, \&
  {Silverman}}]{Speagle14}
{Speagle}, J.~S., {Steinhardt}, C.~L., {Capak}, P.~L., \& {Silverman}, J.~D.
  2014, \apjs, 214, 15

\bibitem[{{Spina} {et~al.}(2022){Spina}, {Magrini}, \& {Cunha}}]{Spina22}
{Spina}, L., {Magrini}, L., \& {Cunha}, K. 2022, Universe, 8, 87

\bibitem[{{Stanway} \& {Eldridge}(2018)}]{StanwayEldridge18}
{Stanway}, E.~R. \& {Eldridge}, J.~J. 2018, \mnras, 479, 75

\bibitem[{{Stanway} {et~al.}(2016){Stanway}, {Eldridge}, \&
  {Becker}}]{Stanway16}
{Stanway}, E.~R., {Eldridge}, J.~J., \& {Becker}, G.~D. 2016, \mnras, 456, 485

\bibitem[{{Stasi{\'n}ska} \& {Izotov}(2003)}]{StasinskaIzotov03}
{Stasi{\'n}ska}, G. \& {Izotov}, Y. 2003, \aap, 397, 71

\bibitem[{{Steidel} {et~al.}(2016){Steidel}, {Strom}, {Pettini}, {Rudie},
  {Reddy}, \& {Trainor}}]{Steidel16}
{Steidel}, C.~C., {Strom}, A.~L., {Pettini}, M., {et~al.} 2016, \apj, 826, 159

\bibitem[{{Strolger} {et~al.}(2020){Strolger}, {Rodney}, {Pacifici}, {Narayan},
  \& {Graur}}]{Strolger20}
{Strolger}, L.-G., {Rodney}, S.~A., {Pacifici}, C., {Narayan}, G., \& {Graur},
  O. 2020, \apj, 890, 140

\bibitem[{{Strom} {et~al.}(2022){Strom}, {Rudie}, {Steidel}, \&
  {Trainor}}]{Strom22}
{Strom}, A.~L., {Rudie}, G.~C., {Steidel}, C.~C., \& {Trainor}, R.~F. 2022,
  \apj, 925, 116

\bibitem[{{Strom} {et~al.}(2018){Strom}, {Steidel}, {Rudie}, {Trainor}, \&
  {Pettini}}]{Strom18}
{Strom}, A.~L., {Steidel}, C.~C., {Rudie}, G.~C., {Trainor}, R.~F., \&
  {Pettini}, M. 2018, \apj, 868, 117

\bibitem[{{Sukhbold} {et~al.}(2016){Sukhbold}, {Ertl}, {Woosley}, {Brown}, \&
  {Janka}}]{Sukhbold16}
{Sukhbold}, T., {Ertl}, T., {Woosley}, S.~E., {Brown}, J.~M., \& {Janka}, H.~T.
  2016, \apj, 821, 38

\bibitem[{{Takahashi} {et~al.}(2018){Takahashi}, {Yoshida}, \&
  {Umeda}}]{Takahashi18}
{Takahashi}, K., {Yoshida}, T., \& {Umeda}, H. 2018, \apj, 857, 111

\bibitem[{{Telford} {et~al.}(2016){Telford}, {Dalcanton}, {Skillman}, \&
  {Conroy}}]{Telford16}
{Telford}, O.~G., {Dalcanton}, J.~J., {Skillman}, E.~D., \& {Conroy}, C. 2016,
  \apj, 827, 35

\bibitem[{{Thomas} {et~al.}(2010){Thomas}, {Maraston}, {Schawinski}, {Sarzi},
  \& {Silk}}]{Thomas10}
{Thomas}, D., {Maraston}, C., {Schawinski}, K., {Sarzi}, M., \& {Silk}, J.
  2010, \mnras, 404, 1775

\bibitem[{{Thuan} {et~al.}(2022){Thuan}, {Guseva}, \& {Izotov}}]{Thuan22}
{Thuan}, T.~X., {Guseva}, N.~G., \& {Izotov}, Y.~I. 2022, \mnras, 516, L81

\bibitem[{{Tolstoy} {et~al.}(2009){Tolstoy}, {Hill}, \& {Tosi}}]{Tolstoy09}
{Tolstoy}, E., {Hill}, V., \& {Tosi}, M. 2009, \araa, 47, 371

\bibitem[{{Tominaga} {et~al.}(2007){Tominaga}, {Umeda}, \&
  {Nomoto}}]{Tominaga07}
{Tominaga}, N., {Umeda}, H., \& {Nomoto}, K. 2007, \apj, 660, 516

\bibitem[{{Toonen} {et~al.}(2012){Toonen}, {Nelemans}, \& {Portegies
  Zwart}}]{Toonen12}
{Toonen}, S., {Nelemans}, G., \& {Portegies Zwart}, S. 2012, \aap, 546, A70

\bibitem[{{Topping} {et~al.}(2020){Topping}, {Shapley}, {Reddy}, {Sanders},
  {Coil}, {Kriek}, {Mobasher}, \& {Siana}}]{Topping20}
{Topping}, M.~W., {Shapley}, A.~E., {Reddy}, N.~A., {et~al.} 2020, \mnras, 495,
  4430

\bibitem[{{Tremonti} {et~al.}(2004){Tremonti}, {Heckman}, {Kauffmann},
  {Brinchmann}, {Charlot}, {White}, {Seibert}, {Peng}, {Schlegel}, {Uomoto},
  {Fukugita}, \& {Brinkmann}}]{Tremonti04}
{Tremonti}, C.~A., {Heckman}, T.~M., {Kauffmann}, G., {et~al.} 2004, \apj, 613,
  898

\bibitem[{{Urbaneja} {et~al.}(2008){Urbaneja}, {Kudritzki}, {Bresolin},
  {Przybilla}, {Gieren}, \& {Pietrzy{\'n}ski}}]{Urbaneja08}
{Urbaneja}, M.~A., {Kudritzki}, R.-P., {Bresolin}, F., {et~al.} 2008, \apj,
  684, 118

\bibitem[{{Vink}(2022)}]{Vink22}
{Vink}, J.~S. 2022, \araa, 60, 203

\bibitem[{{Vink} \& {de Koter}(2005{\natexlab{a}})}]{Vink05}
{Vink}, J.~S. \& {de Koter}, A. 2005{\natexlab{a}}, \aap, 442, 587

\bibitem[{{Vink} \& {de Koter}(2005{\natexlab{b}})}]{VinkDeKoter05}
{Vink}, J.~S. \& {de Koter}, A. 2005{\natexlab{b}}, \aap, 442, 587

\bibitem[{{Vink} {et~al.}(2001){Vink}, {de Koter}, \& {Lamers}}]{Vink01}
{Vink}, J.~S., {de Koter}, A., \& {Lamers}, H.~J.~G.~L.~M. 2001, \aap, 369, 574

\bibitem[{{Vink} {et~al.}(2023){Vink}, {Mehner}, {Crowther}, {Fullerton},
  {Garcia}, {Martins}, {Morrell}, {Oskinova}, {St-Louis}, {ud-Doula}, {Sander},
  {Sana}, {Bouret}, {Kubatova}, {Marchant}, {Martins}, {Wofford}, {van Loon},
  {Telford}, {Gotberg}, {Bowman}, {Erba}, {Kalari}, {Abdul-Masih}, {Alkousa},
  {Backs}, {Barbosa}, {Berlanas}, {Bernini-Peron}, {Bestenlehner}, {Blomme},
  {Bodensteiner}, {Brands}, {Evans}, {David-Uraz}, {Driessen}, {Dsilva},
  {Geen}, {Gomez-Gonzalez}, {Grassitelli}, {Hamann}, {Hawcroft}, {Herrero},
  {Higgins}, {Hillier}, {Ignace}, {Istrate}, {Kaper}, {Kee}, {Kehrig},
  {Keszthelyi}, {Klencki}, {de Koter}, {Kuiper}, {Laplace}, {Larkin},
  {Lefever}, {Leitherer}, {Lennon}, {Mahy}, {Maiz Apellaniz}, {Maravelias},
  {Marcolino}, {McLeod}, {de Mink}, {Najarro}, {Oey}, {Parsons}, {Pauli},
  {Pedersen}, {Prinja}, {Ramachandran}, {Ramirez-Tannus}, {Sabhahit},
  {Schootemeijer}, {Reyero Serantes}, {Shenar}, {Stringfellow}, {Sudnik},
  {Tramper}, \& {Wang}}]{Vink23}
{Vink}, J.~S., {Mehner}, A., {Crowther}, P.~A., {et~al.} 2023, arXiv e-prints,
  arXiv:2305.06376

\bibitem[{{Wang} \& {Han}(2012)}]{WangHan12}
{Wang}, B. \& {Han}, Z. 2012, \nar, 56, 122

\bibitem[{{Webbink}(1984)}]{Webbink84}
{Webbink}, R.~F. 1984, \apj, 277, 355

\bibitem[{{Weidner} \& {Kroupa}(2005)}]{WeidnerKroupa05}
{Weidner}, C. \& {Kroupa}, P. 2005, \apj, 625, 754

\bibitem[{{Weinmann} {et~al.}(2011){Weinmann}, {Neistein}, \&
  {Dekel}}]{Weinmann11}
{Weinmann}, S.~M., {Neistein}, E., \& {Dekel}, A. 2011, \mnras, 417, 2737

\bibitem[{{Weisz} {et~al.}(2011){Weisz}, {Dalcanton}, {Williams}, {Gilbert},
  {Skillman}, {Seth}, {Dolphin}, {McQuinn}, {Gogarten}, {Holtzman}, {Rosema},
  {Cole}, {Karachentsev}, \& {Zaritsky}}]{Weisz11}
{Weisz}, D.~R., {Dalcanton}, J.~J., {Williams}, B.~F., {et~al.} 2011, \apj,
  739, 5

\bibitem[{{Wheeler} {et~al.}(1989){Wheeler}, {Sneden}, \&
  {Truran}}]{WheelerSnedenTruran89}
{Wheeler}, J.~C., {Sneden}, C., \& {Truran}, Jr., J.~W. 1989, \araa, 27, 279

\bibitem[{{Whelan} \& {Iben}(1973)}]{Whelan73}
{Whelan}, J. \& {Iben}, Icko, J. 1973, \apj, 186, 1007

\bibitem[{{Wofford} {et~al.}(2021){Wofford}, {Vidal-Garc{\'\i}a}, {Feltre},
  {Chevallard}, {Charlot}, {Stark}, {Herenz}, \& {Hayes}}]{Wofford21}
{Wofford}, A., {Vidal-Garc{\'\i}a}, A., {Feltre}, A., {et~al.} 2021, \mnras,
  500, 2908

\bibitem[{{Woo} {et~al.}(2008){Woo}, {Courteau}, \& {Dekel}}]{Woo08}
{Woo}, J., {Courteau}, S., \& {Dekel}, A. 2008, \mnras, 390, 1453

\bibitem[{{Woosley}(2017)}]{Woosley17}
{Woosley}, S.~E. 2017, \apj, 836, 244

\bibitem[{{Woosley} \& {Heger}(2007)}]{WoosleyHeger07}
{Woosley}, S.~E. \& {Heger}, A. 2007, \physrep, 442, 269

\bibitem[{{Woosley} {et~al.}(2002){Woosley}, {Heger}, \& {Weaver}}]{Woosley02}
{Woosley}, S.~E., {Heger}, A., \& {Weaver}, T.~A. 2002, Reviews of Modern
  Physics, 74, 1015

\bibitem[{{Xiao} {et~al.}(2018){Xiao}, {Stanway}, \& {Eldridge}}]{Xiao18}
{Xiao}, L., {Stanway}, E.~R., \& {Eldridge}, J.~J. 2018, \mnras, 477, 904

\bibitem[{{Xing} {et~al.}(2023){Xing}, {Zhao}, {Liu}, {Heger}, {Han}, {Aoki},
  {Chen}, {Ishigaki}, {Li}, \& {Zhao}}]{Xing23}
{Xing}, Q.-F., {Zhao}, G., {Liu}, Z.-W., {et~al.} 2023, \nat, 618, 712

\bibitem[{{Zapartas} {et~al.}(2017){Zapartas}, {de Mink}, {Izzard}, {Yoon},
  {Badenes}, {G{\"o}tberg}, {de Koter}, {Neijssel}, {Renzo}, {Schootemeijer},
  \& {Shrotriya}}]{Zapartas17}
{Zapartas}, E., {de Mink}, S.~E., {Izzard}, R.~G., {et~al.} 2017, \aap, 601,
  A29

\bibitem[{{Zhang} \& {Zhao}(2005)}]{ZhangZhao05}
{Zhang}, H.~W. \& {Zhao}, G. 2005, \mnras, 364, 712

\bibitem[{{Zhou} {et~al.}(2021){Zhou}, {Shi}, {Zhang}, \& {Wang}}]{Zhou21}
{Zhou}, L., {Shi}, Y., {Zhang}, Z.-Y., \& {Wang}, J. 2021, \aap, 653, L10

\end{thebibliography}

\end{appendix}

\end{document}